\documentclass[journal=jpccck,manuscript=article,layout=onecolumn,
               super=false]{achemso}
\makeatletter
\input{size10.clo}
\makeatother
\setcitestyle{numbers,square,comma}
\SectionNumbersOn

\usepackage[T1]{fontenc}
\usepackage{amsmath}

\usepackage{newtxtext}
\usepackage{newtxmath}
\usepackage[protrusion=true,expansion=true,tracking=false]{microtype}

\usepackage{geometry}
\usepackage{graphicx}
\usepackage{booktabs}
\usepackage{array}
\usepackage{siunitx}
\usepackage{enumitem}
\DeclareSIUnit{\atmosphere}{atm}
\DeclareSIUnit{\kcal}{kcal}

\usepackage{setspace}
\usepackage{float}
\usepackage{caption}
\AtBeginDocument{%
  \floatstyle{plain}%
  \restylefloat{table}%
  \restylefloat{figure}%
  \captionsetup{labelfont=bf, textfont={md,up}, font=small,
                labelsep=period, justification=justified}%
  \captionsetup[table]{labelfont=bf, textfont={md,up}}%
  \captionsetup[figure]{format=ruled}%
  \setlength{\parindent}{0pt}%
  \setlist{topsep=0.35\baselineskip, partopsep=0pt, parsep=0pt,
           itemsep=0.15\baselineskip}%
}

\makeatletter
\AtBeginDocument{%
  \renewcommand\section{\@startsection{section}{1}{\z@}%
    {-1.6ex \@plus -0.4ex \@minus -0.2ex}{0.6ex \@plus 0.2ex}{\normalfont\large\bfseries}}%
  \renewcommand\subsection{\@startsection{subsection}{2}{\z@}%
    {-1.2ex \@plus -0.3ex \@minus -0.2ex}{0.4ex \@plus 0.1ex}{\normalfont\normalsize\bfseries}}%
}
\makeatother
\DeclareCaptionFormat{ruled}{#1#2#3\par\vspace{3pt}\noindent\rule{\linewidth}{0.4pt}}

\usepackage{etoolbox}
\usepackage{multicol}
\makeatletter
\AtBeginDocument{%
  \patchcmd{\@maketitle}{\newpage}{}{}{\typeout{>>> PATCH FAILED: @maketitle newpage}}%
  \pretocmd{\@maketitle}{\singlespacing}{}{\typeout{>>> PATCH FAILED: title spacing}}%
  \patchcmd{\maketitle}{\newpage}{}{}{\typeout{>>> PATCH FAILED: maketitle newpage 1}}%
  \patchcmd{\maketitle}{\newpage}{}{}{\typeout{>>> PATCH FAILED: maketitle newpage 2}}%
}
\makeatother

\author{Jnanadyuti Patra}
\affiliation{Department of Chemical Engineering, National Institute of
Technology Tiruchirappalli, Tiruchirappalli 620015, Tamil Nadu, India}
\author{Shiladitya Sengupta}
\affiliation{Department of Physics, Indian Institute of Technology Roorkee, Roorkee 247667, Uttarakhand, India}
\author{Nandlal Pingua}
\email{nandlal@nitt.edu}
\affiliation{Department of Chemical Engineering, National Institute of
Technology Tiruchirappalli, Tiruchirappalli 620015, Tamil Nadu, India}

\title{Phase-Offset Rectification of Thermo-osmotic Flow in a Structurally
Symmetric Graphene Nanochannel}

\keywords{thermo-osmosis, nanofluidics, graphene, molecular dynamics,
Brownian ratchet, interfacial transport, hydrogen bonding}

\begin{document}

\begin{abstract}
\singlespacing
\hspace*{-\itemindent}%
Thermo-osmotic flow around a closed periodic channel cancels exactly when the
confining surface is homogeneous, and nanofluidic thermal ratchets reported so far
break it with a structural asymmetry. We test whether a
phase offset $\delta$ between two individually symmetric fields suffices: a
sinusoidal hydroxyl coverage on graphene and a wall-temperature modulation of the
same period $\lambda$. Nonequilibrium molecular dynamics (172 runs,
\SI{3825}{\nano\second}) gives a loop velocity sinusoidal in $\delta$, null controls
consistent with zero, and a flow direction that reverses between
$\delta = \lambda/4$ and $3\lambda/4$ in 13 of 15 independent hydroxyl arrangements
($p = 0.001$). With uniform slip and the wall temperature as the drive, a single
rescaling of the linear-response thermo-osmotic coefficient, by $63 \pm 8$,
reproduces all 24 state points across four independent sweeps
($\chi^2 = 12.3$ for 22 degrees of freedom). The measured
local slip, tens of nanometres on pristine graphene and none from 10\%
coverage, reverses the predicted direction even in exact hydrodynamics. Uniform
walls at the coverages the least-covered strips carry lose 96 to 99.7\% of that
slip, and with the measured local slip the exact flow still reverses, by a factor
of 7.6; slip amplification of the drive could restore the direction, but the period
sweep favours uniform slip, leaving its origin open. Over
hydroxyl patches, stronger binding of interfacial water is nearly cancelled by its
lower density, consistent with the reduced coefficient.
\end{abstract}

\vspace{0.6em}
{\singlespacing\small\noindent
\textbf{Keywords:} thermo-osmosis; nanofluidics; graphene; molecular
dynamics; Brownian ratchet; interfacial transport; hydrogen bonding\par}

\singlespacing
\begin{multicols}{2}
\setlength{\parskip}{0.55\baselineskip}
\raggedcolumns

\section{Introduction}

Fluid transport at the nanoscale departs from continuum hydrodynamics in ways
that have become quantitatively accessible only in the last two
decades~\cite{Bocquet2010,BocquetTabeling2014,Kavokine2021}. Measured flow rates
through carbon nanotubes exceed the no-slip Hagen--Poiseuille prediction by
orders of magnitude~\cite{Majumder2005,Holt2006,Secchi2016}, a discrepancy traced
to the exceptionally low friction of water on graphitic
carbon~\cite{Falk2010,Thomas2008,Kannam2013} and to the breakdown of a local
viscosity description within a few molecular diameters of the
wall~\cite{Bonthuis2013}. The fabrication of channels with atomic-scale
dimensional control~\cite{Radha2016} and of graphene-oxide laminates with
sub-nanometre interlayer spacing~\cite{Nair2012,Joshi2014} has turned these
effects from curiosities into engineering variables, with applications proposed
in desalination~\cite{CohenTanugi2012,Surwade2015,Elimelech2011}, osmotic energy
conversion~\cite{Siria2013,Feng2016,Marbach2019} and thermal management.

Among the driving forces available at this scale, a temperature gradient applied
\emph{along} a solid--liquid interface is attractive because it requires neither
electrodes nor moving parts and can in principle be powered by waste heat. The
resulting flow, thermo-osmosis, originates in the excess enthalpy of the
interfacial fluid layer, and the associated transport coefficient may be
obtained either from equilibrium correlation
functions~\cite{Green1954,Kubo1957,Ganti2017,Anzini2025} or from direct
non-equilibrium simulation~\cite{Fu2018,Ouadfel2024}. Its magnitude in carbon
nanochannels is large enough to be of practical
interest~\cite{Fu2018,Chen2019}, and the same interfacial physics underlies
thermophoresis of suspended colloids and
macromolecules~\cite{Wurger2010,Piazza2008,Duhr2006}.

A closed channel, however, is subject to a strong constraint. Ouadfel \emph{et
al.}\ showed that in a periodic channel with a homogeneous surface, conservation
of flow rate forces the thermo-osmotic and Poiseuille contributions to cancel
exactly, so the net flow vanishes identically regardless of the imposed
gradient~\cite{Ouadfel2024}. The same conclusion follows from the closed-channel
analysis of Bessega \emph{et al.}~\cite{Bessega2026}. Any device that pumps around
a loop must therefore break that cancellation, and the literature is consistent
in how it does so: by building an object that is not symmetric. Conical and
composite nanopores exploit a shape gradient~\cite{Sahebi2022}; an off-centre heat
source exploits an asymmetric thermal boundary condition~\cite{Wang2020,Tsuji2026}; a
stiffness gradient along a nanotube exploits a mechanical
one~\cite{ChenStiff2019}; heterogeneous channels exploit two dissimilar
walls~\cite{Herrero2022}; Janus channels exploit a surface-energy
contrast~\cite{Qi2025}; and the periodic-loop simulations of
ref.~\cite{Ouadfel2024} themselves use an axial two-region chemical
asymmetry. In every case the flow direction is a property of the object, and
reversing the object reverses the flow.

This is precisely the logic of the classical Brownian
ratchet~\cite{Reimann2002,Hanggi2009,Astumian1997,Julicher1997}, in which a
spatially asymmetric potential rectifies unbiased fluctuations. The
B\"{u}ttiker--Landauer construction is the exception that motivates the present
work: there, directed motion arises not from an asymmetric potential but from a
\emph{spatial offset} between a potential landscape and a temperature
landscape~\cite{Buttiker1987,Landauer1988,Magnasco1993}, each of which may be
individually symmetric. That construction has been realised in working fluids at
the millimetre scale, where a periodic temperature field drives a net flow
through Marangoni stresses or through directed bubble growth in nucleate
boiling~\cite{Sugioka2022a,Sugioka2022b,Sugioka2024}. Those realisations are
macroscopic, and the flow they rectify is an interfacial or two-phase one; a
granular motor has since confirmed B\"{u}ttiker's quantitative prediction with
macroscopic particles rather than a fluid~\cite{Devauchelle2026}. At the
nanoscale, a phase offset between two periodic patterns, a surface-charge
modulation and a wall undulation, has been shown to rectify ion transport in a
nanochannel~\cite{Petersen2026}; there the flow is driven by an applied pressure
or field, and the offset selects the response rather than supplying the drive.
Transposing the idea to a single-phase continuum fluid in a nanochannel, where
thermo-osmosis is the response available and the offset is the only asymmetry
present, is the question this paper addresses.

Concretely, consider two fields imposed on the same channel: a sinusoidal
chemical patterning of the wall, and a sinusoidal modulation of the wall
temperature at the same period. Neither field possesses a handedness: each is
invariant under reflection about its own maxima. Their \emph{relative phase}
$\delta$ is not: displacing one field axially relative to the other selects a
direction. If a steady flow appears it cannot be attributed to the geometry,
because the geometry is mirror-symmetric throughout; it must be carried by
$\delta$ alone, and must reverse when $\delta \rightarrow -\delta$.

We report non-equilibrium molecular dynamics simulations testing this
proposition in a hydroxyl-functionalised graphene slit. The predictions
and falsification conditions were committed to version control before any
production run began, and the flux estimators with their unit tests before any
data existed; the outcome is reported against those conditions as written, and
the analyses devised afterwards are identified as such
(Section~\ref{sec:prereg}). Four results follow.
The flow is sinusoidal in the offset and reverses direction with it in thirteen
of fifteen independent hydroxyl arrangements. Its direction is opposite to the
one predicted with the slip lengths assumed before the campaign, and no measured
input, the slip measured at the coverages the pattern carries included, accounts
for it. With the measured slip of the patterned wall treated as uniform, a single
rescaling of the thermo-osmotic coefficient reconciles every sweep of the campaign
with the theory, and identifies the wall temperature as the field that drives the
flow.
And an analysis of the interfacial excess shows why the coefficient is small:
over the hydroxyls, stronger binding of water is compensated by lower density.

\section{Theory}

\subsection{Loop velocity in a patterned channel}

For a depth-averaged, laminar, periodic channel with a locally varying
thermo-osmotic coefficient $M(x)$ and hydraulic conductance $K(x)$, steady-state
mass conservation requires a uniform volumetric flow rate around the loop.
Balancing the local interfacial driving term against the local Poiseuille
response and averaging over one period gives
\begin{equation}
  v \;=\; -\,\frac{\bigl\langle M\,\partial_x T \,/\, (T K)\bigr\rangle}
                  {\bigl\langle K^{-1} \bigr\rangle},
  \label{eq:loop}
\end{equation}
where $\langle\cdot\rangle$ denotes an average over one period and $x$ is the
axial coordinate. Equation~\eqref{eq:loop} makes the cancellation theorem
transparent: for a homogeneous surface $M$ and $K$ are constants, the numerator
reduces to $M\langle \partial_x T / T\rangle / K$, and the average of
$\partial_x \ln T$ around a closed loop vanishes. No net flow is possible,
however large the local gradient is made~\cite{Ouadfel2024,Bessega2026}.

\subsection{Harmonic overlap and the selection rule}

Writing the response function $R \equiv M/(TK)$ and expanding both $R$ and the
temperature field in Fourier series over the period $L$ converts
eq.~\eqref{eq:loop} into an overlap sum,
\begin{equation}
  v \;=\; \bigl\langle K^{-1}\bigr\rangle^{-1}
          \sum_{n} \frac{\pi n}{L}\, A_n C_n
          \sin\!\left(\psi_n - \phi_n\right),
  \label{eq:overlap}
\end{equation}
in which $A_n$ and $\phi_n$ are the amplitude and phase of the $n$th harmonic of
the temperature field and $C_n$, $\psi_n$ those of $R$. A harmonic contributes
only if it is present in \emph{both} fields and if the two carry a nonzero phase
difference. This selection rule governs the design in three ways.

First, it confers robustness against thermostat artefacts. Localised
thermostatting zones are broadband sources: they inject power at every harmonic,
not only the intended one. Because the chemical pattern is deliberately
single-harmonic, contaminating the thermal field at $n = 1,2,3$ with amplitudes
comparable to the drive shifts the predicted flux by only 0.3\%. The chemical
pattern acts as a matched filter.

Second, it forbids an entire class of designs. An $N$-period chemical pattern
opposed to a single hot/cold pair populates $C_N$ and $A_1$, which do not
overlap; the sum vanishes identically by parity. An incommensurate arrangement
therefore cannot pump, however strong the gradient. This is not a peripheral
observation: it closed an earlier design for this study, and it furnishes
control C4 below, which is a test of the theory rather than of the apparatus.

Third, the explicit factor $\pi n / L$ predicts that at fixed box length the flux
grows in proportion to the number of periods. This prediction is logically
independent of the phase dependence, since it concerns the structure of
eq.~\eqref{eq:overlap} rather than the linearity of the underlying response,
and it is tested separately in Section~\ref{sec:teeth}.

\subsection{Primary hypothesis}

Specialising eq.~\eqref{eq:overlap} to matched single-harmonic fields displaced
by $\delta$ leaves one surviving term,
\begin{equation}
  v(\delta) \;=\; -\,v_{\max}\,\sin\!\left(\frac{2\pi\delta}{\lambda}\right),
  \label{eq:h1}
\end{equation}
with nodes at $\delta = 0$ and $\lambda/2$, extrema of opposite sign at
$\lambda/4$ and $3\lambda/4$, and odd symmetry about the origin. The hypothesis
is falsifiable in a strong sense: it constrains not a number but a functional
form across the whole range of $\delta$, with two enforced zeros and a sign
reversal. An artefact reproducing the entire curve would be a remarkable
artefact. The sign of $v_{\max}$ is not fixed by symmetry. Because
$R = M/(TK)$, a coverage that lowers both the thermo-osmotic coefficient and the
hydraulic conductance modulates $R$ by roughly the difference of the two, so the
direction of the flux at a given offset depends on which of them varies more
strongly across the pattern.

\section{Methods}

\subsection{Channel geometry}

\begin{figure*}[tp]
  \centering
  \includegraphics[width=\textwidth,alt={Two molecular renders. (a) A hydroxyl group grafted on an sp3 carbon of the inner graphene sheet, with a TIP4P/2005 water molecule donating a hydrogen bond to the hydroxyl oxygen; each atom type is labelled, and the massless charge site M of the water molecule is visible through a translucent oxygen. (b) One 10.1 nm chemical period of the simulation cell, centred on a hydroxyl-rich region and cut open at the mid-plane: the lower wall with its grafted hydroxyl groups in front, the water and the upper wall behind.}]{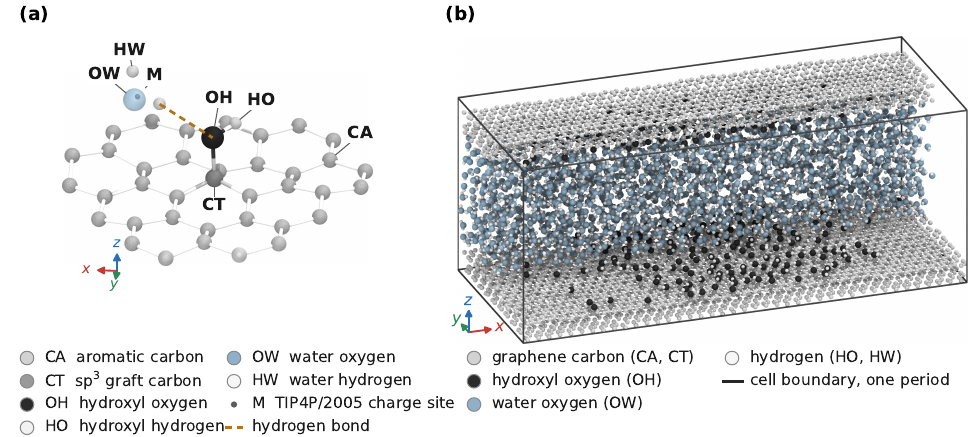}
  \caption{Atomistic model. (a) The building block, taken from the simulated
  structure at the end of production: an $\mathrm{sp}^3$ graft carbon (CT) bearing a
  hydroxyl group (OH, HO) in the inner graphene sheet (CA), and a TIP4P/2005 water
  molecule (OW, HW) donating a hydrogen bond to the hydroxyl oxygen. The negative
  charge of the water molecule sits on the massless site M,
  \SI{0.1546}{\angstrom} from the oxygen along the H--O--H bisector; the oxygen is
  drawn translucent so that M is visible. (b) One chemical period
  ($\lambda = \SI{10.1}{\nano\meter}$) of the $40.3 \times 4.26 \times
  \SI{4.01}{\nano\meter}$ cell, centred on a hydroxyl-rich region and cut open at
  the mid-plane in $y$: in front of the cut only the lower wall is shown, exposing
  its grafted hydroxyl groups; behind it are the water and the upper wall. Atoms
  are numbered by LAMMPS type as in Table~\ref{tab:ff}.}
  \label{fig:model}
\end{figure*}

The simulation cell is a fully periodic graphene slit of dimensions
$403.36 \times 42.60 \times 40.05$~\si{\angstrom} containing \num{68094} atoms:
\num{26240} carbon atoms in four sheets, one AB-stacked bilayer per wall,
\num{1328} hydroxyl groups grafted to the inner sheets, and \num{13066} water
molecules. The clear separation between inner surfaces is
$h = \SI{30}{\angstrom}$. Figure~\ref{fig:model} shows the building block of the
model and one chemical period of the cell.

The axial box length is not a free parameter. It is set to \num{164} zigzag
lattice repeats, the value nearest \SI{400}{\angstrom} that is simultaneously an
integer number of graphene repeats and divisible by the tooth count, so that all
four chemical periods are lattice-identical and the periodic image joins without
strain. A box merely close to commensurate produces spurious stress and four
subtly different teeth, which would confound a measurement whose entire signal is
a phase relationship. The resulting chemical period is $\lambda =
\SI{100.84}{\angstrom}$, giving $\lambda/h = 3.36$.

Hydroxyl coverage follows $\xi(x) = N_{\mathrm{O}}/N_{\mathrm{C}}$, varying
sinusoidally between 0 and 0.20 with mean 0.10. Graft sites are placed by seeded
rejection sampling that forbids occupation of nearest-neighbour carbon atoms,
reflecting the steric constraint on vicinal hydroxylation of a honeycomb
lattice~\cite{Dreyer2010,Lerf1998}. That constraint caps attainable coverage near
$\xi = 0.33$, a limit central to the interpretation in
Section~\ref{sec:discussion}. Grafted carbons are $\mathrm{sp}^3$ and displaced
out of plane. The cell is neutral to \SI{3.7e-14}{\elementarycharge}.

\subsection{Force field}
\label{sec:ff}

Water is represented by the rigid four-site TIP4P/2005 model~\cite{Abascal2005},
chosen because it reproduces the density maximum, the equation of state and the
transport coefficients of liquid water across a wide range of state points more
faithfully than three-site alternatives, and because its performance has been
documented systematically against
them~\cite{Vega2011,Jorgensen1983,Berendsen1987}. Since the observable here is an
interfacial enthalpy excess, a model misrepresenting the temperature dependence
of the hydrogen-bond network would bias the quantity of interest directly. The
negative charge resides on a massless M-site displaced \SI{0.1546}{\angstrom}
from the oxygen along the H--O--H bisector. In the LAMMPS implementation that
displacement is applied internally by the pair style, so the data file must
assign the M-site charge to the oxygen; leaving the oxygen neutral gives every
molecule a net charge of $+1.1128\,e$ and renders the electrostatics of the
entire cell meaningless.

The wall is described by OPLS-AA throughout, pristine and hydroxylated carbon
alike~\cite{Jorgensen1996}. A reactive or bond-order potential was considered and
rejected deliberately. Such potentials describe the phonon spectrum of pristine
graphene more faithfully, but they cannot represent the $\mathrm{sp}^3$ carbon at
a graft site, and the graft site is precisely where the interfacial enthalpy
excess driving $M$ is generated. Because the design imposes $T(x)$ through dense
thermostatting zones, the intrinsic thermal conductivity of the wall is largely
overridden and phonon fidelity buys little here. Atom types, charges and
Lennard-Jones parameters are collected in Table~\ref{tab:ff}, and each type is
identified in Figure~\ref{fig:model}a.

\begin{table*}[t]
\centering

\small
\begin{tabular}{lccccl}
\toprule
Type & Mass & $q$ & $\sigma$ & $\epsilon$ & Role \\
\midrule
CA & 12.011  & $ 0.0000$ & 3.5500 & 0.0700 & aromatic carbon \\
CT & 12.011  & $+0.1500$ & 3.5000 & 0.0660 & $\mathrm{sp}^3$ graft carbon \\
OH & 15.999  & $-0.5850$ & 3.1200 & 0.1700 & hydroxyl oxygen \\
HO &  1.008  & $+0.4350$ & 0.0000 & 0.0000 & hydroxyl hydrogen \\
OW & 15.9994 & $-1.1128$ & 3.1589 & 0.1852 & water oxygen, carries M-site charge \\
HW &  1.008  & $+0.5564$ & 0.0000 & 0.0000 & water hydrogen \\
\bottomrule
\end{tabular}

\vspace{0.6em}
\begin{tabular}{lrr@{\hspace{2.2em}}lrr}
\toprule
Bond & $K_r$ & $r_0$ & Angle & $K_\theta$ & $\theta_0$ \\
\midrule
C--C (sheet) & 469.0 & 1.420 & C--C--C & 63.0 & 120.00 \\
C--O & 320.0 & 1.410 & C--C--O & 50.0 & 109.50 \\
O--H, hydroxyl & rigid & 0.945 & C--O--H & 55.0 & 108.50 \\
O--H, water & rigid & 0.9572 & H--O--H, water & rigid & 104.52 \\
\bottomrule
\end{tabular}
\caption{Force-field parameters. Water is TIP4P/2005~\cite{Abascal2005}; the wall
is OPLS-AA~\cite{Jorgensen1996}. The water oxygen carries the M-site charge, as
the LAMMPS TIP4P implementation requires. Mass in g\,mol$^{-1}$, charge in
units of $e$, $\sigma$ in \AA, $\epsilon$ in kcal\,mol$^{-1}$. Bonded terms are
harmonic in the LAMMPS convention $E = K_r (r - r_0)^2$ and
$E = K_\theta(\theta - \theta_0)^2$, with $K_r$ in kcal\,mol$^{-1}$\,\AA$^{-2}$, $r_0$
in \AA, $K_\theta$ in kcal\,mol$^{-1}$\,rad$^{-2}$ and $\theta_0$ in degrees;
\emph{rigid} marks terms constrained by SHAKE. A harmonic improper term,
$K = \SI{2.5}{\kcal\per\mole}$\,rad$^{-2}$ about $0^{\circ}$, holds each sheet
carbon in the plane of its three neighbours.}
\label{tab:ff}
\end{table*}

Unlike-atom Lennard-Jones parameters follow the geometric mixing rule of
OPLS-AA, with one exception. The carbon--water-oxygen interaction, which sets the
wetting behaviour and hence both the slip length and the interfacial enthalpy,
uses $\epsilon_{\mathrm{CO}} = \SI{0.1140}{\kcal\per\mole}$, the geometric mean of
the carbon and oxygen well depths (0.1139) rounded to three figures, and $\sigma_{\mathrm{CO}} =
\SI{3.2800}{\angstrom}$, below the geometric mean of \SI{3.349}{\angstrom}, for both
carbon types. It is the parameter most consequential for the result and the one
most often adjusted in the literature to reproduce a target contact
angle~\cite{Werder2003}; its effect on the direction of the flow is tested in
Section~\ref{sec:direction} and remains among the limitations, while the
sensitivity of the interfacial excess to the
interaction cutoff is reported in Section~\ref{sec:validation}.

Dispersion and real-space electrostatics are truncated at \SI{12}{\angstrom} and
\SI{9}{\angstrom} respectively, with long-range electrostatics evaluated by a
particle--particle particle--mesh Ewald
solver~\cite{Hockney1988,Darden1993,Essmann1995} at a relative force tolerance of
\num{e-4}. No long-range dispersion tail correction is applied in the confined
cell, where the assumption of isotropic bulk density beyond the cutoff on which
such a correction relies is invalid; the bulk reference calculations of
Section~\ref{sec:validation} do apply it. The O--H bonds of water and of the
hydroxyl groups and the H--O--H angle of water are held rigid with
SHAKE~\cite{Ryckaert1977} at a relative tolerance of \num{e-5}, removing the
fastest vibrational modes and licensing the \SI{2}{\femto\second} timestep used
with a velocity-Verlet integrator~\cite{Verlet1967}. The remaining bonded terms of
the wall are harmonic (Table~\ref{tab:ff}), with a harmonic improper term that
keeps the sheets planar, and all intramolecular non-bonded pairs are excluded. All
simulations were carried out with LAMMPS (22 July 2025, update
5)~\cite{Plimpton1995,Thompson2022}. Molecular images were rendered with VMD
1.9.4~\cite{Humphrey1996}, ray-traced with OSPRay~\cite{Wald2017} or VMD's internal
Tachyon renderer, and graphs were plotted with gnuplot 6.0.

\subsection{Langevin dynamics and the thermostatting scheme}
\label{sec:langevin}

\begin{figure*}[tp]
  \centering
  \includegraphics[width=\textwidth,alt={(a) Side view of the whole simulation cell at an offset of a quarter period, with a slab of water between the two graphene walls; wall carbon is coloured by the measured temperature of its thermostat zone and hydroxyl oxygens are dark. (b) Three top views of the lower wall at offsets of zero, a quarter and a half period. The hydroxyl pattern is the same in all three while the temperature pattern shifts; dashed lines mark the coverage maxima and triangles the temperature maxima.}]{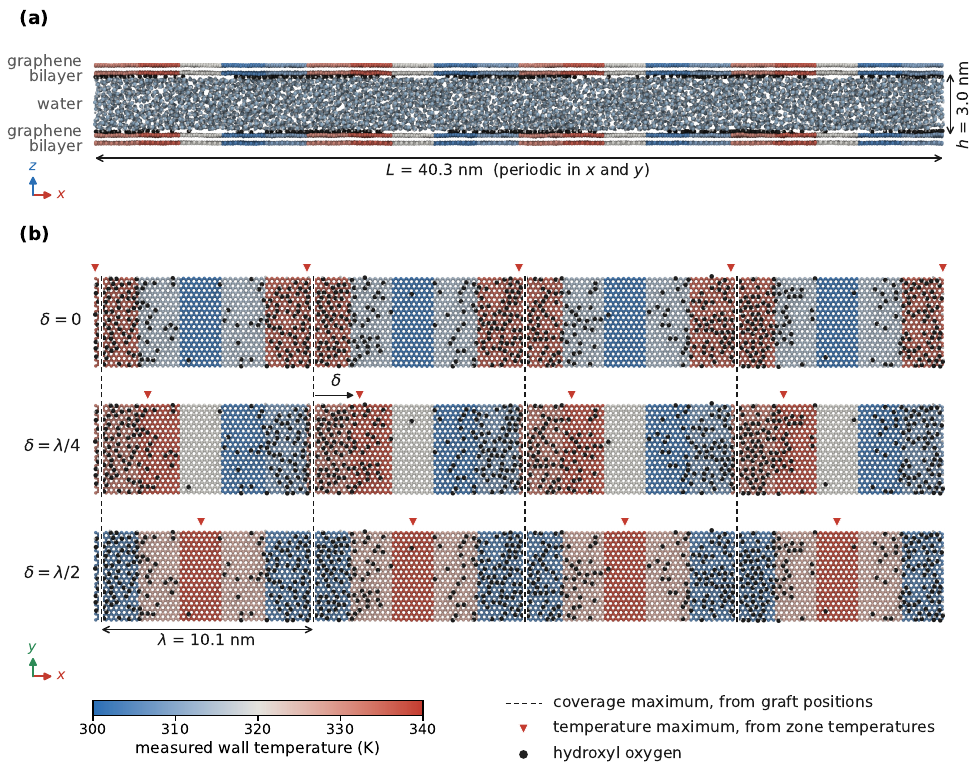}
  \caption{Simulation cell and the phase offset. (a) The whole cell viewed along
  $y$ at $\delta = \lambda/4$, with water shown in a \SI{1.4}{\nano\meter} slab; wall
  carbon is coloured by the measured temperature of its thermostat zone and
  hydroxyl oxygens are dark. (b) The inner surface of the lower wall at
  $\delta = 0$, $\lambda/4$ and $\lambda/2$. The hydroxyl arrangement is identical in
  the three strips, as it is in every run of a replica; only the thermal pattern
  moves. Dashed lines mark the coverage maxima, measured from the graft positions,
  and triangles the temperature maxima, measured from the zone temperatures; their
  separation reproduces $\delta$ to within $0.03\lambda$. Temperatures are averages
  over \SI{50}{\nano\second} and five replicas; the realised modulation of the
  wall is \SI{16.947+-0.035}{\kelvin} against a nominal \SI{20}{\kelvin}
  (Section~\ref{sec:discretisation}).}
  \label{fig:cell}
\end{figure*}

The thermal pattern is imposed by twenty independent Langevin thermostats acting
on disjoint axial slabs of wall carbon. Each thermostatted atom obeys
\begin{equation}
  m_i \ddot{\mathbf{r}}_i \;=\; \mathbf{F}_i
      \;-\; \frac{m_i}{\tau_d}\,\dot{\mathbf{r}}_i
      \;+\; \boldsymbol{\eta}_i(t),
  \label{eq:langevin}
\end{equation}
where $\mathbf{F}_i$ is the conservative force, $\tau_d$ the damping time and
$\boldsymbol{\eta}_i$ a Gaussian white noise whose variance is fixed by the
fluctuation--dissipation theorem,
\begin{equation}
  \bigl\langle \eta_{i\alpha}(t)\,\eta_{j\beta}(t')\bigr\rangle
  \;=\; \frac{2 m_i k_{\mathrm{B}} T^{\mathrm{tgt}}}{\tau_d}\,
        \delta_{ij}\,\delta_{\alpha\beta}\,\delta(t-t').
  \label{eq:fdt}
\end{equation}
This is the standard stochastic thermostat of Schneider and Stoll and of Grest
and Kremer~\cite{Schneider1978,Grest1986}. Each zone carries its own target,
sampled from $T(x) = T_0 + \Delta T \cos\!\left[2\pi N (x-\delta)/L\right]$ with
$T_0 = \SI{320}{\kelvin}$. Figure~\ref{fig:cell} shows the pattern this realises
on the wall together with the hydroxyl coverage, at three offsets.
The damping time is $\tau_d =
\SI{100}{\femto\second}$: short enough that a zone tracks its target against the
heat flux drawn by the fluid, long enough that the stochastic force does not
dominate the conservative dynamics of the lattice.

Three features of this scheme are deliberate and load-bearing.

\emph{Water is never thermostatted.} This is the single most important
methodological constraint in the study. A lab-frame thermostat applied to the
fluid does not merely set its temperature: the friction term in
eq.~\eqref{eq:langevin} is a momentum sink referenced to the laboratory frame,
so it injects momentum whenever the fluid moves relative to that frame and
manufactures a streaming velocity indistinguishable from genuine pumping. The
fluid here is integrated by unmodified Newtonian dynamics and reaches its
temperature solely by contact with the wall. The constraint is enforced by
static analysis of the rendered input decks rather than by intention, and every
production deck is machine-checked before submission.

\emph{A Langevin thermostat is preferred to a Nos\'{e}--Hoover
chain}~\cite{Nose1984,Hoover1985} because the thermostat required here is
\emph{spatially resolved}: twenty independent temperatures must be maintained
simultaneously in adjacent slabs a few nanometres apart. A deterministic
extended-system thermostat couples all degrees of freedom in its group to a
single fictitious variable and is poorly suited to enforcing a prescribed
spatial profile, whereas a stochastic local thermostat imposes the profile atom
by atom. The alternative of a momentum-swapping
scheme~\cite{MullerPlathe1997} imposes a heat \emph{flux} rather than a
temperature field, which is the wrong boundary condition for a design whose
independent variable is the phase of $T(x)$.

\emph{Momentum bookkeeping.} Each Langevin fix conserves the total random
impulse applied to its group, so the stochastic term injects no net momentum.
The friction term is not removed and cannot be: it is what couples the wall to
the laboratory frame and pins it there. The consequence is that total axial
momentum of the cell is not conserved and must not be used as a diagnostic. In a
successful run the water carries essentially all of it, so a gate demanding zero
total momentum would reject exactly the outcome being sought. We therefore gate
on the \emph{wall} drift velocity, which must vanish, and report the total as a
diagnostic without a pass--fail criterion.

Because the thermostat applies to carbon rather than to the fluid, the static
group assignment that LAMMPS requires is physically sound: bonded lattice atoms
vibrate about fixed sites and never migrate between zones. The same construction
applied to the fluid would be incorrect, since molecules would cross zone
boundaries carrying the wrong target temperature with them.

\subsection{Discretisation of the thermal field}
\label{sec:discretisation}

Representing a continuous $T(x)$ by $z$ discrete zones per period attenuates the
imposed fundamental by $\mathrm{sinc}(\pi/z)$. With five zones per period this is
$0.9357$, so a nominal \SI{20}{\kelvin} amplitude becomes an achieved first
harmonic of \SI{18.714}{\kelvin}; every prediction quoted here uses the achieved
value rather than the nominal one. The staircase also introduces higher
harmonics, which the selection rule of eq.~\eqref{eq:overlap} suppresses.

This attenuation depends on zones \emph{per period}, not on the total zone
count, which matters for the sweep of Section~\ref{sec:teeth}: holding the number
of zones fixed while varying the tooth count would make the drive amplitude a
function of the independent variable, and at two zones per period, exactly the
Nyquist limit, the staircase degenerates into a square wave whose fundamental
is $4/\pi$ times nominal and which carries strong third and fifth harmonics the
model does not contain. The zone count is therefore held proportional to the
tooth count throughout.

The realised wall modulation, the first harmonic of the per-zone temperatures
averaged over the 40 runs of the phase sweep, is \SI{16.947+-0.035}{\kelvin} and lies
within $0.002\lambda$ of the intended phase, a transfer ratio of $0.906$ relative to
the imposed harmonic. The deficit is inter-zone conduction, the
fin-equation physics of a conductive wall bridging adjacent thermostats, and
not a defect of the imposition.

\subsection{Protocol and uncertainty}

Each run comprises conjugate-gradient minimisation, \SI{0.5}{\nano\second} of
equilibration at \SI{320}{\kelvin} with all atoms thermostatted, and
\SI{25}{\nano\second} of production during which only the wall zones are
thermostatted. Production is divided into \SI{12.5}{\nano\second} segments that
hand off through complete configuration files rather than restart files, because
a restart reloads the thermostat group definitions and collides with the zone
definition file on resumption.

Primary and control state points use five independent replicas; supporting
sweeps use three and are labelled as such wherever they appear. Replicas differ
only in the random seed of the thermostat: within a replica index the built
structure is bit-identical across all phase points, so the phase offset is the
only quantity distinguishing them. This removes build-to-build variation in the
hydroxyl arrangement as a confound, which matters because the signal sought is
comparable in size to the scatter such variation would introduce.

Uncertainties quoted throughout are 95\% confidence intervals from the Student
$t$ distribution \emph{across replicas} ($t = 2.776$ at $n = 5$, $4.303$ at
$n = 3$). Within-replica block averaging~\cite{Flyvbjerg1989} confirms that each
trajectory is individually converged, and is never used as the error on the
mean. Equilibration is detected per run by the automated marginal-variance
method of Chodera~\cite{Chodera2016}. Fitted amplitudes are tested with the
replica as the unit, because runs sharing a replica index share the hydroxyl
arrangement: the fit is made to each replica and the $t$-interval taken across
replicas. A bootstrap interval~\cite{Efron1979}, computed when the phase sweep was first
reduced, is also reported; it resamples the eight offsets rather than the
replicas, which are the independent unit, and so understates the uncertainty. The ordering of the phase dependence is tested by permutation, and
comparisons between offsets are paired by replica.

The complete campaign comprises 172 independent runs,
\SI{3825}{\nano\second} of production and 582 node-days on 48-core nodes, at a
measured throughput of \SI{6.57}{\nano\second\per\day} per node
(Table~S1 of the Supporting Information, which follows the references).

\subsection{Flux estimators}

The loop velocity is computed by three independent routes, which are required to
agree; disagreement is treated as a defect to be located rather than a spread to
be averaged. The first is the centre-of-mass velocity, the mass-weighted axial
velocity of the water group, sampled every \num{1000} steps to give \num{12500}
samples per run. The second integrates the mass-weighted axial velocity over bins
along the transport axis, which also resolves the cross-channel structure of the
flow. The third counts the net number of molecules crossing a fixed plane per unit
time, from stored positions alone.

The third route is the most stringent, and is the reason it is included despite
its cost. It never evaluates a velocity: an error in a velocity accumulator, a
group silently losing molecules, or a biased sampling cadence cannot survive it,
because none of those quantities enters the estimate. Its agreement with the
centre-of-mass route is therefore a far stronger statement than agreement
between two velocity-based estimators would be. It fails in one case, located and
explained: on pristine graphene under the larger body force the water moves nearly
half the box between stored frames, so the count cannot tell forward from backward
motion, and the other two routes, which agree, are used there (Supporting
Information, Section~S3).

\subsection{Interfacial structure and energetics}
\label{sec:analysis}

Full-system configurations, which include the wall, are written at the end of
equilibration and of each production segment of the 40 primary runs: 120
configurations in all. Hydrogen bonds are counted with O$\cdots$O $<$
\SI{3.5}{\angstrom} and H--O$\cdots$O $<$
$30^{\circ}$~\cite{Luzar1996a,Luzar1996b} and under three further
criteria spanning 3.2--\SI{3.5}{\angstrom} and 25--$35^{\circ}$, with water and
hydroxyl oxygens alike as donors and acceptors and each hydrogen assigned to its
nearest oxygen. The interaction energy of each water molecule is evaluated with the
production force field, including the TIP4P/2005 charge site, using a
\SI{12}{\angstrom} cutoff between neutral groups (water molecules, hydroxyl groups
and carbon atoms) in place of the particle--mesh sum; half of each water--water
pair energy is assigned to each molecule and all of the wall interaction to the
water. The energy coefficients reported below are converged in this cutoff:
raising it to 16 and \SI{19.5}{\angstrom} changes the binding part by at most
\SI{0.009}{\kcal\per\mole\per\angstrom}, the density part by 0.002 and the
complete excess by 0.007, about a ninth of its 95\% interval. The first water layer ends at the first minimum of the water density,
\SI{4.5}{\angstrom} from the inner carbon plane, and water more than
\SI{11}{\angstrom} from both planes is the bulk reference. Every $n = 4$
coefficient is referenced to the phase of the hydroxyl coverage measured from the
graft positions of the same configuration, and the production values at the eight
offsets are fitted for each replica as $c(\delta) = a + b\,e^{i\theta}$, where
$\theta$ is the phase of the temperature maximum relative to the coverage maximum:
$a$ is the part of the modulation locked to the chemistry and $b$ the part locked
to the thermal pattern. The equilibration configurations, at uniform temperature,
test $a$ directly.

\subsection{Hydrodynamics of the patterned channel}
\label{sec:stokesmethod}

The loop flow produced by a given distribution of slip and thermo-osmotic drive is
computed exactly for a periodic slit with spatially varying Navier slip. By symmetry
the stream function is odd about the mid-plane, and each Fourier mode solves the
biharmonic equation in closed form. The slip condition
$u + b(x)\,\partial u/\partial n = u_{\mathrm{s}}(x)$, with $u_{\mathrm{s}}$ the
thermo-osmotic slip velocity, couples the modes and is enforced by collocation at
1601 points per period, with no net pressure drop around the loop. The solver
reproduces uniform slip exactly, slip-corrected Poiseuille flow, the lubrication
model in the long-period limit, and the shear-free stripe
results~\cite{Philip1972,LaugaStone2003} in both the unconfined and the thin-channel
limits. The hydrodynamic wall is placed at the edge of the water density,
\SI{2.5}{\angstrom} from the inner carbon plane, with a channel height of
\SI{2.45}{\nano\meter}, and slip lengths are measured from it. Every local slip
length, the pristine one included, is taken from the measured conductance by the
uniform-slip relation $K \propto 1 + 3b/H$, so that the calculation reproduces the
measured conductance contrast.

\subsection{Pre-registration}
\label{sec:prereg}

The hypothesis and its predicted values, the five controls, the falsification
conditions, the background subtraction, the replica counts, the statistical
procedure and the detection floor were committed to version control before any
production run was launched, and the flux estimators with their unit tests
before any production data existed. Every estimator carries a test against a
synthetic trajectory with a known analytic answer.
Section~\ref{sec:falsify} reports the outcome against those conditions as
written, including the two that were triggered.

Runs added after the first data fall into two kinds. For five further hydroxyl
arrangements, the conductance of uniform walls, the carbon--water
$\sigma_{\mathrm{CO}}$ at its geometric mean and walls at the coverages the
near-pristine strips carry, the reading of every outcome was registered before
any of the output existed. The body-force runs of Section~\ref{sec:stall} and
the re-measured \SI{40}{\kelvin} point of Section~\ref{sec:amplitude} were
specified before they ran but without a registered reading. Analyses devised
after the data, and testing no registered prediction, include the bootstrap
interval and the permutation test of Section~\ref{sec:phase}, the global test
of Section~\ref{sec:global}, the full-system interfacial analysis of
Section~\ref{sec:mechanism}, and the exact hydrodynamics of
Section~\ref{sec:stokesmethod}, whose application to the low-coverage walls was
nonetheless registered before those runs.

\section{Results}

\subsection{Model and protocol validation}
\label{sec:validation}

Four quantities were validated before production, each because an error in it
would bias the interfacial enthalpy excess that constitutes the signal.

The bulk equation of state of the water model was reproduced across eight
temperatures, within 0.2\% of experiment from 265 to \SI{360}{\kelvin} and within
1.18\% in the supercooled liquid at \SI{250}{\kelvin}, consistent with the documented
behaviour of TIP4P/2005~\cite{Abascal2005,Vega2011}; at this precision the density
maximum, near \SI{278}{\kelvin} for the model, is not resolved. The Lennard-Jones cutoff was tested by
paired re-evaluation of identical configurations at
\SIlist{10;12;14}{\angstrom}, which isolates the cutoff dependence from
run-to-run noise: the interfacial excess at \SI{12}{\angstrom} differs from that
at \SI{14}{\angstrom} by 1.1\%, below the 5\% tolerance fixed in advance, and
\SI{12}{\angstrom} was adopted.

The molecule count is not a free parameter and cannot be corrected during a run.
The loop is fully periodic with no reservoir, so the number of molecules in a
fixed box sets the pressure directly, and no constant-pressure stage can relax
it without breaking lattice commensuration. The count was therefore calibrated by
sweeping it and locating the crossing of \SI{1}{\atmosphere} normal pressure,
giving \num{3267(20)} molecules at $h = \SI{30}{\angstrom}$, equivalent to an
excluded layer of \SI{3.500+-0.026}{\angstrom} per wall. The same procedure at
$h = \SI{25}{\angstrom}$ gives \SI{3.318+-0.068}{\angstrom}: the excluded layer
is weakly gap-dependent and is measured, not interpolated, for every gap used.
Both values lie near the carbon--oxygen contact distance rather than at half of
it, which is the assumption an uncalibrated estimate would make.

The scalar pressure of the whole cell, \SI{-2.30e4}{\atmosphere}, a net tension,
and a channel-core water density 7.6\% below bulk remain unexplained. The
pressure was recorded only as a scalar, not resolved into its wall and fluid
contributions, and is not attributed here. We state both rather than
eliding them; neither quantity enters the phase-offset signal, which is a
difference between state points sharing an identical structure.

\begin{figure*}[tp]
  \centering
  \includegraphics[width=\textwidth,alt={Two plots of loop velocity against the phase offset between the chemical and thermal patterns, over one period. (a) On the scale of the pre-registered prediction. (b) On the scale of the measurement: replica means with 95 percent intervals follow a sinusoid of amplitude about 0.026 m/s, shown with the measured detection floor as a shaded band and the linear-response model as a dotted curve.}]{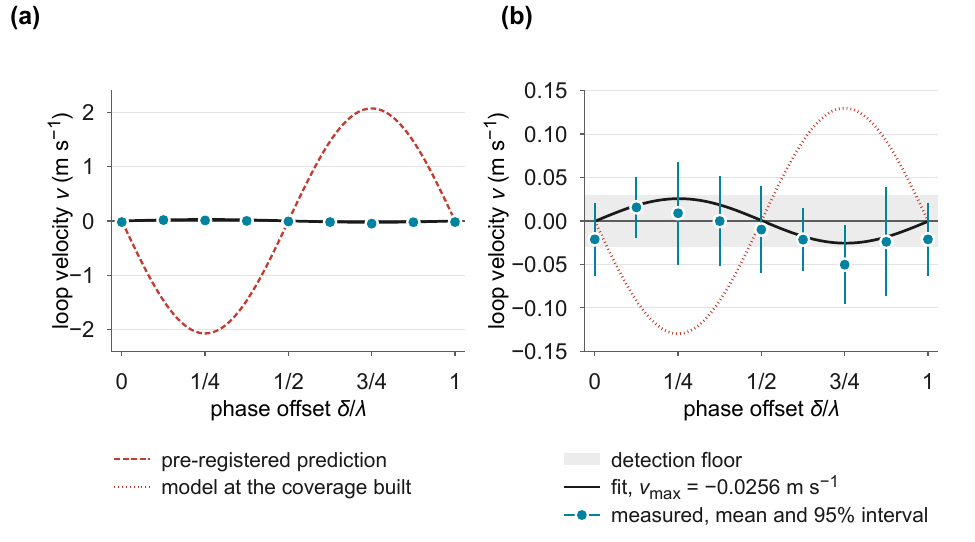}
  \caption{Loop velocity against the chemical--thermal phase offset $\delta$.
  (a) On the scale of the pre-registered prediction, eq.~\eqref{eq:h1} with
  $v_{\max} = \SI{2.07}{\meter\per\second}$. (b) On the scale of the measurement.
  Points are means over five independent replicas with 95\% $t$-intervals, after
  subtraction of the C3 background; the shaded band is the detection floor
  measured from the data. The solid curve is the fit of eq.~\eqref{eq:h1} with a
  cosine term, $v_{\max} = \SI{-0.0256}{\meter\per\second}$. The dotted curve in
  (b) is the linear-response model at the coverage built, driven by the measured
  wall temperatures (Table~\ref{tab:model}): it has the functional form of the
  measurement and the opposite sign, with $v_{\max} = \SI{+0.13}{\meter\per\second}$.
  The points at $\delta = 0$ and $\lambda/2$
  are the null controls C1 and C2.}
  \label{fig:phase}
\end{figure*}

\subsection{Phase dependence of the loop velocity}
\label{sec:phase}

Figure~\ref{fig:phase} shows the measured loop velocity across eight phase
offsets spanning one full period. Fitting eq.~\eqref{eq:h1} together with a cosine
term gives $v_{\max} = \SI{-0.0256}{\meter\per\second}$. Fitted to each replica
separately, the amplitude is $-0.063$, $-0.005$, $-0.027$, $-0.016$ and
\SI{-0.018}{\meter\per\second}, all of one sign, with a 95\% $t$-interval of
$-0.053$ to $+0.002$ ($p = 0.063$); the plane-crossing route gives $-0.025$
($p = 0.069$). The bootstrap interval over offsets, $-0.0378$ to $-0.0144$, is
narrower than this and is not relied on. Fitted to the two
\SI{12.5}{\nano\second} production segments separately, the replica amplitudes
agree to $+0.002 \pm 0.049$, so the flux is stationary. The cosine coefficient is $-0.0005$, consistent with the pure-sine form
eq.~\eqref{eq:h1} demands, and fitted with a constant term the unsubtracted
velocities give an offset of \SI{-0.0014}{\meter\per\second}: no component of the
flux fails to reverse with phase, which is what a thermostat artefact would
produce.

The fitted amplitude is negative. Equation~\eqref{eq:h1} is written so that the
linear-response model of this channel predicts $v_{\max} > 0$, flow in $-x$ at
$\delta = \lambda/4$ and in $+x$ at $3\lambda/4$. The measured flow runs in $+x$
at $\lambda/4$ and in $-x$ at $3\lambda/4$: the functional form is that of
eq.~\eqref{eq:h1}, and the direction is the opposite.
Section~\ref{sec:direction} tests the reversal at a second coverage and traces
its origin.

No individual offset in Figure~\ref{fig:phase} is convincingly nonzero on its
own, and we do not claim otherwise. The evidence resides in the ordering of the
eight points. A permutation test, devised after the data once the registered
$R^2$ criterion proved uninformative (Section~\ref{sec:falsify}), reassigns the
measured values to offsets at random; over \num{20000} shuffles it places the
observed degree of
sinusoidal organisation, the fitted amplitude with its phase left free, at
$p = 0.012$; enumerating all \num{40320} orderings gives 0.011. That test is
invariant under any rescaling of the amplitude, which matters here because the
realised signal is far smaller than anticipated, and it is unaffected by the
background subtraction described below, a constant that cannot change the
amplitude of a sinusoid sampled over a full period. Restricted to the phase of eq.~\eqref{eq:h1}, the ordering is
significant at $p = 0.0012$ for an amplitude of either sign, and at $p = 0.9996$
in the predicted direction; shuffling the offsets within each replica, which
respects the shared hydroxyl arrangement, gives $p < 10^{-4}$. The amplitude of
this sweep alone is therefore resolved in its ordering but not across independent
structures; the reversal of Section~\ref{sec:direction} is the structure-level
test.

\subsection{Null controls}

\begin{table*}[t]
\centering

\begin{tabular}{cllcc}
\toprule
ID & Configuration & Role & COM velocity & Plane crossing \\
\midrule
C1 & $\delta = 0$, in phase          & load-bearing artefact null & $-0.0097 \pm 0.0243$ & $-0.0087$ \\
C2 & $\delta = \lambda/2$, antiphase & second enforced node       & $+0.0015 \pm 0.0362$ & $+0.0017$ \\
C3 & uniform coverage                & thermostat-placement null  & $+0.0115 \pm 0.0339$ & $+0.0096$ \\
C4 & single hot/cold pair            & parity test of eq.~\eqref{eq:overlap} & $-0.0014 \pm 0.0437$ & $-0.0018$ \\
C5 & isothermal                      & second-law null            & $+0.0116 \pm 0.0490$ & $+0.0115$ \\
\bottomrule
\end{tabular}
\caption{Null controls, measured by two independent flux routes. All five are
consistent with zero. Uncertainties are 95\% $t$-intervals over five replicas;
velocities in \si{\meter\per\second}.}
\label{tab:controls}
\end{table*}

Table~\ref{tab:controls} collects the five null controls. All are consistent
with zero in both flux routes. Their roles differ and are worth distinguishing.

C5, the isothermal channel, must vanish by the second law; it is reported for
completeness and because, remarkably, no prior study of a patterned
thermo-osmotic channel appears to report it. C1 is the load-bearing control.
Unlike C5 it is fully driven: every thermostat and every gradient operates
exactly as in the primary measurement, the system is far from equilibrium, and
nothing but the phase offset distinguishes it from the $\lambda/4$ case. A
thermostat-placement artefact would appear here at full strength. It does not.
C3 isolates the same question differently, holding the thermal pattern fixed
while removing the chemical one. C4 tests the parity argument of
eq.~\eqref{eq:overlap} directly, and its null confirms that an incommensurate
design cannot pump.

Because localised thermostatting zones are broadband sources, the pre-registered
analysis subtracts the C3 flux from every claimed value with its interval
propagated. Since C3 is itself null this removes no bias and adds noise; it
leaves the fitted amplitude unchanged and is retained because it is the
committed analysis.

\subsection{The detection floor, measured rather than assumed}

The detection floor registered in advance, \SI{0.14}{\meter\per\second},
followed from an assumed velocity autocorrelation time of $\tau \approx
\SI{30}{\pico\second}$. We measured the statistical inefficiency directly across
all 65 primary runs at \num{12500} samples each, obtaining $g =
\num{1.613+-0.015}$ (standard deviation 0.059 across runs) and hence $\tau \approx \SI{0.6}{\pico\second}$, fifty times
shorter than assumed. The centre-of-mass velocity of a thirteen-thousand-molecule
system decorrelates far faster than the estimate allowed.

The resulting floor at five replicas of \SI{25}{\nano\second} is
\SI{0.030}{\meter\per\second}, a factor 4.7 below the pre-registered value, and
the measured per-point intervals (\SIrange{0.010}{0.053}{\meter\per\second}) are
consistent with it. This correction is not incidental: at the pre-registered
floor the signal reported here would have been formally unresolvable and the
study would have terminated in a null. It also retrospectively justifies the
\SI{25}{\nano\second} run length, which had been inherited from the design
rather than derived.

\begin{table*}[t]
\centering
\small
\begin{tabular}{lrrr}
\toprule
Slip length, pristine to hydroxylated & \multicolumn{3}{c}{Predicted $v_{\max}$ (\si{\meter\per\second})} \\
\cmidrule(lr){2-4}
 & nominal $T(x)$ & measured wall $T$ & measured water $T$ \\
\midrule
60 to \SI{1}{\nano\meter}, the model's mapping & $+0.152$ & $+0.130$ & $+0.088$ \\
3 to \SI{0.05}{\nano\meter} & $-0.114$ & $-0.096$ & $-0.066$ \\
uniform, any value & $-1.641$ & $-1.391$ & $-0.942$ \\
\addlinespace
measured & \multicolumn{3}{c}{$-0.0256$, 95\% interval $-0.0378$ to $-0.0144$} \\
\bottomrule
\end{tabular}
\caption{Linear-response prediction of $v_{\max}$ in the convention of
eq.~\eqref{eq:h1}, at the coverage built ($\xi = 0$ to 0.20, four periods), for three
treatments of slip and three temperature fields. The thermo-osmotic coefficient
follows the model's mapping throughout, \SI{e-6}{} to
\SI{e-7}{\meter\squared\per\second} between pristine and fully hydroxylated
graphene. \emph{Nominal} is the sinusoid of amplitude \SI{18.71}{\kelvin} that the
zone staircase imposes; \emph{measured wall} uses the replica-averaged zone
temperatures; \emph{measured water} uses the channel-averaged water temperature
profile, whose modulation is \SI{7.3}{\kelvin}. Section~\ref{sec:global} selects
the wall temperature as the driving field and excludes the water temperature. A uniform slip length, including
the sub-nanometre value measured in Section~\ref{sec:stall}, leaves the
conductance unmodulated, so every uniform value gives the same prediction.}
\label{tab:model}
\end{table*}

\begin{figure*}[tp]
  \centering
  \includegraphics[width=\textwidth,alt={(a) Loop velocity at offsets of a quarter and three quarters of a period against the maximum hydroxyl coverage of the pattern, 0.10, 0.20 and 0.30, with 95 percent intervals; the two offsets give opposite directions. (b) Rectification amplitude against maximum coverage, with a weighted straight line fitted through the origin.}]{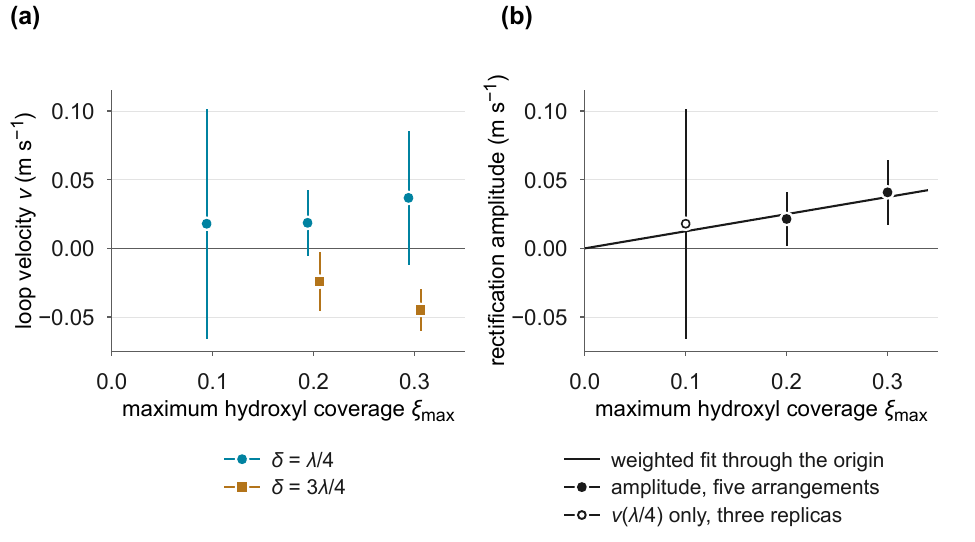}
  \caption{Direction and amplitude of the flux at three hydroxyl coverages.
  (a) Loop velocity at $\delta = \lambda/4$ and $3\lambda/4$ against the maximum
  coverage $\xi_{\max}$ of the pattern, from raw centre-of-mass velocities; means
  and 95\% $t$-intervals over ten arrangements at $\xi_{\max} = 0.20$ and five at
  0.30, and over three replicas at $\xi_{\max} = 0.10$, where only $\lambda/4$ was
  measured. (b) Rectification amplitude, half the difference between the two
  offsets within each hydroxyl arrangement, with 95\% $t$-intervals across
  arrangements; the line is a weighted fit through the origin to the two resolved
  amplitudes, of slope \SI{0.125}{\meter\per\second} per unit $\xi_{\max}$. A positive amplitude is flow in $+x$ at $\delta = \lambda/4$, opposite
  to the prediction of the model.}
  \label{fig:coverage}
\end{figure*}

\subsection{Direction of the flow, and its dependence on coverage}
\label{sec:direction}

The reversal is the most direct test of the effect, and it is resolved at the
level of independent structures. Each replica index fixes one hydroxyl
arrangement, shared by all offsets of that replica, so half the difference between
$v(\lambda/4)$ and $v(3\lambda/4)$ within one replica is an amplitude free of both
the background and the arrangement. At $\xi_{\max} = 0.20$ the five arrangements
of the phase sweep give \SI{+0.030+-0.039}{\meter\per\second}, all positive
($p = 0.10$). Five further arrangements were therefore built at the same coverage
and run at the two extrema; they give \SI{+0.013+-0.029}{\meter\per\second}, three of
five positive, the two negative amplitudes of $-0.009$ and \SI{-0.010}{\meter\per\second} being a
third of the per-run standard deviation. The ten together give
\SI{+0.021+-0.020}{\meter\per\second} ($p = 0.036$), eight of ten positive. At
$\xi_{\max} = 0.30$, close to the highest coverage the grafting permits, five
arrangements give \SI{+0.041+-0.024}{\meter\per\second}, all positive
($p = 0.0088$), and $v(3\lambda/4) = \SI{-0.0448+-0.0150}{\meter\per\second}$ is
resolved on its own (Figure~\ref{fig:coverage}a). Over all fifteen arrangements the
reversal is \SI{+0.028+-0.014}{\meter\per\second} ($p = 0.0010$), with thirteen of
fifteen in the same sense (sign test $p = 0.0074$), and the plane-crossing route
gives $+0.020$, $+0.040$ and $+0.027$ for the three sets. A positive amplitude is
flow in $+x$ at $\lambda/4$, the direction of the primary fit. The three-replica
point at $\xi_{\max} = 0.10$, measured at $\lambda/4$ only, is
\SI{+0.018+-0.084}{\meter\per\second}, of the same sign and unresolved, and all six
tooth counts of Section~\ref{sec:teeth}, measured at $\lambda/4$, are positive. The
amplitude is larger at the higher coverage, consistent with growth in proportion to
coverage at \SI{0.12}{\meter\per\second} per unit $\xi_{\max}$
(Figure~\ref{fig:coverage}b), although the two amplitudes are not resolved from each
other.

The direction is not yet shown to be robust to the carbon--water interaction. The
$\sigma_{\mathrm{CO}}$ of \SI{3.28}{\angstrom} used throughout lies below the
geometric mean of \SI{3.349}{\angstrom}. At that mean, ten phase-sweep arrangements
give a reversal of \SI{+0.017+-0.017}{\meter\per\second}, eight of ten positive
($p = 0.053$), against \SI{+0.021+-0.020}{\meter\per\second} for the same ten at the
production value. Paired arrangement by arrangement, the change,
\SI{-0.005+-0.031}{\meter\per\second} ($p = 0.74$), is not resolved in either
direction, and six of ten keep their sign, short of the eight required by the
criterion fixed before these runs (Supporting Information, Table~S17). Doubling the
arrangements halved the interval without deciding it.

The direction the model predicts is set by a near-cancellation. Through
$R = M/(TK)$, eq.~\eqref{eq:overlap} responds to the difference between the
relative modulations of $M$ and $K$. The model's coverage mapping lowers $M$ from
\SI{e-6}{} to \SI{e-7}{\meter\squared\per\second} and the slip length from 60 to
\SI{1}{\nano\meter} between pristine and fully hydroxylated graphene, so that over
$\xi = 0$ to 0.20 the thermo-osmotic coefficient is modulated by $-9.89\%$ and the
conductance by $-10.81\%$, both in phase with the coverage. The response function
is left with a modulation of only $+0.92\%$, carrying the phase of $K$ rather than
of $M$, and the sign of the predicted flux belongs to the slip mapping.
Table~\ref{tab:model} shows the consequence. With the mapping, the model predicts
$v_{\max}$ between $+0.088$ and \SI{+0.152}{\meter\per\second}, depending on the
temperature field that drives it. With any uniform slip length the conductance is
unmodulated, the sign reverses to the one observed, and the magnitude rises to
between 0.94 and \SI{1.64}{\meter\per\second}; the change of sign occurs where the
pristine slip length falls below \SI{5.4}{\nano\meter} at a fixed ratio of
pristine to hydroxylated slip.

The conductance of uniform surfaces was therefore measured directly, with the same
isothermal body force on walls of uniform coverage 0, 0.10, 0.20 and 0.30. The
mapping's form is wrong, but not in a way that rescues a local picture. Pristine
graphene conducts $177 \pm 21$ times the $\xi = 0.20$ surface, its response still
linear at \SI{40}{\meter\per\second}. Its slip length is \SI{45+-21}{\nano\meter}
from the velocity profile, which is plug-like and poorly curved, and
\SI{72}{\nano\meter} from the conductance ratio, close to the \SI{60}{\nano\meter}
the model assumed. The three hydroxylated surfaces are indistinguishable from one
another, with the no-slip plane about an {\aa}ngstr{\"o}m inside the edge of the
water density ($b = -0.08$ to \SI{-0.10}{\nano\meter}), so ten percent coverage
already removes the slip entirely (Supporting Information, Section~S8). The mapping
fails at its hydroxylated end, where it assumed \SI{1}{\nano\meter}.
Applied locally, such a $K(\xi)$ is modulated across the pattern far more strongly
than $M$, and the model again predicts the pre-registered direction; it becomes
undecided only if the pristine slip is confined to coverage below about 1\%,
which the conductance measured at those coverages rules out (below).
Because the least-covered strips are only two to three nanometres wide, narrower
than the pristine slip length, the local description was checked against the exact
two-dimensional Stokes flow of the patterned channel with the measured local slip
(Section~\ref{sec:stokesmethod}; Supporting Information, Section~S10). At
$h/\lambda = 0.24$ the channel is thin enough that the exact flow stays close to the
local one. The flow driven by the slip contrast exceeds the thermo-osmotic flow by a
factor of 1.0 to 12.8, depending on how the slip falls between $\xi = 0$ and 0.10,
and the predicted direction remains the pre-registered one. The patterned wall's
small overall slip, below \SI{0.2}{\nano\meter} (Section~\ref{sec:stall}), follows
from conductances adding in series, not from any suppression of slip on the pristine
strips, and the conductance of the patterned channel, 1.18 (95\%: 0.56--2.57) times
that at uniform $\xi = 0.20$, is consistent with every fall-off considered, which
predict 1.05 to 1.85.

The same slip acts on the drive as well as on the conductance. On a slipping wall an
interfacially driven slip velocity is amplified by $(1 + b/\ell)$, where $\ell$ is
the distance from the hydrodynamic wall to the centroid of the interfacial enthalpy
excess~\cite{AjdariBocquet2006}. On the pristine strips this amplification competes
with the conductance enhancement $(1 + 6b/h)$, so the direction turns on $\ell$
against roughly $h/6$. The intrinsic, no-slip thermo-osmotic contrast of the pattern
is small: the depth-weighted energy excess varies across it by only
$+2.4 \pm 2.5\%$ (Section~\ref{sec:mechanism}). With that contrast the exact
calculation gives the observed direction only for $\ell$ below a threshold of 0.24 to
\SI{0.38}{\nano\meter}, depending on where the contrast lies within its interval and
on how the slip falls off. The centroid of the potential-energy excess over the
near-pristine strips lies \SI{0.26}{\nano\meter} from the hydrodynamic wall, but
anywhere between 0.10 and \SI{0.36}{\nano\meter} as the integration range is varied.
Amplification would also raise the predicted magnitude, by a factor of 1.5 to 9 at
$\ell = \SI{0.26}{\nano\meter}$, and it would bend the period-number scaling. As $N$
grows the pristine strips narrow further below their slip length, and with the
measured contrast the amplified flow per period at $N = 10$ falls to 0.4 to 0.6
times its four-period value at
$\ell = \SI{0.10}{\nano\meter}$, and to between $+0.15$ and $-0.42$ times at
\SI{0.26}{\nano\meter}. The measured sweep shows no such fall
(Section~\ref{sec:teeth}). Replacing the proportional law by these shapes in the
global test of Section~\ref{sec:global} raises $\chi^2$ above its uniform-slip value
of 12.3 by 4.7 to 7.9 at \SI{0.10}{\nano\meter} and by 12.8 to 27.0 at
\SI{0.26}{\nano\meter}, for the same number of parameters (Supporting Information,
Section~S10). This comparison was made after the data.

The direction is therefore not derived from the measured inputs, and two accounts
were left open: slip amplification, or near-pristine strips that do not slip as a
uniform pristine wall does. The second is now measurable, and was measured. Counting
hydroxyls along the pattern, the strictly hydroxyl-free region is only 0.5 to
\SI{1}{\nano\meter} wide rather than the two to three nanometres of the least-covered
strip; the rest of that strip sits at $\xi \approx 0.006$ to 0.05. Uniform walls at
those coverages were run on the isothermal body-force design, three replicas of
\SI{10}{\nano\second} each (Supporting Information, Table~S15). The collapse of the
slip is steep: relative to the $\xi = 0.20$ surface the conductance falls from 177
times at $\xi = 0$ to 8.0, 2.9 and 1.57 times at realised coverages of 0.0062,
0.0188 and 0.0500, so one hydroxyl per \SI{5}{\angstrom} of strip removes 96\% of the
pristine slip and eight remove 99.7\%. By $K \propto 1 + 3b/H$ the slip lengths are
72, 2.9, 0.78 and \SI{0.23}{\nano\meter}: still a fifth to twice the channel
half-height of \SI{1.23}{\nano\meter}, so the collapse is not complete where the
strips are.

Feeding that measured $b(\xi)$ into the exact Stokes flow, with the realised
per-slice coverage of each of the ten reversal structures, leaves the direction
reversed. The slip-driven flow exceeds the thermo-osmotic one by a factor of 7.6 and
carries the opposite sign: $V_K/V_M = -7.6$, with a 95\% interval of $-9.3$ to
$-5.2$; three constructions of the local slip map and
the upper case for the hydroxylated reference give between $-7.6$ and $-7.9$
(Supporting Information, Table~S16). This reading was registered before the runs.
The low-coverage form of the second account is therefore excluded: the near-pristine
strips do slip two orders of magnitude less than pristine graphene, and it is still
far too much. What remains is slip amplification, which reconciles the direction with
the period sweep only near the lower end of the measured range of $\ell$, about
\SI{0.1}{\nano\meter}, and even there less well than uniform slip does -- a direct
enthalpy-excess profile, rather than the potential-energy proxy used here, would test
it -- or a friction raised at the strip edges, which nothing here measures. No
measured input accounts for the observed direction.
Section~\ref{sec:global} fixes the magnitude from every sweep at once, treating the
slip as uniform: a boundary condition that fits every sweep, and that the measured
local slip contradicts.

\begin{figure*}[tp]
  \centering
  \includegraphics[width=\textwidth,alt={(a) Loop velocity at a quarter-period offset against the number of chemical periods in a box of fixed length. (b) Loop velocity against the imposed wall-temperature modulation, with the measured detection floor shaded. Both panels show replica means with 95 percent intervals and weighted fits constrained through the origin.}]{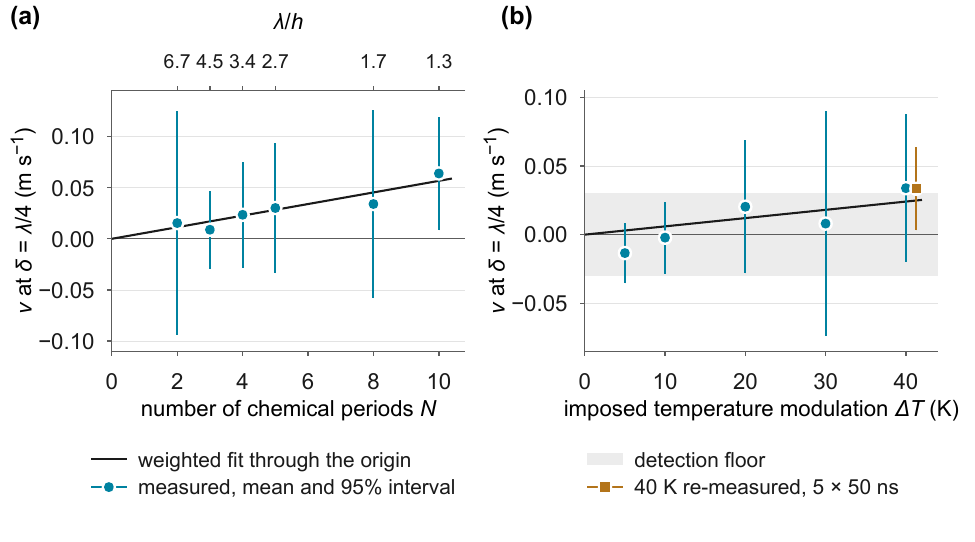}
  \caption{Scaling of the loop velocity at $\delta = \lambda/4$. (a) Against the
  number of chemical periods $N$ at fixed box length, testing the $\pi n/L$ factor
  of eq.~\eqref{eq:overlap}; the upper axis gives the corresponding $\lambda/h$,
  below about 3 of which the lubrication approximation is expected to degrade.
  (b) Against the imposed temperature modulation $\Delta T$; the shaded band is the
  measured detection floor. Points are means with 95\% $t$-intervals over three
  replicas (five at $N = 4$ in (a) and at \SI{20}{\kelvin} in (b)); lines are
  weighted fits constrained through the origin, as a proportionality requires. The
  square in (b) is the \SI{40}{\kelvin} point re-measured with five replicas of
  \SI{50}{\nano\second}, drawn slightly offset and fitted as a separate point; it is
  the only point of that sweep above unit signal-to-noise. The sweep was sized
  against the pre-registered amplitude of \SI{2.07}{\meter\per\second}; its slope is
  resolved, but it is reported as unable to test linearity.}
  \label{fig:scaling}
\end{figure*}

\subsection{Scaling with the number of chemical periods}
\label{sec:teeth}

Equation~\eqref{eq:overlap} predicts $v \propto N$ at fixed $L$. Varying the
tooth count from 2 to 10, shown in Figure~\ref{fig:scaling}a, gives a weighted slope
through the origin of \SI{0.00567+-0.00096}{\meter\per\second} per period,
$5.9\sigma$ from zero, with $\chi^2/\mathrm{dof} = 0.30$ ($p = 0.955$), a
bootstrap interval of \numrange{0.00407}{0.00714} excluding zero, and a Spearman
rank correlation $\rho(N,v) = 0.943$ ($p = 0.005$). The two flux
routes agree at every point, and Section~\ref{sec:global} shows that the measured
wall temperatures account for the scaling quantitatively.

This is the most statistically secure result of the campaign, and it succeeds
where the amplitude sweep of Section~\ref{sec:amplitude} does not for a
straightforward reason: at ten periods the signal reaches
\SI{0.064}{\meter\per\second}, twice the detection floor, whereas the amplitude
sweep never leaves the floor. It also tests a logically distinct proposition,
the harmonic structure of eq.~\eqref{eq:overlap} rather than the linearity of
the response, and therefore does not substitute for the measurement
Section~\ref{sec:amplitude} was intended to provide.

Two checks are embedded in the sweep. Holding zones proportional to periods
requires that zones sharing a target temperature share a single Langevin fix,
which differs from the one-fix-per-zone arrangement used in the primary
measurement, and one might reasonably worry that the grouping itself biases the
result. The four-period point of this sweep reproduces the primary geometry under
the alternative grouping and returns \SI{0.0235+-0.0518}{\meter\per\second}
against \SI{0.0203+-0.0486}{\meter\per\second}, a discrepancy of $|z| = 0.15$.
No grouping artefact is detectable.

Separately, the lubrication approximation underlying eq.~\eqref{eq:loop} is
expected to degrade once the period becomes comparable to the channel height.
Residuals against the proportional law at $\lambda/h = 2.69$, $1.68$ and $1.34$
are $+0.12$, $-0.53$ and $+0.56$ standard errors respectively. We report no
evidence of breakdown at this precision, which is a weaker claim than the absence
of breakdown: three-replica intervals would not resolve a modest departure, and
the sub-continuum corrections expected in this
regime~\cite{Kavokine2021,Bonthuis2013} may simply be smaller than the noise.

\subsection{Drive amplitude: a resolved slope, untested linearity}
\label{sec:amplitude}

Linear response requires $v \propto \Delta T$. Because the original sweep proved
underpowered, its strongest point was re-measured with five replicas of
\SI{50}{\nano\second} and entered in the fit as a separate point beside the
original, never pooled with it. The re-measured point,
\SI{+0.033+-0.030}{\meter\per\second} at \SI{40}{\kelvin}, agrees with the original
three-replica value of \SI{+0.034+-0.054}{\meter\per\second} and is the only point of
the sweep above unit signal-to-noise. The slope through the origin, shown in
Figure~\ref{fig:scaling}b, is
\SI{0.000602+-0.000179}{\meter\per\second\per\kelvin}, $3.4\sigma$ from zero
(bootstrap 95\% interval 0.00031--0.00090), against the
\SI{0.001017}{\meter\per\second\per\kelvin} that proportionality through the
\SI{20}{\kelvin} measurement of the primary sweep implies.

The flux therefore grows with the drive, but we do not claim to have tested
linearity, and the reason is a design failure rather than a physical finding. The
sweep was sized against the pre-registered amplitude of
\SI{2.07}{\meter\per\second}, eighty times the one realised; at the realised
amplitude three replicas of \SI{20}{\nano\second} give 95\% intervals of 0.022 to
\SI{0.082}{\meter\per\second}. The reduced chi-squared of the proportional fit is
2.79, which taken alone would suggest the law fails. It does not support that
reading: the standard errors are themselves estimates from three to five
replicas, and simulating the null with the law exactly true and the same replica
counts gives a chi-squared at least this large 27\% of the time. The proportional
law is neither rejected nor established by these data.

One consequence follows and is observed throughout this paper: no extrapolation
to experimentally accessible gradients is offered. The gradients imposed here are
of order \SI{e9}{\kelvin\per\meter}, and converting a result obtained at that
scale into a prediction at laboratory scale requires precisely the linearity that
has not been demonstrated.

\begin{figure*}[tp]
  \centering
  \includegraphics[width=\textwidth,alt={(a) Measured loop velocity at all 24 thermally driven state points against the rescaled linear-response prediction, with colour and marker shape showing the sweep; the points lie along the line of equality. (b) The period-number sweep against the same model driven by the wall temperature, solid, or by the channel-averaged water temperature, dotted.}]{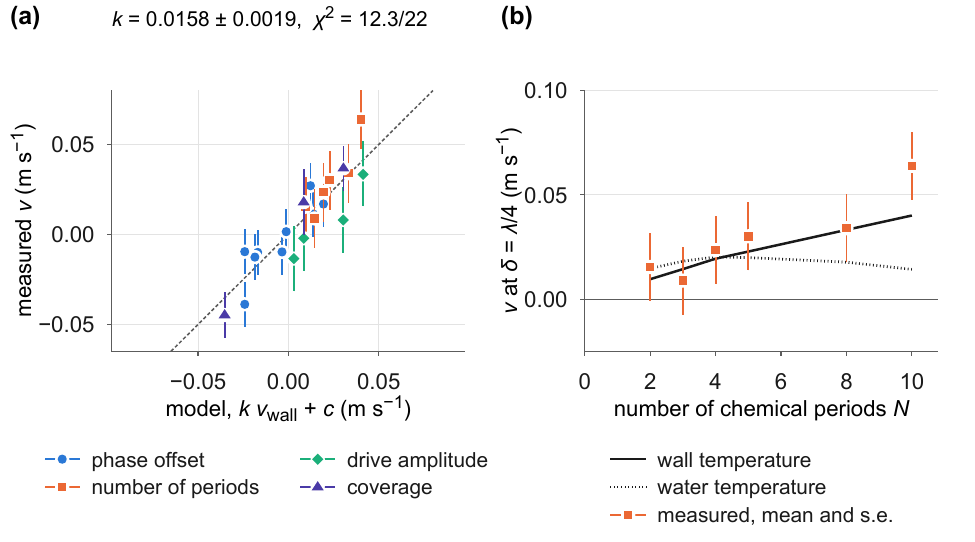}
  \caption{One transport coefficient for the whole campaign. (a) Measured loop
  velocity at all 24 thermally driven state points against the linear-response
  model with the patterned wall's measured slip taken as uniform, driven by each
  point's measured wall temperatures,
  after a single fitted rescaling $k$ and offset $c$. Colour and marker shape
  encode the sweep; error bars are standard errors from the per-run noise pooled
  over all 98 runs. The dashed line is equality. (b) The period-number sweep at
  $\delta = \lambda/4$ against the same model driven by the wall temperature (solid)
  or by the channel-averaged water temperature (dotted), each rescaled by its own
  global fit.}
  \label{fig:global}
\end{figure*}

\subsection{A single transport coefficient reconciles every sweep}
\label{sec:global}

If the harmonic-overlap theory is right and only the size of the thermo-osmotic
coefficient is wrong, one rescaling of that coefficient must reproduce every
thermally driven state point of the campaign, in sign and in size, whatever was
varied. This is tested directly, though the test was devised after the data. For
each of the 24 such state points (the eight
offsets and the two extrema on five further hydroxyl arrangements, six period
numbers, four further drive amplitudes and the re-measured \SI{40}{\kelvin} point,
and three points at the other two coverages) eq.~\eqref{eq:loop} is evaluated with
the coverage profile that point was built with, the measured slip of the patterned
wall, taken as uniform, and the temperature field that point realised, and the
measured velocities are fitted as $v = k\,v_{\mathrm{model}} + c$. Because
three-replica variance estimates are themselves noisy, each point's uncertainty is
taken from the per-run standard deviation pooled over all 98 runs,
\SI{0.0283}{\meter\per\second} at \SI{25}{\nano\second}, scaled by run length and
replica count.

Driven by the measured wall temperatures, the fit returns $k = 0.0158 \pm 0.0019$
and $c = \SI{-0.0024+-0.0032}{\meter\per\second}$, with $\chi^2 = 12.3$ for 22
degrees of freedom ($p = 0.95$; Figure~\ref{fig:global}a). The offset is
consistent with zero and every sweep falls on the same line: fitted separately, the
phase, period-number and amplitude sweeps give $k = 0.0156 \pm 0.0037$,
$0.0243 \pm 0.0100$ and $0.0185 \pm 0.0067$, and the two coverages together
$0.0173 \pm 0.0028$. One number reconciles the campaign with the theory, and the
direction of every state point follows without adjustment once the measured slip
of the patterned wall is treated as uniform: the boundary condition that
Section~\ref{sec:direction} shows the measured local slip contradicts. Because a uniform conductance cancels from
eq.~\eqref{eq:loop}, that number scales the modulation of $M$ alone: the model's
modulation amplitude of \SI{9.0e-8}{\meter\squared\per\second} at $\xi = 0$ to 0.20
becomes \SI{1.4+-0.2e-9}{\meter\squared\per\second}, smaller by a factor of
$63 \pm 8$.
The measured phase of the temperature maximum follows the nominal offset to within
$0.03\lambda$ at the wall and $0.06\lambda$ in the water at every $\delta$, so no
correction to the offset enters.

The period-number sweep also identifies the temperature that drives the flow. The
channel-averaged water temperature is modulated less at short periods, where
conduction through the fluid smooths it, and a model driven by it predicts a flux
that barely changes between two and ten periods; driven by the wall temperatures,
which the zone construction holds fixed, the model predicts a nearly proportional
rise (Figure~\ref{fig:global}b). The measured flux rises. Fitted to the
period-number sweep alone, the water-temperature model returns
$k = -0.07 \pm 0.06$, the wrong sign, while the wall-temperature model returns the
global value. Thermo-osmotic slip is set in the first molecular layers, whose
temperature follows the wall, and the channel average is not the relevant field.

\subsection{Interfacial structure and the magnitude of the flux}
\label{sec:mechanism}

The global test fixes the size of the thermo-osmotic contrast; the trajectories
show why it is small. Thermo-osmosis originates in the excess enthalpy of the
interfacial fluid over the bulk, weighted by distance from the
wall~\cite{Fu2018,Ganti2017}, and the chemical pattern can
modulate that excess through the bonding, the energy and the density of the water.
Each is measured in the 120 full-system configurations described in
Section~\ref{sec:analysis}. Of what follows, only the water-only hydrogen-bond
profile was a registered observable; the full-system analysis was devised after
the data.

The hydrogen-bond profile computed from the water-only trajectories, a donated
water--water count averaged over the channel height, gives 1.693 bonds per
molecule and a per-run $n = 4$ magnitude of 1.5\% to 2.6\% across criteria. That
construction cannot register bonds to hydroxyl groups, averages over a height of
which the first water layer holds 18\% of the molecules, and, as a magnitude,
discards the sign of the modulation and absorbs the thermal pattern. Restricted to
it, the full-system analysis reproduces 1.694 bonds per molecule and the same
magnitudes to within 0.2 percentage points. Table~\ref{tab:mechanism} collects the
separated components, from which three findings follow.

\emph{The thermal pattern modulates hydrogen bonding on its own.} The component
locked to the wall temperature is negative, as it must be, since bonds break where
the wall is hot. For the original measure it lies between $-2.6\%$ and $-1.1\%$
across criteria, as large as or larger than the chemical component, and it sets
the ordering of the per-run magnitudes across criteria. That ordering is
therefore not evidence that stricter criteria resolve more of the chemical
perturbation.

\begin{figure*}[tp]
  \centering
  \includegraphics[width=\textwidth,alt={(a) Top view of the first layer of water on the lower wall at a quarter-period offset, with hydrogen bonds drawn from donor hydrogen to acceptor oxygen, and brackets marking windows centred on the coverage maximum and minimum. (b) Hydrogen bonds per first-layer water molecule in those two windows, by bonding partner. (c) The chemistry-locked coefficient of the interfacial potential-energy excess, split into a binding part and a density part that nearly cancel, with their sum.}]{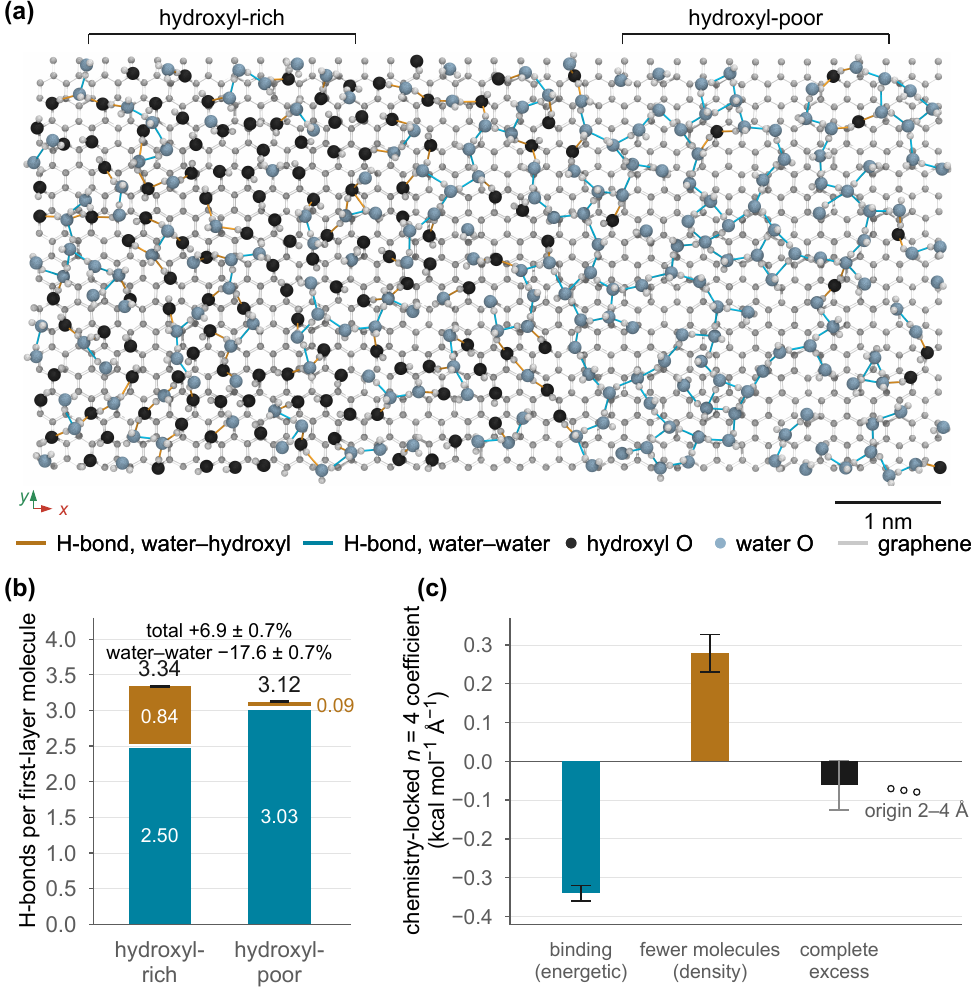}
  \caption{Hydrogen bonding and energetics of interfacial water over
  hydroxyl-rich and hydroxyl-poor wall. (a) The first water layer, with oxygens
  within \SI{4.5}{\angstrom} of the inner carbon plane, on the lower wall of one
  configuration at the end of production ($\delta = \lambda/4$), viewed from the
  fluid. Hydrogen bonds (O$\cdots$O $<$ \SI{3.5}{\angstrom}, H--O$\cdots$O $<$
  $30^{\circ}$) between displayed molecules are drawn from donor hydrogen to
  acceptor oxygen; bonds between hydroxyl groups are omitted. Brackets mark
  $\lambda/4$-wide windows centred on the coverage maximum and minimum measured
  from the graft positions. (b) Hydrogen bonds per first-layer water molecule in
  those windows, divided by partner and averaged over 80 production
  configurations; error bars and the quoted differences are 95\% $t$-intervals
  across five replicas. (c) Chemistry-locked $n = 4$ coefficient of the
  depth-weighted potential-energy excess of water within \SI{1.5}{\nano\meter} of
  the wall, divided into its binding part and its density part, with their sum;
  $z$ is measured from the inner carbon plane, and open markers give the sum with
  the origin moved to 2, 3 and \SI{4}{\angstrom}. Error bars are 95\% intervals
  across replicas.}
  \label{fig:interface}
\end{figure*}

\emph{Over the hydroxyls, water exchanges one bonding partner for another.} Locked
to the chemistry, the original measure is negative, between $-1.3\%$ and $-1.0\%$:
the pattern reduces water--water bonding. Figure~\ref{fig:interface} shows why.
First-layer water over the hydroxyl-rich wall forms 0.84 hydrogen bonds per
molecule with hydroxyl groups, against 0.09 over the hydroxyl-poor wall, and 2.50
with other water molecules, against 3.03. The total rises by only
$6.9 \pm 0.7\%$ while water--water bonding falls by
$17.6 \pm 0.7\%$: bonds to the hydroxyls largely replace bonds within the
water network rather than adding to them. As an $n = 4$ harmonic, the total
first-layer modulation locked to the chemistry is 2.9--4.4\%
across criteria, and the uniform-temperature equilibration configurations give
3.3--4.6\%, confirming that it is set by the chemistry.

\emph{Stronger binding is compensated by lower density.} Over the hydroxyl-rich
wall, first-layer water loses \SI{1.43}{\kcal\per\mole} of interaction with other
water molecules ($-7.88$ against $-9.31$) and gains \SI{3.98}{\kcal\per\mole} with
the hydroxyl groups ($-4.45$ against $-0.48$), while its dispersion interaction
with the carbon is unchanged at \SI{-1.53}{\kcal\per\mole}. Relative to core water,
at \SI{-10.86+-0.01}{\kcal\per\mole}, each first-layer molecule is therefore bound
more strongly by \SI{3.00+-0.04}{\kcal\per\mole} over the hydroxyl-rich wall
against \SI{0.47+-0.03}{\kcal\per\mole} over the hydroxyl-poor wall. The layer
holds fewer molecules there: its density is modulated by $-16.9 \pm 0.5\%$ locked
to the chemistry. The quantity that sets $M$ is not a per-molecule energy but a
per-volume excess weighted by distance from the wall,
$\int z\,[\rho(z)e(z) - \rho_{\mathrm{b}}e_{\mathrm{b}}]\,\mathrm{d}z$, which
contains both effects: a binding part,
$\int z\,\rho(z)[e(z) - e_{\mathrm{b}}]\,\mathrm{d}z$, and a density part,
$e_{\mathrm{b}}\int z\,[\rho(z) - \rho_{\mathrm{b}}]\,\mathrm{d}z$. Evaluated over
water within \SI{1.5}{\nano\meter} of each wall, with $z$ measured from the inner
carbon plane, the chemistry-locked $n = 4$ coefficient of the binding part is
\SI{-0.340+-0.020}{\kcal\per\mole\per\angstrom} and that of the density part
\SI{+0.279+-0.048}{\kcal\per\mole\per\angstrom}. Their sum, the complete excess, is
\SI{-0.062+-0.064}{\kcal\per\mole\per\angstrom}: consistent with zero,
$18 \pm 18\%$ of the binding part alone (Figure~\ref{fig:interface}c), and
$+2.4 \pm 2.5\%$ of the mean depth-weighted excess, \SI{-2.53}{\kcal\per\mole\per\angstrom}. The
cancellation does not depend on where $z$ is measured from. Moving the origin to 2,
3 or \SI{4}{\angstrom} from the carbon plane changes the binding part from $-0.162$
to $+0.017$, through a change of sign, while the complete excess stays between
$-0.070$ and $-0.079$, with intervals of $\pm 0.05$; the uniform-temperature
configurations give $-0.11 \pm 0.12$.

\emph{The compensation is consistent with the small coefficient but does not
measure it.} A chemical pattern that perturbs interfacial water strongly while
leaving its depth-weighted energy excess nearly unmodulated is what a
thermo-osmotic contrast 63 times below a linear coverage mapping requires. The
analysis stops short of computing that contrast: interaction energies with a
truncated electrostatic sum are not enthalpies, the kinetic and pressure--volume
parts of the excess are not resolved, and the enthalpy route to $M$ is known to be
delicate~\cite{Ganti2017}. No value of $M$ is claimed from it.

\begin{table*}[t]
\centering
\small
\setlength{\tabcolsep}{3.6pt}
\begin{tabular}{llrrrr}
\toprule
Quantity & Water counted & \multicolumn{2}{c}{Chemistry-locked} &
Uniform $T$ & $T(x)$-locked \\
\cmidrule(lr){3-4}
 & & \SI{3.5}{\angstrom}, $30^{\circ}$ & all criteria & & \\
\midrule
\multicolumn{2}{l}{$M$ presumed by the model} & $+9.89$ & -- & -- & -- \\
\addlinespace
\multicolumn{6}{l}{\emph{hydrogen-bond population per molecule}} \\
water--water donated & whole channel & $-1.22 \pm 0.06$ & $-1.28$ to $-1.00$ & $-1.05 \pm 0.15$ & $-1.51 \pm 0.10$ \\
water--water donated & first layer & $-12.5 \pm 0.3$ & $-12.7$ to $-11.6$ & $-11.8 \pm 0.5$ & $-1.95 \pm 0.14$ \\
all partners & whole channel & $+0.56 \pm 0.07$ & $+0.53$ to $+0.69$ & $+0.73 \pm 0.19$ & $-1.51 \pm 0.08$ \\
all partners & first two layers & $+1.11 \pm 0.11$ & $+1.00$ to $+1.35$ & $+1.32 \pm 0.19$ & $-1.59 \pm 0.13$ \\
all partners & first layer & $+3.47 \pm 0.39$ & $+2.88$ to $+4.35$ & $+3.88 \pm 0.31$ & $-1.79 \pm 0.13$ \\
\addlinespace
\multicolumn{6}{l}{\emph{water density}} \\
first layer & water-only trajectories & $-16.9 \pm 0.5$ & -- & -- & $1.8 \pm 0.2$ \\
\bottomrule
\end{tabular}
\caption{Modulation at the chemical period, in percent of the mean, from 120
full-system configurations. \emph{Chemistry-locked} is the in-phase component $a$,
referenced to the measured hydroxyl coverage, from the production configurations;
\emph{uniform $T$} is the same quantity in the equilibration configurations, where
the wall is at a uniform \SI{320}{\kelvin}; \emph{$T(x)$-locked} is the thermal
component $b$, whose negative sign means fewer bonds where the wall is hottest.
Hydrogen bonds use O$\cdots$O $<$ \SI{3.5}{\angstrom} and H--O$\cdots$O $<$
$30^{\circ}$, with 95\% $t$-intervals across five replicas; the fourth column spans
four criteria, with cutoffs of 3.2--\SI{3.5}{\angstrom} and angles of
25--$35^{\circ}$. The first layer ends at the first minimum of the water density,
\SI{4.5}{\angstrom} from the inner carbon plane, and the second at
\SI{8.6}{\angstrom}. The density row uses the replica-resolved density maps of the
water-only trajectories (52 configurations per run), and its thermal entry is the
magnitude of $b$. Donated water--water bonds over the whole channel is the
construction of the water-only trajectory profile, whose per-run magnitude, with
neither sign nor thermal separation, is 1.5--2.6\% across criteria
($1.71 \pm 0.77\%$ at \SI{3.5}{\angstrom} and $30^{\circ}$, standard deviation over
40 runs).}
\label{tab:mechanism}
\end{table*}

A further candidate explanation of the small flux was tested and rejected. If the flow reversed sign
across the channel height, partial cancellation would reduce the net flux without
any change in the interfacial physics. In a single replica the cross-channel
velocity profile does appear to cross zero within the fluid. Averaged over
replicas, however, only 1 of 49 bins is significantly negative, where
approximately 2.5 false positives are expected by chance at the 95\% level, and
the ratio of net to gross flux rises to 0.929. The apparent reversal is per-bin
noise and is not claimed.

\subsection{Hydraulic mobility and stall pressure}
\label{sec:stall}

Because the measured flux is small, a directly swept flux--back-pressure
characteristic is not feasible: resolving the curve near its zero crossing would
require some 3900 node-days, nearly seven times the entire campaign.
Under linear response, however, the thermal and body-force drives superpose, so
the hydraulic mobility may be measured where the velocity is large and the stall
point computed rather than observed.

A body force applied to the water alone also drags the wall, which the Langevin
friction holds only at a steady drift proportional to the force, here
\SI{486}{\meter\per\second} per unit per-atom force, or 6.5--10.7\% of the water
velocity; under the thermal drive the same drift is below
\SI{e-4}{\meter\per\second}. Mobility, slip and stall concern the water moving
past its wall, so velocities are taken relative to it (Supporting Information,
Table~S12). Three isothermal body forces then move the water at
\SIlist{0.1236;0.2634;0.6009}{\meter\per\second}, giving a mobility of
$\num{6001+-85}$~m\,s$^{-1}$ per unit applied force with $R^2 = 0.989$; in the
laboratory frame the mobility would be 8\% larger. The unconstrained intercept is
\SI{0.0027+-0.0104}{\meter\per\second}, consistent with zero: an isothermal
channel at zero applied force does not flow, so the superposition licensing the
substitution holds rather than being assumed. Combining the mobility with the
reversal amplitude of the ten arrangements at this coverage
(Section~\ref{sec:direction}) and its 95\% interval gives
\begin{equation*}
  \Delta p_{\mathrm{stall}} = \SI{-1.0+-0.9}{\mega\pascal},
\end{equation*}
whose sign follows the rectification direction and whose interval only just
excludes zero; the uncertainty is inherited almost entirely from the thermal flux.
With the five replicas of the phase sweep alone, $|v_{\max}|$ gives
\SI{-1.2+-1.3}{\mega\pascal}.

The velocity profiles give the slip directly. Divided by their force, taken
relative to the wall and pooled over the nine runs, they fit a parabola with a
viscosity of \SI{0.68+-0.16}{\milli\pascal\second}, consistent with TIP4P/2005 near
\SI{320}{\kelvin}, and a slip length of \SI{0.19+-0.11}{\nano\meter} with the wall at
the edge of the water density (Supporting Information, Figure~S3). With that
viscosity the mobility is 1.25 to 1.33 times the no-slip value, depending on the
height assigned to the fluid, where the slip lengths of the model's coverage
mapping, 48 to \SI{60}{\nano\meter} at this coverage, would make it about a hundred
times. Because moving the reference plane by an {\aa}ngstr{\"o}m changes a slip
length of this size by as much, it is best read as an upper bound of about
\SI{0.2}{\nano\meter}. Set against the tens of nanometres reported for pristine
graphitic surfaces~\cite{Falk2010,Secchi2016,Kannam2013}, this indicates that the
hydroxyl patches set the hydrodynamic boundary condition of the whole interface.

\subsection{Pre-registered conditions}
\label{sec:falsify}

Two of the five pre-registered falsification conditions were triggered, and both
are reported here as written. A third outcome, which the conditions do not
address but the pre-registered values do, is reported alongside them.

Condition 3, $v_{\max} < \SI{0.5}{\meter\per\second}$, fired: the measured
amplitude, \SI{-0.0256}{\meter\per\second}, lies below the threshold in magnitude
and is opposite in sign. The pre-registration directs that the discrepancy be
reported and its origin sought, which Sections~\ref{sec:direction}
and~\ref{sec:mechanism} do.

The direction of the flux is not a falsification condition, but the
pre-registered values fix it, placing the flow in $-x$ at $\delta = \lambda/4$. The
measured flow runs in $+x$ there, in both sweeps that measure the two extrema, and
this is reported with the same weight as the conditions that fired.

Condition 2, a sinusoid fit with $R^2 < 0.9$, also fired, at $R^2 = 0.415$. We
report the trigger, and we report why the threshold cannot discriminate at the
amplitude realised. The coefficient of determination is not scale-free: it
conflates whether the shape is correct with whether the signal exceeds the noise,
and the threshold was fixed expecting an amplitude eighty times larger.
Simulating eq.~\eqref{eq:h1} as exactly true at the measured amplitude, with each
point's replica noise and the error of the background subtracted from all of them,
gives $P(R^2 \geq 0.9) = 0.05$, so that the pre-registered threshold is met only
5\% of the time when the hypothesis is correct, and places the observed value well
within expectation, $P(R^2 \leq 0.415) = 0.36$. The permutation test on the phase ordering, devised after the data, is the
scale-free analogue of the same question and is reported alongside the
condition, with both numbers given. A
pre-registered threshold that proves uninformative at the realised signal level
is a finding about the design; replacing it silently with a criterion that passes
would not be.

\section{Discussion}
\label{sec:discussion}

\subsection{What the measurements establish}

Two claims are supported by the data. Phase-offset rectification occurs in a
channel structurally symmetric throughout: the loop velocity is sinusoidal in the
offset with significant phase ordering, all five null controls vanish, three
independent estimators agree, and the direction reverses between
$\delta = \lambda/4$ and $3\lambda/4$ in thirteen of fifteen independent hydroxyl
arrangements ($p = 0.001$). The harmonic-overlap theory describes the flux
quantitatively once its inputs are measured and the slip is treated as uniform: with
the wall temperature as the driving field, a single rescaling of the thermo-osmotic
contrast reproduces all 24 thermally driven state points across four independent
sweeps, including the $v \propto N$ scaling.

Six claims are not supported. The amplitude of the primary phase sweep is not
resolved across independent structures on its own ($p = 0.063$); the effect rests
instead on the reversal across fifteen structures and on the global test. The
magnitude of the thermo-osmotic contrast is fitted rather than predicted, and is 63
times below the linear coverage mapping. The direction is not shown to be robust to
the carbon--water $\sigma_{\mathrm{CO}}$, whose geometric-mean value leaves the
reversal unresolved on ten arrangements ($+0.017 \pm 0.017$~m\,s$^{-1}$). The amplitude sweep resolves a slope but
does not test linear response in the drive, so no extrapolation to laboratory
gradients is offered. The stall pressure is signed by the rectification but only
marginally resolved in magnitude. And the direction is not derived from the measured
slip: the local slip contrast predicts the opposite, slip amplification of the drive
restores the observed direction only for an interfacial excess close to the wall and
then fits the period sweep less well, and the effectively uniform boundary condition
that fits every sweep is contradicted by the slip measured at the coverages the
near-pristine strips carry.

\subsection{Why the prediction fails, and what follows for design}

Both failures trace to the transport coefficients the model assumed, not to the
rectification mechanism, and the direction and the magnitude trace to different
coefficients. The direction is decided by how coverage modulates the hydraulic
conductance relative to the thermo-osmotic coefficient, because
eq.~\eqref{eq:overlap} responds to $R = M/(TK)$. The model took slip lengths of 60
and \SI{1}{\nano\meter} for pristine and fully hydroxylated graphene, under which
$K$ falls across the pattern slightly faster than $M$ and inverts the phase of $R$.
Measured on uniform surfaces, the slip falls far more steeply, from tens of
nanometres on pristine graphene, close to the model's value, to nothing at ten
percent coverage, and acting on the conductance that contrast inverts $R$ more strongly
still, in exact hydrodynamics as in the local model. One effect could restore the
observed direction: the same slip amplifying the thermo-osmotic drive on the pristine
strips, which needs the interfacial enthalpy excess within a fraction of a nanometre
of the wall and fits the period sweep less well. The other candidate, narrow pristine
strips that do not slip as a uniform pristine wall does, was tested and fails
(Section~\ref{sec:direction}), and a friction raised at the strip edges, which no
run here isolates, remains. A design that relies on the phase of $R$
must therefore establish how slip behaves on the pattern itself, not only on its
constituent surfaces: the slip contrast common in carbon
nanofluidics~\cite{Falk2010,Secchi2016} can otherwise cancel or reverse the pump.

The magnitude is decided by $M$. The model interpolated $M$ linearly between
surfaces varied uniformly in a different system~\cite{Ouadfel2024}, which presumes
that 20\% hydroxyl coverage changes the interfacial excess 20\% as much as full
hydroxylation. The trajectories show a subtler interface. Water over the
hydroxyls trades water--water hydrogen bonds for bonds to the hydroxyl groups, is
bound more strongly by \SI{2.5}{\kcal\per\mole}, and is 17\% less dense, and in the
depth-weighted energy excess the stronger binding and the lower density nearly
cancel. A chemically strong perturbation can be thermodynamically weak, which is
what the 63-fold reduction of the fitted contrast requires. For design, the lever is
therefore not how strongly a graft perturbs water but whether its effects on
binding and on density add or cancel.

Coverage offers limited headroom. The amplitude grows with hydroxyl coverage, but
the steric prohibition on hydroxylating adjacent carbon
atoms~\cite{Dreyer2010,Lerf1998} caps coverage near $\xi = 0.33$, and the
proportional fit of Figure~\ref{fig:coverage}b gives about
\SI{0.04}{\meter\per\second} there at the present drive. A design requiring a
larger contrast must obtain it another way: through a different graft chemistry, a
different substrate such as boron nitride or a transition-metal
dichalcogenide~\cite{Secchi2016,Feng2016}, or through a mechanism not depending on
interfacial enthalpy contrast at all. The slip measurements point the same way:
ten percent hydroxyl coverage is enough to turn graphene, which slips tens of
nanometres, into a surface that behaves hydrodynamically like an ordinary
wetted solid, forfeiting the very superlubricity that makes carbon nanofluidics
attractive~\cite{Falk2010,Secchi2016}.

\subsection{Relation to prior work}

The distinction from the existing literature is specific rather than
incremental. Where refs.~\cite{Sahebi2022,Wang2020,ChenStiff2019,Herrero2022,Qi2025}
break the cancellation theorem of ref.~\cite{Ouadfel2024} by constructing an
asymmetric
object, the present work breaks it with a relative phase between two symmetric
fields. The mechanism is that of the B\"{u}ttiker--Landauer
ratchet~\cite{Buttiker1987,Landauer1988} realised in a continuum fluid rather than
for a single Brownian particle, and it predicts a signature a structural ratchet
does not: the direction reverses continuously and sinusoidally as the offset is
varied, passing through two exact nulls. That signature, rather than the
magnitude or the direction predicted by the parameterised model, is what the
present measurements confirm.

Against the quantitative literature the comparison is less favourable and is
stated as such. Simulations report fast thermo-osmotic flow through carbon
nanotube membranes~\cite{Fu2018}, and thermally driven flow through graphene
channels at rates equivalent to pressure-driven flow at hundreds of
bar~\cite{Chen2019}, far above what is measured here. Those systems, however,
exploit an open geometry between reservoirs or an unbroken pristine surface and are
not subject to the loop cancellation this design must defeat. The price of rectifying within a closed symmetric loop is that only the
modulated part of an already modest transport coefficient contributes.

\subsection{Limitations}

Five limitations bear on the interpretation. The thermo-osmotic contrast is fitted
rather than computed: the energetic analysis of Section~\ref{sec:mechanism} resolves
only the potential-energy part of the interfacial excess, with a group-truncated
electrostatic sum that is converged in its cutoff but is not the particle--mesh
energy, and a direct determination of $M(\xi)$ at fixed uniform coverages, by
the equilibrium correlation route of refs.~\cite{Ganti2017,Anzini2025}, would turn
the global fit of Section~\ref{sec:global} into a prediction. The direction rests on
how slip behaves on the narrow near-pristine strips. With the local slip of uniform
walls it is reversed, and it stays reversed when the slip measured at the strips' own
coverages, $\xi = 0.006$ to 0.05, is used in exact hydrodynamics, which rules out
the collapse of the strips' slip as the explanation
(Section~\ref{sec:direction}). Slip amplification restores the direction only for an
interfacial excess close to the wall, measured here through the potential energy
alone, and then fits the period sweep less well than the effectively uniform boundary
condition the global test assumes. A direct enthalpy-excess profile over pristine
graphene would test that, and the friction at a strip edge, which no run here
isolates, remains the other possibility. The carbon--water
$\sigma_{\mathrm{CO}}$, which sets wetting, slip and the interfacial excess together,
was tested at its geometric-mean value on ten arrangements, and the reversal there
is still not resolved: $+0.017 \pm 0.017$~m\,s$^{-1}$ with eight of ten positive
($p = 0.053$), against $+0.021 \pm 0.020$ for the same ten structures at the
production value, a paired difference of $-0.005 \pm 0.031$ ($p = 0.74$) and six of
ten keeping their sign. Doubling the number of arrangements tightened the interval
on that difference by a factor of 1.9 and moved none of the pre-registered
thresholds, so a free-energy calibration of $\sigma_{\mathrm{CO}}$ against the
wetting of the patterned surface, rather than more arrangements, is what would
decide whether the direction depends on it.
The residual pressure and core density deficit noted in
Section~\ref{sec:validation} remain unexplained, though neither enters a
difference between state points sharing a structure. And the water model, while
well validated in bulk~\cite{Abascal2005,Vega2011}, is non-polarisable, so the
response of the interfacial hydrogen-bond network to a strongly heterogeneous
surface is represented approximately; a polarisable treatment would test whether
the measured interfacial response is itself model-limited.

\section{Conclusions}

Directional water transport can be rectified in a graphene nanochannel that is
mirror-symmetric throughout, using only the phase offset between a chemical and a
thermal modulation of matched period. The loop velocity is sinusoidal in the offset,
with phase ordering significant at $p = 0.012$ by permutation, all five null
controls vanishing and three independent flux estimators agreeing to within 2.4\%,
and its direction reverses between $\delta = \lambda/4$ and $3\lambda/4$ in thirteen
of fifteen independent hydroxyl arrangements ($p = 0.001$).

The harmonic-overlap theory describes this flux once its inputs are measured,
except for its direction. With the wall temperature as the driving field and the
slip treated as uniform, a single rescaling of the thermo-osmotic contrast, by a
factor of $63 \pm 8$, reproduces all 24 thermally driven state points across sweeps
of offset, period number, drive amplitude and coverage ($\chi^2 = 12.3$ for 22
degrees of freedom), whereas the channel-averaged water temperature fails the
period-number sweep. But pristine graphene slips tens of nanometres and ten percent
hydroxyl coverage removes that slip, and acting on the conductance that contrast
predicts the opposite direction, in exact hydrodynamics as in the local model. The
observed direction is recovered if the same slip amplifies the thermo-osmotic
drive enough, which requires the enthalpy excess of pristine graphene within 0.24 to
\SI{0.38}{\nano\meter} of the wall and, to preserve $v \propto N$, nearer
\SI{0.1}{\nano\meter}. The alternative, that the narrow pristine strips do not slip
as a uniform pristine wall does, was tested and fails: at the coverages those strips
carry, uniform walls lose 96 to 99.7\% of the pristine slip, and what is left still
reverses the direction by a factor of 7.6. A friction raised at the strip edges,
which no run here isolates, is the remaining possibility; no measured input accounts
for the direction. The small contrast has a molecular
origin: over the hydroxyl patches interfacial water binds more strongly but is less
dense, and the two nearly cancel in the depth-weighted energy excess.

The rectification mechanism and the structure of its theory are therefore
established, and what remains to be predicted is the thermo-osmotic contrast of the
patterned surface, together with the boundary condition that decides its direction:
the global test is reconciled by a slip that is effectively uniform, and the slip
measured coverage by coverage is not.
Computing it directly, and finding
a surface chemistry whose effects on the binding and the density of interfacial
water add rather than cancel, are the natural next steps.

\begin{acknowledgement}
The authors acknowledge the National Supercomputing Mission (NSM) for providing
the computing resources of PARAM Porul at the National Institute of Technology
Tiruchirappalli, which is implemented by C-DAC and supported by the Ministry of
Electronics and Information Technology (MeitY) and the Department of Science and
Technology (DST), Government of India. Every simulation in this work, 582
node-days in all, ran on that facility. The authors thank the Department of
Chemical Engineering, National Institute of Technology Tiruchirappalli, for the
facilities and support that made the work possible. Molecular images were
rendered with VMD, developed by the Theoretical and Computational Biophysics
Group in the Beckman Institute for Advanced Science and Technology at the
University of Illinois at Urbana-Champaign.
\end{acknowledgement}

\end{multicols}

\bibliography{manuscript-references}

\providecommand{\latin}[1]{#1}
\makeatletter
\providecommand{\doi}
  {\begingroup\let\do\@makeother\dospecials
  \catcode`\{=1 \catcode`\}=2 \doi@aux}
\providecommand{\doi@aux}[1]{\endgroup\texttt{#1}}
\makeatother
\providecommand*\mcitethebibliography{\thebibliography}
\csname @ifundefined\endcsname{endmcitethebibliography}
  {\let\endmcitethebibliography\endthebibliography}{}
\begin{mcitethebibliography}{79}
\providecommand*\natexlab[1]{#1}
\providecommand*\mciteSetBstSublistMode[1]{}
\providecommand*\mciteSetBstMaxWidthForm[2]{}
\providecommand*\mciteBstWouldAddEndPuncttrue
  {\def\EndOfBibitem{\unskip.}}
\providecommand*\mciteBstWouldAddEndPunctfalse
  {\let\EndOfBibitem\relax}
\providecommand*\mciteSetBstMidEndSepPunct[3]{}
\providecommand*\mciteSetBstSublistLabelBeginEnd[3]{}
\providecommand*\EndOfBibitem{}
\mciteSetBstSublistMode{f}
\mciteSetBstMaxWidthForm{subitem}{(\alph{mcitesubitemcount})}
\mciteSetBstSublistLabelBeginEnd
  {\mcitemaxwidthsubitemform\space}
  {\relax}
  {\relax}

\bibitem[Bocquet and Charlaix(2010)Bocquet, and Charlaix]{Bocquet2010}
Bocquet,~L.; Charlaix,~E. Nanofluidics, from Bulk to Interfaces. \emph{Chem.
  Soc. Rev.} \textbf{2010}, \emph{39}, 1073--1095\relax
\mciteBstWouldAddEndPuncttrue
\mciteSetBstMidEndSepPunct{\mcitedefaultmidpunct}
{\mcitedefaultendpunct}{\mcitedefaultseppunct}\relax
\EndOfBibitem
\bibitem[Bocquet and Tabeling(2014)Bocquet, and Tabeling]{BocquetTabeling2014}
Bocquet,~L.; Tabeling,~P. Physics and Technological Aspects of Nanofluidics.
  \emph{Lab Chip} \textbf{2014}, \emph{14}, 3143--3158\relax
\mciteBstWouldAddEndPuncttrue
\mciteSetBstMidEndSepPunct{\mcitedefaultmidpunct}
{\mcitedefaultendpunct}{\mcitedefaultseppunct}\relax
\EndOfBibitem
\bibitem[Kavokine \latin{et~al.}(2021)Kavokine, Netz, and
  Bocquet]{Kavokine2021}
Kavokine,~N.; Netz,~R.~R.; Bocquet,~L. Fluids at the Nanoscale: From Continuum
  to Subcontinuum Transport. \emph{Annu. Rev. Fluid Mech.} \textbf{2021},
  \emph{53}, 377--410\relax
\mciteBstWouldAddEndPuncttrue
\mciteSetBstMidEndSepPunct{\mcitedefaultmidpunct}
{\mcitedefaultendpunct}{\mcitedefaultseppunct}\relax
\EndOfBibitem
\bibitem[Majumder \latin{et~al.}(2005)Majumder, Chopra, Andrews, and
  Hinds]{Majumder2005}
Majumder,~M.; Chopra,~N.; Andrews,~R.; Hinds,~B.~J. Enhanced Flow in Carbon
  Nanotubes. \emph{Nature} \textbf{2005}, \emph{438}, 44\relax
\mciteBstWouldAddEndPuncttrue
\mciteSetBstMidEndSepPunct{\mcitedefaultmidpunct}
{\mcitedefaultendpunct}{\mcitedefaultseppunct}\relax
\EndOfBibitem
\bibitem[Holt \latin{et~al.}(2006)Holt, Park, Wang, Stadermann, Artyukhin,
  Grigoropoulos, Noy, and Bakajin]{Holt2006}
Holt,~J.~K.; Park,~H.~G.; Wang,~Y.; Stadermann,~M.; Artyukhin,~A.~B.;
  Grigoropoulos,~C.~P.; Noy,~A.; Bakajin,~O. Fast Mass Transport through
  Sub-2-Nanometer Carbon Nanotubes. \emph{Science} \textbf{2006}, \emph{312},
  1034--1037\relax
\mciteBstWouldAddEndPuncttrue
\mciteSetBstMidEndSepPunct{\mcitedefaultmidpunct}
{\mcitedefaultendpunct}{\mcitedefaultseppunct}\relax
\EndOfBibitem
\bibitem[Secchi \latin{et~al.}(2016)Secchi, Marbach, Nigu{\`e}s, Stein, Siria,
  and Bocquet]{Secchi2016}
Secchi,~E.; Marbach,~S.; Nigu{\`e}s,~A.; Stein,~D.; Siria,~A.; Bocquet,~L.
  Massive Radius-Dependent Flow Slippage in Carbon Nanotubes. \emph{Nature}
  \textbf{2016}, \emph{537}, 210--213\relax
\mciteBstWouldAddEndPuncttrue
\mciteSetBstMidEndSepPunct{\mcitedefaultmidpunct}
{\mcitedefaultendpunct}{\mcitedefaultseppunct}\relax
\EndOfBibitem
\bibitem[Falk \latin{et~al.}(2010)Falk, Sedlmeier, Joly, Netz, and
  Bocquet]{Falk2010}
Falk,~K.; Sedlmeier,~F.; Joly,~L.; Netz,~R.~R.; Bocquet,~L. Molecular Origin of
  Fast Water Transport in Carbon Nanotube Membranes: Superlubricity versus
  Curvature Dependent Friction. \emph{Nano Lett.} \textbf{2010}, \emph{10},
  4067--4073\relax
\mciteBstWouldAddEndPuncttrue
\mciteSetBstMidEndSepPunct{\mcitedefaultmidpunct}
{\mcitedefaultendpunct}{\mcitedefaultseppunct}\relax
\EndOfBibitem
\bibitem[Thomas and McGaughey(2008)Thomas, and McGaughey]{Thomas2008}
Thomas,~J.~A.; McGaughey,~A. J.~H. Reassessing Fast Water Transport through
  Carbon Nanotubes. \emph{Nano Lett.} \textbf{2008}, \emph{8}, 2788--2793\relax
\mciteBstWouldAddEndPuncttrue
\mciteSetBstMidEndSepPunct{\mcitedefaultmidpunct}
{\mcitedefaultendpunct}{\mcitedefaultseppunct}\relax
\EndOfBibitem
\bibitem[Kannam \latin{et~al.}(2013)Kannam, Todd, Hansen, and
  Daivis]{Kannam2013}
Kannam,~S.~K.; Todd,~B.~D.; Hansen,~J.~S.; Daivis,~P.~J. How Fast Does Water
  Flow in Carbon Nanotubes? \emph{J. Chem. Phys.} \textbf{2013}, \emph{138},
  094701\relax
\mciteBstWouldAddEndPuncttrue
\mciteSetBstMidEndSepPunct{\mcitedefaultmidpunct}
{\mcitedefaultendpunct}{\mcitedefaultseppunct}\relax
\EndOfBibitem
\bibitem[Bonthuis and Netz(2013)Bonthuis, and Netz]{Bonthuis2013}
Bonthuis,~D.~J.; Netz,~R.~R. Beyond the Continuum: How Molecular Solvent
  Structure Affects Electrostatics and Hydrodynamics at Solid--Electrolyte
  Interfaces. \emph{J. Phys. Chem. B} \textbf{2013}, \emph{117},
  11397--11413\relax
\mciteBstWouldAddEndPuncttrue
\mciteSetBstMidEndSepPunct{\mcitedefaultmidpunct}
{\mcitedefaultendpunct}{\mcitedefaultseppunct}\relax
\EndOfBibitem
\bibitem[Radha \latin{et~al.}(2016)Radha, Esfandiar, Wang, Rooney, Gopinadhan,
  Keerthi, Mishchenko, Janardanan, Blake, Fumagalli, Lozada-Hidalgo, Garaj,
  Haigh, Grigorieva, Wu, and Geim]{Radha2016}
Radha,~B.; Esfandiar,~A.; Wang,~F.~C.; Rooney,~A.~P.; Gopinadhan,~K.;
  Keerthi,~A.; Mishchenko,~A.; Janardanan,~A.; Blake,~P.; Fumagalli,~L.
  \latin{et~al.}  Molecular Transport through Capillaries Made with
  Atomic-Scale Precision. \emph{Nature} \textbf{2016}, \emph{538},
  222--225\relax
\mciteBstWouldAddEndPuncttrue
\mciteSetBstMidEndSepPunct{\mcitedefaultmidpunct}
{\mcitedefaultendpunct}{\mcitedefaultseppunct}\relax
\EndOfBibitem
\bibitem[Nair \latin{et~al.}(2012)Nair, Wu, Jayaram, Grigorieva, and
  Geim]{Nair2012}
Nair,~R.~R.; Wu,~H.~A.; Jayaram,~P.~N.; Grigorieva,~I.~V.; Geim,~A.~K.
  Unimpeded Permeation of Water through Helium-Leak-Tight Graphene-Based
  Membranes. \emph{Science} \textbf{2012}, \emph{335}, 442--444\relax
\mciteBstWouldAddEndPuncttrue
\mciteSetBstMidEndSepPunct{\mcitedefaultmidpunct}
{\mcitedefaultendpunct}{\mcitedefaultseppunct}\relax
\EndOfBibitem
\bibitem[Joshi \latin{et~al.}(2014)Joshi, Carbone, Wang, Kravets, Su,
  Grigorieva, Wu, Geim, and Nair]{Joshi2014}
Joshi,~R.~K.; Carbone,~P.; Wang,~F.~C.; Kravets,~V.~G.; Su,~Y.;
  Grigorieva,~I.~V.; Wu,~H.~A.; Geim,~A.~K.; Nair,~R.~R. Precise and Ultrafast
  Molecular Sieving through Graphene Oxide Membranes. \emph{Science}
  \textbf{2014}, \emph{343}, 752--754\relax
\mciteBstWouldAddEndPuncttrue
\mciteSetBstMidEndSepPunct{\mcitedefaultmidpunct}
{\mcitedefaultendpunct}{\mcitedefaultseppunct}\relax
\EndOfBibitem
\bibitem[Cohen-Tanugi and Grossman(2012)Cohen-Tanugi, and
  Grossman]{CohenTanugi2012}
Cohen-Tanugi,~D.; Grossman,~J.~C. Water Desalination across Nanoporous
  Graphene. \emph{Nano Lett.} \textbf{2012}, \emph{12}, 3602--3608\relax
\mciteBstWouldAddEndPuncttrue
\mciteSetBstMidEndSepPunct{\mcitedefaultmidpunct}
{\mcitedefaultendpunct}{\mcitedefaultseppunct}\relax
\EndOfBibitem
\bibitem[Surwade \latin{et~al.}(2015)Surwade, Smirnov, Vlassiouk, Unocic,
  Veith, Dai, and Mahurin]{Surwade2015}
Surwade,~S.~P.; Smirnov,~S.~N.; Vlassiouk,~I.~V.; Unocic,~R.~R.; Veith,~G.~M.;
  Dai,~S.; Mahurin,~S.~M. Water Desalination Using Nanoporous Single-Layer
  Graphene. \emph{Nat. Nanotechnol.} \textbf{2015}, \emph{10}, 459--464\relax
\mciteBstWouldAddEndPuncttrue
\mciteSetBstMidEndSepPunct{\mcitedefaultmidpunct}
{\mcitedefaultendpunct}{\mcitedefaultseppunct}\relax
\EndOfBibitem
\bibitem[Elimelech and Phillip(2011)Elimelech, and Phillip]{Elimelech2011}
Elimelech,~M.; Phillip,~W.~A. The Future of Seawater Desalination: Energy,
  Technology, and the Environment. \emph{Science} \textbf{2011}, \emph{333},
  712--717\relax
\mciteBstWouldAddEndPuncttrue
\mciteSetBstMidEndSepPunct{\mcitedefaultmidpunct}
{\mcitedefaultendpunct}{\mcitedefaultseppunct}\relax
\EndOfBibitem
\bibitem[Siria \latin{et~al.}(2013)Siria, Poncharal, Biance, Fulcrand, Blase,
  Purcell, and Bocquet]{Siria2013}
Siria,~A.; Poncharal,~P.; Biance,~A.-L.; Fulcrand,~R.; Blase,~X.;
  Purcell,~S.~T.; Bocquet,~L. Giant Osmotic Energy Conversion Measured in a
  Single Transmembrane Boron Nitride Nanotube. \emph{Nature} \textbf{2013},
  \emph{494}, 455--458\relax
\mciteBstWouldAddEndPuncttrue
\mciteSetBstMidEndSepPunct{\mcitedefaultmidpunct}
{\mcitedefaultendpunct}{\mcitedefaultseppunct}\relax
\EndOfBibitem
\bibitem[Feng \latin{et~al.}(2016)Feng, Graf, Liu, Ovchinnikov, Dumcenco,
  Heiranian, Nandigana, Aluru, Kis, and Radenovic]{Feng2016}
Feng,~J.; Graf,~M.; Liu,~K.; Ovchinnikov,~D.; Dumcenco,~D.; Heiranian,~M.;
  Nandigana,~V.; Aluru,~N.~R.; Kis,~A.; Radenovic,~A. Single-Layer {MoS$_2$}
  Nanopores as Nanopower Generators. \emph{Nature} \textbf{2016}, \emph{536},
  197--200\relax
\mciteBstWouldAddEndPuncttrue
\mciteSetBstMidEndSepPunct{\mcitedefaultmidpunct}
{\mcitedefaultendpunct}{\mcitedefaultseppunct}\relax
\EndOfBibitem
\bibitem[Marbach and Bocquet(2019)Marbach, and Bocquet]{Marbach2019}
Marbach,~S.; Bocquet,~L. Osmosis, from Molecular Insights to Large-Scale
  Applications. \emph{Chem. Soc. Rev.} \textbf{2019}, \emph{48},
  3102--3144\relax
\mciteBstWouldAddEndPuncttrue
\mciteSetBstMidEndSepPunct{\mcitedefaultmidpunct}
{\mcitedefaultendpunct}{\mcitedefaultseppunct}\relax
\EndOfBibitem
\bibitem[Green(1954)]{Green1954}
Green,~M.~S. Markoff Random Processes and the Statistical Mechanics of
  Time-Dependent Phenomena. {II}. Irreversible Processes in Fluids. \emph{J.
  Chem. Phys.} \textbf{1954}, \emph{22}, 398--413\relax
\mciteBstWouldAddEndPuncttrue
\mciteSetBstMidEndSepPunct{\mcitedefaultmidpunct}
{\mcitedefaultendpunct}{\mcitedefaultseppunct}\relax
\EndOfBibitem
\bibitem[Kubo(1957)]{Kubo1957}
Kubo,~R. Statistical-Mechanical Theory of Irreversible Processes. {I}. General
  Theory and Simple Applications to Magnetic and Conduction Problems. \emph{J.
  Phys. Soc. Jpn.} \textbf{1957}, \emph{12}, 570--586\relax
\mciteBstWouldAddEndPuncttrue
\mciteSetBstMidEndSepPunct{\mcitedefaultmidpunct}
{\mcitedefaultendpunct}{\mcitedefaultseppunct}\relax
\EndOfBibitem
\bibitem[Ganti \latin{et~al.}(2017)Ganti, Liu, and Frenkel]{Ganti2017}
Ganti,~R.; Liu,~Y.; Frenkel,~D. Molecular Simulation of Thermo-osmotic Slip.
  \emph{Phys. Rev. Lett.} \textbf{2017}, \emph{119}, 038002\relax
\mciteBstWouldAddEndPuncttrue
\mciteSetBstMidEndSepPunct{\mcitedefaultmidpunct}
{\mcitedefaultendpunct}{\mcitedefaultseppunct}\relax
\EndOfBibitem
\bibitem[Anzini \latin{et~al.}(2025)Anzini, Filiberti, and Parola]{Anzini2025}
Anzini,~P.; Filiberti,~Z.; Parola,~A. Temperature-Driven Flows in Nanochannels:
  Theory and Simulations. \emph{J. Chem. Phys.} \textbf{2025}, \emph{162},
  094501\relax
\mciteBstWouldAddEndPuncttrue
\mciteSetBstMidEndSepPunct{\mcitedefaultmidpunct}
{\mcitedefaultendpunct}{\mcitedefaultseppunct}\relax
\EndOfBibitem
\bibitem[Fu \latin{et~al.}(2018)Fu, Merabia, and Joly]{Fu2018}
Fu,~L.; Merabia,~S.; Joly,~L. Understanding Fast and Robust Thermo-osmotic
  Flows through Carbon Nanotube Membranes: Thermodynamics Meets Hydrodynamics.
  \emph{J. Phys. Chem. Lett.} \textbf{2018}, \emph{9}, 2086--2092\relax
\mciteBstWouldAddEndPuncttrue
\mciteSetBstMidEndSepPunct{\mcitedefaultmidpunct}
{\mcitedefaultendpunct}{\mcitedefaultseppunct}\relax
\EndOfBibitem
\bibitem[Ouadfel \latin{et~al.}(2024)Ouadfel, Merabia, Yamaguchi, and
  Joly]{Ouadfel2024}
Ouadfel,~M.; Merabia,~S.; Yamaguchi,~Y.; Joly,~L. Equilibrium and
  Non-Equilibrium Molecular Dynamics Simulation of Thermo-Osmosis: Enhanced
  Effects on Polarised Graphene Surfaces. \emph{Mol. Phys.} \textbf{2024},
  \emph{122}, e2392016\relax
\mciteBstWouldAddEndPuncttrue
\mciteSetBstMidEndSepPunct{\mcitedefaultmidpunct}
{\mcitedefaultendpunct}{\mcitedefaultseppunct}\relax
\EndOfBibitem
\bibitem[Chen \latin{et~al.}(2019)Chen, Jiang, Liu, Liu, Liu, and Hu]{Chen2019}
Chen,~B.; Jiang,~H.; Liu,~H.; Liu,~K.; Liu,~X.; Hu,~X. Thermal-Driven Flow
  inside Graphene Channels for Water Desalination. \emph{2D Mater.}
  \textbf{2019}, \emph{6}, 035018\relax
\mciteBstWouldAddEndPuncttrue
\mciteSetBstMidEndSepPunct{\mcitedefaultmidpunct}
{\mcitedefaultendpunct}{\mcitedefaultseppunct}\relax
\EndOfBibitem
\bibitem[W{\"u}rger(2010)]{Wurger2010}
W{\"u}rger,~A. Thermal Non-equilibrium Transport in Colloids. \emph{Rep. Prog.
  Phys.} \textbf{2010}, \emph{73}, 126601\relax
\mciteBstWouldAddEndPuncttrue
\mciteSetBstMidEndSepPunct{\mcitedefaultmidpunct}
{\mcitedefaultendpunct}{\mcitedefaultseppunct}\relax
\EndOfBibitem
\bibitem[Piazza and Parola(2008)Piazza, and Parola]{Piazza2008}
Piazza,~R.; Parola,~A. Thermophoresis in Colloidal Suspensions. \emph{J. Phys.:
  Condens. Matter} \textbf{2008}, \emph{20}, 153102\relax
\mciteBstWouldAddEndPuncttrue
\mciteSetBstMidEndSepPunct{\mcitedefaultmidpunct}
{\mcitedefaultendpunct}{\mcitedefaultseppunct}\relax
\EndOfBibitem
\bibitem[Duhr and Braun(2006)Duhr, and Braun]{Duhr2006}
Duhr,~S.; Braun,~D. Why Molecules Move along a Temperature Gradient.
  \emph{Proc. Natl. Acad. Sci. U.S.A.} \textbf{2006}, \emph{103},
  19678--19682\relax
\mciteBstWouldAddEndPuncttrue
\mciteSetBstMidEndSepPunct{\mcitedefaultmidpunct}
{\mcitedefaultendpunct}{\mcitedefaultseppunct}\relax
\EndOfBibitem
\bibitem[Bessega \latin{et~al.}(2026)Bessega, Anzini, and Parola]{Bessega2026}
Bessega,~M.; Anzini,~P.; Parola,~A. Thermo-osmotic Pressure in Closed
  Nanochannels. \emph{J. Chem. Phys.} \textbf{2026}, \emph{164}, 064503\relax
\mciteBstWouldAddEndPuncttrue
\mciteSetBstMidEndSepPunct{\mcitedefaultmidpunct}
{\mcitedefaultendpunct}{\mcitedefaultseppunct}\relax
\EndOfBibitem
\bibitem[Sahebi and Azimian(2022)Sahebi, and Azimian]{Sahebi2022}
Sahebi,~M.; Azimian,~A.~R. Nanoscale Fluid Pumping Using a Symmetric
  Temperature Gradient: A Molecular Dynamics Study. \emph{Nanoscale Microscale
  Thermophys. Eng.} \textbf{2022}, \emph{26}, 84--94\relax
\mciteBstWouldAddEndPuncttrue
\mciteSetBstMidEndSepPunct{\mcitedefaultmidpunct}
{\mcitedefaultendpunct}{\mcitedefaultseppunct}\relax
\EndOfBibitem
\bibitem[Wang \latin{et~al.}(2020)Wang, Liu, Jing, Mohamad, and
  Prezhdo]{Wang2020}
Wang,~X.; Liu,~M.; Jing,~D.; Mohamad,~A.; Prezhdo,~O. Net Unidirectional Fluid
  Transport in Locally Heated Nanochannel by Thermo-osmosis. \emph{Nano Lett.}
  \textbf{2020}, \emph{20}, 8965--8971\relax
\mciteBstWouldAddEndPuncttrue
\mciteSetBstMidEndSepPunct{\mcitedefaultmidpunct}
{\mcitedefaultendpunct}{\mcitedefaultseppunct}\relax
\EndOfBibitem
\bibitem[Tsuji \latin{et~al.}(2026)Tsuji, Suzuki, Taguchi, Ishida, and
  Teshima]{Tsuji2026}
Tsuji,~T.; Suzuki,~S.; Taguchi,~S.; Ishida,~H.; Teshima,~H. Optothermal
  Actuation of Unidirectional Thermo-osmotic Flows. 2026;
  arXiv:2606.27735\relax
\mciteBstWouldAddEndPuncttrue
\mciteSetBstMidEndSepPunct{\mcitedefaultmidpunct}
{\mcitedefaultendpunct}{\mcitedefaultseppunct}\relax
\EndOfBibitem
\bibitem[Chen \latin{et~al.}(2019)Chen, Cheng, Zhang, and Zhang]{ChenStiff2019}
Chen,~L.; Cheng,~Y.; Zhang,~X.; Zhang,~Z. Spontaneous Directional Motion of
  Water Molecules in Single-Walled Carbon Nanotubes with a Stiffness Gradient.
  \emph{Nanoscale Adv.} \textbf{2019}, \emph{1}, 1175--1180\relax
\mciteBstWouldAddEndPuncttrue
\mciteSetBstMidEndSepPunct{\mcitedefaultmidpunct}
{\mcitedefaultendpunct}{\mcitedefaultseppunct}\relax
\EndOfBibitem
\bibitem[Herrero \latin{et~al.}(2022)Herrero, De~San~F{\'e}liciano, Merabia,
  and Joly]{Herrero2022}
Herrero,~C.; De~San~F{\'e}liciano,~M.; Merabia,~S.; Joly,~L. Fast and Versatile
  Thermo-osmotic Flows with a Pinch of Salt. \emph{Nanoscale} \textbf{2022},
  \emph{14}, 626--631\relax
\mciteBstWouldAddEndPuncttrue
\mciteSetBstMidEndSepPunct{\mcitedefaultmidpunct}
{\mcitedefaultendpunct}{\mcitedefaultseppunct}\relax
\EndOfBibitem
\bibitem[Qi \latin{et~al.}(2025)Qi, Li, Li, Wang, Chen, and Xi]{Qi2025}
Qi,~K.; Li,~Z.; Li,~H.; Wang,~J.; Chen,~Y.; Xi,~G. Self-Sustained Flow in Janus
  Nanochannels Based on Thermo-Osmosis. \emph{J. Phys. Chem. B} \textbf{2025},
  \emph{129}, 8043--8049\relax
\mciteBstWouldAddEndPuncttrue
\mciteSetBstMidEndSepPunct{\mcitedefaultmidpunct}
{\mcitedefaultendpunct}{\mcitedefaultseppunct}\relax
\EndOfBibitem
\bibitem[Reimann(2002)]{Reimann2002}
Reimann,~P. Brownian Motors: Noisy Transport far from Equilibrium. \emph{Phys.
  Rep.} \textbf{2002}, \emph{361}, 57--265\relax
\mciteBstWouldAddEndPuncttrue
\mciteSetBstMidEndSepPunct{\mcitedefaultmidpunct}
{\mcitedefaultendpunct}{\mcitedefaultseppunct}\relax
\EndOfBibitem
\bibitem[H{\"a}nggi and Marchesoni(2009)H{\"a}nggi, and Marchesoni]{Hanggi2009}
H{\"a}nggi,~P.; Marchesoni,~F. Artificial Brownian Motors: Controlling
  Transport on the Nanoscale. \emph{Rev. Mod. Phys.} \textbf{2009}, \emph{81},
  387--442\relax
\mciteBstWouldAddEndPuncttrue
\mciteSetBstMidEndSepPunct{\mcitedefaultmidpunct}
{\mcitedefaultendpunct}{\mcitedefaultseppunct}\relax
\EndOfBibitem
\bibitem[Astumian(1997)]{Astumian1997}
Astumian,~R.~D. Thermodynamics and Kinetics of a {B}rownian Motor.
  \emph{Science} \textbf{1997}, \emph{276}, 917--922\relax
\mciteBstWouldAddEndPuncttrue
\mciteSetBstMidEndSepPunct{\mcitedefaultmidpunct}
{\mcitedefaultendpunct}{\mcitedefaultseppunct}\relax
\EndOfBibitem
\bibitem[J{\"u}licher \latin{et~al.}(1997)J{\"u}licher, Ajdari, and
  Prost]{Julicher1997}
J{\"u}licher,~F.; Ajdari,~A.; Prost,~J. Modeling Molecular Motors. \emph{Rev.
  Mod. Phys.} \textbf{1997}, \emph{69}, 1269--1282\relax
\mciteBstWouldAddEndPuncttrue
\mciteSetBstMidEndSepPunct{\mcitedefaultmidpunct}
{\mcitedefaultendpunct}{\mcitedefaultseppunct}\relax
\EndOfBibitem
\bibitem[B{\"u}ttiker(1987)]{Buttiker1987}
B{\"u}ttiker,~M. Transport as a Consequence of State-Dependent Diffusion.
  \emph{Z. Phys. B} \textbf{1987}, \emph{68}, 161--167\relax
\mciteBstWouldAddEndPuncttrue
\mciteSetBstMidEndSepPunct{\mcitedefaultmidpunct}
{\mcitedefaultendpunct}{\mcitedefaultseppunct}\relax
\EndOfBibitem
\bibitem[Landauer(1988)]{Landauer1988}
Landauer,~R. Motion out of Noisy States. \emph{J. Stat. Phys.} \textbf{1988},
  \emph{53}, 233--248\relax
\mciteBstWouldAddEndPuncttrue
\mciteSetBstMidEndSepPunct{\mcitedefaultmidpunct}
{\mcitedefaultendpunct}{\mcitedefaultseppunct}\relax
\EndOfBibitem
\bibitem[Magnasco(1993)]{Magnasco1993}
Magnasco,~M.~O. Forced Thermal Ratchets. \emph{Phys. Rev. Lett.} \textbf{1993},
  \emph{71}, 1477--1481\relax
\mciteBstWouldAddEndPuncttrue
\mciteSetBstMidEndSepPunct{\mcitedefaultmidpunct}
{\mcitedefaultendpunct}{\mcitedefaultseppunct}\relax
\EndOfBibitem
\bibitem[Sugioka and Someya(2022)Sugioka, and Someya]{Sugioka2022a}
Sugioka,~H.; Someya,~Y. Rapid increase in strength of vortices and a
  directional net flow due to {M}arangoni force in {B}\"{u}ttiker--{L}andauer
  ratchet. \emph{Phys. Fluids} \textbf{2022}, \emph{34}, 064113\relax
\mciteBstWouldAddEndPuncttrue
\mciteSetBstMidEndSepPunct{\mcitedefaultmidpunct}
{\mcitedefaultendpunct}{\mcitedefaultseppunct}\relax
\EndOfBibitem
\bibitem[Sugioka and Yoshijima(2022)Sugioka, and Yoshijima]{Sugioka2022b}
Sugioka,~H.; Yoshijima,~H. Sustainable high-pressure light-driven water pump
  with a spiral tube structure and {B}\"{u}ttiker--{L}andauer ratchet.
  \emph{Phys. Fluids} \textbf{2022}, \emph{34}, 114121\relax
\mciteBstWouldAddEndPuncttrue
\mciteSetBstMidEndSepPunct{\mcitedefaultmidpunct}
{\mcitedefaultendpunct}{\mcitedefaultseppunct}\relax
\EndOfBibitem
\bibitem[Sugioka and Miyauchi(2024)Sugioka, and Miyauchi]{Sugioka2024}
Sugioka,~H.; Miyauchi,~A. Generation of a net flow using a two-phase
  {B}\"{u}ttiker--{L}andauer ratchet pump in a nucleate boiling region.
  \emph{J. Phys. Soc. Jpn.} \textbf{2024}, \emph{93}, 064402\relax
\mciteBstWouldAddEndPuncttrue
\mciteSetBstMidEndSepPunct{\mcitedefaultmidpunct}
{\mcitedefaultendpunct}{\mcitedefaultseppunct}\relax
\EndOfBibitem
\bibitem[Devauchelle \latin{et~al.}(2026)Devauchelle, Popovi\'{c}, Szymczak,
  Abramian, and Lazarus]{Devauchelle2026}
Devauchelle,~O.; Popovi\'{c},~P.; Szymczak,~P.; Abramian,~A.; Lazarus,~A.
  Granular {B}\"{u}ttiker--{L}andauer Motor. \emph{Phys. Rev. Lett.}
  \textbf{2026}, \emph{137}, 097101\relax
\mciteBstWouldAddEndPuncttrue
\mciteSetBstMidEndSepPunct{\mcitedefaultmidpunct}
{\mcitedefaultendpunct}{\mcitedefaultseppunct}\relax
\EndOfBibitem
\bibitem[Petersen \latin{et~al.}(2026)Petersen, Golchin, Im, and
  de~Barros]{Petersen2026}
Petersen,~T.; Golchin,~P.; Im,~J.; de~Barros,~F. P.~J. Electrokinetic effects
  on flow and ion transport in charge-patterned corrugated nanochannels.
  \emph{Phys. Rev. Fluids} \textbf{2026}, \emph{11}, 114501\relax
\mciteBstWouldAddEndPuncttrue
\mciteSetBstMidEndSepPunct{\mcitedefaultmidpunct}
{\mcitedefaultendpunct}{\mcitedefaultseppunct}\relax
\EndOfBibitem
\bibitem[Dreyer \latin{et~al.}(2010)Dreyer, Park, Bielawski, and
  Ruoff]{Dreyer2010}
Dreyer,~D.~R.; Park,~S.; Bielawski,~C.~W.; Ruoff,~R.~S. The Chemistry of
  Graphene Oxide. \emph{Chem. Soc. Rev.} \textbf{2010}, \emph{39},
  228--240\relax
\mciteBstWouldAddEndPuncttrue
\mciteSetBstMidEndSepPunct{\mcitedefaultmidpunct}
{\mcitedefaultendpunct}{\mcitedefaultseppunct}\relax
\EndOfBibitem
\bibitem[Lerf \latin{et~al.}(1998)Lerf, He, Forster, and Klinowski]{Lerf1998}
Lerf,~A.; He,~H.; Forster,~M.; Klinowski,~J. Structure of Graphite Oxide
  Revisited. \emph{J. Phys. Chem. B} \textbf{1998}, \emph{102},
  4477--4482\relax
\mciteBstWouldAddEndPuncttrue
\mciteSetBstMidEndSepPunct{\mcitedefaultmidpunct}
{\mcitedefaultendpunct}{\mcitedefaultseppunct}\relax
\EndOfBibitem
\bibitem[Abascal and Vega(2005)Abascal, and Vega]{Abascal2005}
Abascal,~J. L.~F.; Vega,~C. A General Purpose Model for the Condensed Phases of
  Water: TIP4P/2005. \emph{J. Chem. Phys.} \textbf{2005}, \emph{123},
  234505\relax
\mciteBstWouldAddEndPuncttrue
\mciteSetBstMidEndSepPunct{\mcitedefaultmidpunct}
{\mcitedefaultendpunct}{\mcitedefaultseppunct}\relax
\EndOfBibitem
\bibitem[Vega and Abascal(2011)Vega, and Abascal]{Vega2011}
Vega,~C.; Abascal,~J. L.~F. Simulating Water with Rigid Non-polarizable Models:
  A General Perspective. \emph{Phys. Chem. Chem. Phys.} \textbf{2011},
  \emph{13}, 19663--19688\relax
\mciteBstWouldAddEndPuncttrue
\mciteSetBstMidEndSepPunct{\mcitedefaultmidpunct}
{\mcitedefaultendpunct}{\mcitedefaultseppunct}\relax
\EndOfBibitem
\bibitem[Jorgensen \latin{et~al.}(1983)Jorgensen, Chandrasekhar, Madura, Impey,
  and Klein]{Jorgensen1983}
Jorgensen,~W.~L.; Chandrasekhar,~J.; Madura,~J.~D.; Impey,~R.~W.; Klein,~M.~L.
  Comparison of Simple Potential Functions for Simulating Liquid Water.
  \emph{J. Chem. Phys.} \textbf{1983}, \emph{79}, 926--935\relax
\mciteBstWouldAddEndPuncttrue
\mciteSetBstMidEndSepPunct{\mcitedefaultmidpunct}
{\mcitedefaultendpunct}{\mcitedefaultseppunct}\relax
\EndOfBibitem
\bibitem[Berendsen \latin{et~al.}(1987)Berendsen, Grigera, and
  Straatsma]{Berendsen1987}
Berendsen,~H. J.~C.; Grigera,~J.~R.; Straatsma,~T.~P. The Missing Term in
  Effective Pair Potentials. \emph{J. Phys. Chem.} \textbf{1987}, \emph{91},
  6269--6271\relax
\mciteBstWouldAddEndPuncttrue
\mciteSetBstMidEndSepPunct{\mcitedefaultmidpunct}
{\mcitedefaultendpunct}{\mcitedefaultseppunct}\relax
\EndOfBibitem
\bibitem[Jorgensen \latin{et~al.}(1996)Jorgensen, Maxwell, and
  Tirado-Rives]{Jorgensen1996}
Jorgensen,~W.~L.; Maxwell,~D.~S.; Tirado-Rives,~J. Development and Testing of
  the OPLS All-Atom Force Field on Conformational Energetics and Properties of
  Organic Liquids. \emph{J. Am. Chem. Soc.} \textbf{1996}, \emph{118},
  11225--11236\relax
\mciteBstWouldAddEndPuncttrue
\mciteSetBstMidEndSepPunct{\mcitedefaultmidpunct}
{\mcitedefaultendpunct}{\mcitedefaultseppunct}\relax
\EndOfBibitem
\bibitem[Werder \latin{et~al.}(2003)Werder, Walther, Jaffe, Halicioglu, and
  Koumoutsakos]{Werder2003}
Werder,~T.; Walther,~J.~H.; Jaffe,~R.~L.; Halicioglu,~T.; Koumoutsakos,~P. On
  the Water--Carbon Interaction for Use in Molecular Dynamics Simulations of
  Graphite and Carbon Nanotubes. \emph{J. Phys. Chem. B} \textbf{2003},
  \emph{107}, 1345--1352\relax
\mciteBstWouldAddEndPuncttrue
\mciteSetBstMidEndSepPunct{\mcitedefaultmidpunct}
{\mcitedefaultendpunct}{\mcitedefaultseppunct}\relax
\EndOfBibitem
\bibitem[Hockney and Eastwood(1988)Hockney, and Eastwood]{Hockney1988}
Hockney,~R.~W.; Eastwood,~J.~W. \emph{Computer Simulation Using Particles};
  Adam Hilger: Bristol, 1988\relax
\mciteBstWouldAddEndPuncttrue
\mciteSetBstMidEndSepPunct{\mcitedefaultmidpunct}
{\mcitedefaultendpunct}{\mcitedefaultseppunct}\relax
\EndOfBibitem
\bibitem[Darden \latin{et~al.}(1993)Darden, York, and Pedersen]{Darden1993}
Darden,~T.; York,~D.; Pedersen,~L. Particle Mesh {E}wald: An $N\log(N)$ Method
  for Ewald Sums in Large Systems. \emph{J. Chem. Phys.} \textbf{1993},
  \emph{98}, 10089--10092\relax
\mciteBstWouldAddEndPuncttrue
\mciteSetBstMidEndSepPunct{\mcitedefaultmidpunct}
{\mcitedefaultendpunct}{\mcitedefaultseppunct}\relax
\EndOfBibitem
\bibitem[Essmann \latin{et~al.}(1995)Essmann, Perera, Berkowitz, Darden, Lee,
  and Pedersen]{Essmann1995}
Essmann,~U.; Perera,~L.; Berkowitz,~M.~L.; Darden,~T.; Lee,~H.; Pedersen,~L.~G.
  A Smooth Particle Mesh {E}wald Method. \emph{J. Chem. Phys.} \textbf{1995},
  \emph{103}, 8577--8593\relax
\mciteBstWouldAddEndPuncttrue
\mciteSetBstMidEndSepPunct{\mcitedefaultmidpunct}
{\mcitedefaultendpunct}{\mcitedefaultseppunct}\relax
\EndOfBibitem
\bibitem[Ryckaert \latin{et~al.}(1977)Ryckaert, Ciccotti, and
  Berendsen]{Ryckaert1977}
Ryckaert,~J.-P.; Ciccotti,~G.; Berendsen,~H. J.~C. Numerical Integration of the
  Cartesian Equations of Motion of a System with Constraints: Molecular
  Dynamics of $n$-Alkanes. \emph{J. Comput. Phys.} \textbf{1977}, \emph{23},
  327--341\relax
\mciteBstWouldAddEndPuncttrue
\mciteSetBstMidEndSepPunct{\mcitedefaultmidpunct}
{\mcitedefaultendpunct}{\mcitedefaultseppunct}\relax
\EndOfBibitem
\bibitem[Verlet(1967)]{Verlet1967}
Verlet,~L. Computer ``Experiments'' on Classical Fluids. I. Thermodynamical
  Properties of Lennard-Jones Molecules. \emph{Phys. Rev.} \textbf{1967},
  \emph{159}, 98--103\relax
\mciteBstWouldAddEndPuncttrue
\mciteSetBstMidEndSepPunct{\mcitedefaultmidpunct}
{\mcitedefaultendpunct}{\mcitedefaultseppunct}\relax
\EndOfBibitem
\bibitem[Plimpton(1995)]{Plimpton1995}
Plimpton,~S. Fast Parallel Algorithms for Short-Range Molecular Dynamics.
  \emph{J. Comput. Phys.} \textbf{1995}, \emph{117}, 1--19\relax
\mciteBstWouldAddEndPuncttrue
\mciteSetBstMidEndSepPunct{\mcitedefaultmidpunct}
{\mcitedefaultendpunct}{\mcitedefaultseppunct}\relax
\EndOfBibitem
\bibitem[Thompson \latin{et~al.}(2022)Thompson, Aktulga, Berger, Bolintineanu,
  Brown, Crozier, in~'t Veld, Kohlmeyer, Moore, Nguyen, Shan, Stevens,
  Tranchida, Trott, and Plimpton]{Thompson2022}
Thompson,~A.~P.; Aktulga,~H.~M.; Berger,~R.; Bolintineanu,~D.~S.; Brown,~W.~M.;
  Crozier,~P.~S.; in~'t Veld,~P.~J.; Kohlmeyer,~A.; Moore,~S.~G.; Nguyen,~T.~D.
  \latin{et~al.}  {LAMMPS} -- A Flexible Simulation Tool for Particle-Based
  Materials Modeling at the Atomic, Meso, and Continuum Scales. \emph{Comput.
  Phys. Commun.} \textbf{2022}, \emph{271}, 108171\relax
\mciteBstWouldAddEndPuncttrue
\mciteSetBstMidEndSepPunct{\mcitedefaultmidpunct}
{\mcitedefaultendpunct}{\mcitedefaultseppunct}\relax
\EndOfBibitem
\bibitem[Humphrey \latin{et~al.}(1996)Humphrey, Dalke, and
  Schulten]{Humphrey1996}
Humphrey,~W.; Dalke,~A.; Schulten,~K. {VMD}: Visual Molecular Dynamics.
  \emph{J. Mol. Graphics} \textbf{1996}, \emph{14}, 33--38\relax
\mciteBstWouldAddEndPuncttrue
\mciteSetBstMidEndSepPunct{\mcitedefaultmidpunct}
{\mcitedefaultendpunct}{\mcitedefaultseppunct}\relax
\EndOfBibitem
\bibitem[Wald \latin{et~al.}(2017)Wald, Johnson, Amstutz, Brownlee, Knoll,
  Jeffers, G{\"u}nther, and Navratil]{Wald2017}
Wald,~I.; Johnson,~G.~P.; Amstutz,~J.; Brownlee,~C.; Knoll,~A.; Jeffers,~J.;
  G{\"u}nther,~J.; Navratil,~P. {OSPRay} -- A {CPU} Ray Tracing Framework for
  Scientific Visualization. \emph{IEEE Trans. Vis. Comput. Graph.}
  \textbf{2017}, \emph{23}, 931--940\relax
\mciteBstWouldAddEndPuncttrue
\mciteSetBstMidEndSepPunct{\mcitedefaultmidpunct}
{\mcitedefaultendpunct}{\mcitedefaultseppunct}\relax
\EndOfBibitem
\bibitem[Schneider and Stoll(1978)Schneider, and Stoll]{Schneider1978}
Schneider,~T.; Stoll,~E. Molecular-Dynamics Study of a Three-Dimensional
  One-Component Model for Distortive Phase Transitions. \emph{Phys. Rev. B}
  \textbf{1978}, \emph{17}, 1302--1322\relax
\mciteBstWouldAddEndPuncttrue
\mciteSetBstMidEndSepPunct{\mcitedefaultmidpunct}
{\mcitedefaultendpunct}{\mcitedefaultseppunct}\relax
\EndOfBibitem
\bibitem[Grest and Kremer(1986)Grest, and Kremer]{Grest1986}
Grest,~G.~S.; Kremer,~K. Molecular Dynamics Simulation for Polymers in the
  Presence of a Heat Bath. \emph{Phys. Rev. A} \textbf{1986}, \emph{33},
  3628--3631\relax
\mciteBstWouldAddEndPuncttrue
\mciteSetBstMidEndSepPunct{\mcitedefaultmidpunct}
{\mcitedefaultendpunct}{\mcitedefaultseppunct}\relax
\EndOfBibitem
\bibitem[Nos{\'e}(1984)]{Nose1984}
Nos{\'e},~S. A Unified Formulation of the Constant Temperature Molecular
  Dynamics Methods. \emph{J. Chem. Phys.} \textbf{1984}, \emph{81},
  511--519\relax
\mciteBstWouldAddEndPuncttrue
\mciteSetBstMidEndSepPunct{\mcitedefaultmidpunct}
{\mcitedefaultendpunct}{\mcitedefaultseppunct}\relax
\EndOfBibitem
\bibitem[Hoover(1985)]{Hoover1985}
Hoover,~W.~G. Canonical Dynamics: Equilibrium Phase-Space Distributions.
  \emph{Phys. Rev. A} \textbf{1985}, \emph{31}, 1695--1697\relax
\mciteBstWouldAddEndPuncttrue
\mciteSetBstMidEndSepPunct{\mcitedefaultmidpunct}
{\mcitedefaultendpunct}{\mcitedefaultseppunct}\relax
\EndOfBibitem
\bibitem[M{\"u}ller-Plathe(1997)]{MullerPlathe1997}
M{\"u}ller-Plathe,~F. A Simple Nonequilibrium Molecular Dynamics Method for
  Calculating the Thermal Conductivity. \emph{J. Chem. Phys.} \textbf{1997},
  \emph{106}, 6082--6085\relax
\mciteBstWouldAddEndPuncttrue
\mciteSetBstMidEndSepPunct{\mcitedefaultmidpunct}
{\mcitedefaultendpunct}{\mcitedefaultseppunct}\relax
\EndOfBibitem
\bibitem[Flyvbjerg and Petersen(1989)Flyvbjerg, and Petersen]{Flyvbjerg1989}
Flyvbjerg,~H.; Petersen,~H.~G. Error Estimates on Averages of Correlated Data.
  \emph{J. Chem. Phys.} \textbf{1989}, \emph{91}, 461--466\relax
\mciteBstWouldAddEndPuncttrue
\mciteSetBstMidEndSepPunct{\mcitedefaultmidpunct}
{\mcitedefaultendpunct}{\mcitedefaultseppunct}\relax
\EndOfBibitem
\bibitem[Chodera(2016)]{Chodera2016}
Chodera,~J.~D. A Simple Method for Automated Equilibration Detection in
  Molecular Simulations. \emph{J. Chem. Theory Comput.} \textbf{2016},
  \emph{12}, 1799--1805\relax
\mciteBstWouldAddEndPuncttrue
\mciteSetBstMidEndSepPunct{\mcitedefaultmidpunct}
{\mcitedefaultendpunct}{\mcitedefaultseppunct}\relax
\EndOfBibitem
\bibitem[Efron(1979)]{Efron1979}
Efron,~B. Bootstrap Methods: Another Look at the Jackknife. \emph{Ann. Stat.}
  \textbf{1979}, \emph{7}, 1--26\relax
\mciteBstWouldAddEndPuncttrue
\mciteSetBstMidEndSepPunct{\mcitedefaultmidpunct}
{\mcitedefaultendpunct}{\mcitedefaultseppunct}\relax
\EndOfBibitem
\bibitem[Luzar and Chandler(1996)Luzar, and Chandler]{Luzar1996a}
Luzar,~A.; Chandler,~D. Hydrogen-Bond Kinetics in Liquid Water. \emph{Nature}
  \textbf{1996}, \emph{379}, 55--57\relax
\mciteBstWouldAddEndPuncttrue
\mciteSetBstMidEndSepPunct{\mcitedefaultmidpunct}
{\mcitedefaultendpunct}{\mcitedefaultseppunct}\relax
\EndOfBibitem
\bibitem[Luzar and Chandler(1996)Luzar, and Chandler]{Luzar1996b}
Luzar,~A.; Chandler,~D. Effect of Environment on Hydrogen Bond Dynamics in
  Liquid Water. \emph{Phys. Rev. Lett.} \textbf{1996}, \emph{76},
  928--931\relax
\mciteBstWouldAddEndPuncttrue
\mciteSetBstMidEndSepPunct{\mcitedefaultmidpunct}
{\mcitedefaultendpunct}{\mcitedefaultseppunct}\relax
\EndOfBibitem
\bibitem[Philip(1972)]{Philip1972}
Philip,~J.~R. Flows Satisfying Mixed No-Slip and No-Shear Conditions. \emph{Z.
  Angew. Math. Phys.} \textbf{1972}, \emph{23}, 353--372\relax
\mciteBstWouldAddEndPuncttrue
\mciteSetBstMidEndSepPunct{\mcitedefaultmidpunct}
{\mcitedefaultendpunct}{\mcitedefaultseppunct}\relax
\EndOfBibitem
\bibitem[Lauga and Stone(2003)Lauga, and Stone]{LaugaStone2003}
Lauga,~E.; Stone,~H.~A. Effective Slip in Pressure-Driven {Stokes} Flow.
  \emph{J. Fluid Mech.} \textbf{2003}, \emph{489}, 55--77\relax
\mciteBstWouldAddEndPuncttrue
\mciteSetBstMidEndSepPunct{\mcitedefaultmidpunct}
{\mcitedefaultendpunct}{\mcitedefaultseppunct}\relax
\EndOfBibitem
\bibitem[Ajdari and Bocquet(2006)Ajdari, and Bocquet]{AjdariBocquet2006}
Ajdari,~A.; Bocquet,~L. Giant Amplification of Interfacially Driven Transport
  by Hydrodynamic Slip: Diffusio-Osmosis and Beyond. \emph{Phys. Rev. Lett.}
  \textbf{2006}, \emph{96}, 186102\relax
\mciteBstWouldAddEndPuncttrue
\mciteSetBstMidEndSepPunct{\mcitedefaultmidpunct}
{\mcitedefaultendpunct}{\mcitedefaultseppunct}\relax
\EndOfBibitem
\end{mcitethebibliography}


\providecommand{\latin}[1]{#1}
\makeatletter
\providecommand{\doi}
  {\begingroup\let\do\@makeother\dospecials
  \catcode`\{=1 \catcode`\}=2 \doi@aux}
\providecommand{\doi@aux}[1]{\endgroup\texttt{#1}}
\makeatother
\providecommand*\mcitethebibliography{\thebibliography}
\csname @ifundefined\endcsname{endmcitethebibliography}
  {\let\endmcitethebibliography\endthebibliography}{}
\begin{mcitethebibliography}{6}
\providecommand*\natexlab[1]{#1}
\providecommand*\mciteSetBstSublistMode[1]{}
\providecommand*\mciteSetBstMaxWidthForm[2]{}
\providecommand*\mciteBstWouldAddEndPuncttrue
  {\def\EndOfBibitem{\unskip.}}
\providecommand*\mciteBstWouldAddEndPunctfalse
  {\let\EndOfBibitem\relax}
\providecommand*\mciteSetBstMidEndSepPunct[3]{}
\providecommand*\mciteSetBstSublistLabelBeginEnd[3]{}
\providecommand*\EndOfBibitem{}
\mciteSetBstSublistMode{f}
\mciteSetBstMaxWidthForm{subitem}{(\alph{mcitesubitemcount})}
\mciteSetBstSublistLabelBeginEnd
  {\mcitemaxwidthsubitemform\space}
  {\relax}
  {\relax}

\bibitem[Abascal and Vega(2005)Abascal, and Vega]{Abascal2005}
Abascal,~J. L.~F.; Vega,~C. A General Purpose Model for the Condensed Phases of
  Water: TIP4P/2005. \emph{J. Chem. Phys.} \textbf{2005}, \emph{123},
  234505\relax
\mciteBstWouldAddEndPuncttrue
\mciteSetBstMidEndSepPunct{\mcitedefaultmidpunct}
{\mcitedefaultendpunct}{\mcitedefaultseppunct}\relax
\EndOfBibitem
\bibitem[Vega and Abascal(2011)Vega, and Abascal]{Vega2011}
Vega,~C.; Abascal,~J. L.~F. Simulating Water with Rigid Non-polarizable Models:
  A General Perspective. \emph{Phys. Chem. Chem. Phys.} \textbf{2011},
  \emph{13}, 19663--19688\relax
\mciteBstWouldAddEndPuncttrue
\mciteSetBstMidEndSepPunct{\mcitedefaultmidpunct}
{\mcitedefaultendpunct}{\mcitedefaultseppunct}\relax
\EndOfBibitem
\bibitem[Flyvbjerg and Petersen(1989)Flyvbjerg, and Petersen]{Flyvbjerg1989}
Flyvbjerg,~H.; Petersen,~H.~G. Error Estimates on Averages of Correlated Data.
  \emph{J. Chem. Phys.} \textbf{1989}, \emph{91}, 461--466\relax
\mciteBstWouldAddEndPuncttrue
\mciteSetBstMidEndSepPunct{\mcitedefaultmidpunct}
{\mcitedefaultendpunct}{\mcitedefaultseppunct}\relax
\EndOfBibitem
\bibitem[Chodera(2016)]{Chodera2016}
Chodera,~J.~D. A Simple Method for Automated Equilibration Detection in
  Molecular Simulations. \emph{J. Chem. Theory Comput.} \textbf{2016},
  \emph{12}, 1799--1805\relax
\mciteBstWouldAddEndPuncttrue
\mciteSetBstMidEndSepPunct{\mcitedefaultmidpunct}
{\mcitedefaultendpunct}{\mcitedefaultseppunct}\relax
\EndOfBibitem
\bibitem[Philip(1972)]{Philip1972}
Philip,~J.~R. Flows Satisfying Mixed No-Slip and No-Shear Conditions. \emph{Z.
  Angew. Math. Phys.} \textbf{1972}, \emph{23}, 353--372\relax
\mciteBstWouldAddEndPuncttrue
\mciteSetBstMidEndSepPunct{\mcitedefaultmidpunct}
{\mcitedefaultendpunct}{\mcitedefaultseppunct}\relax
\EndOfBibitem
\bibitem[Lauga and Stone(2003)Lauga, and Stone]{LaugaStone2003}
Lauga,~E.; Stone,~H.~A. Effective Slip in Pressure-Driven {Stokes} Flow.
  \emph{J. Fluid Mech.} \textbf{2003}, \emph{489}, 55--77\relax
\mciteBstWouldAddEndPuncttrue
\mciteSetBstMidEndSepPunct{\mcitedefaultmidpunct}
{\mcitedefaultendpunct}{\mcitedefaultseppunct}\relax
\EndOfBibitem
\end{mcitethebibliography}

\end{document}


\begin{center}
{\large\bfseries Supporting Information}\\[0.6em]
{\Large\bfseries Phase-Offset Rectification of Thermo-osmotic Flow in a Structurally
Symmetric Graphene Nanochannel}\\[0.9em]
{Jnanadyuti Patra,$^{\dagger}$ Shiladitya Sengupta,$^{\ddagger}$ and Nandlal Pingua$^{\ast,\dagger}$}\\[0.4em]
{\small $^{\dagger}$Department of Chemical Engineering, National Institute of
Technology Tiruchirappalli, Tiruchirappalli 620015, Tamil Nadu, India}\\[0.15em]
{\small $^{\ddagger}$Department of Physics, Indian Institute of Technology Roorkee, Roorkee 247667, Uttarakhand, India}\\[0.45em]
{\small E-mail: \texttt{nandlal@nitt.edu}}
\end{center}

\vspace{0.6em}
\noindent\textbf{\large Table of Contents}
\begin{itemize}[leftmargin=2.6em, labelsep=0.6em, itemsep=0.15em, topsep=0.4em]
  \item[\textbf{Section~S1:}] Campaign design \dotfill\ \srangelast{sec:si-matrix}{end:si-matrix}
  \item[\textbf{Section~S2:}] Water model, fill and cutoff \dotfill\ \srangelast{sec:si-validation}{end:si-validation}
  \item[\textbf{Section~S3:}] Flux estimators and sampling \dotfill\ \srangelast{sec:si-flux}{end:si-flux}
  \item[\textbf{Section~S4:}] Replica- and structure-level results \dotfill\ \srangelast{sec:si-replicas}{end:si-replicas}
  \item[\textbf{Section~S5:}] Subsets of the global test \dotfill\ \srangelast{sec:si-global}{end:si-global}
  \item[\textbf{Section~S6:}] Interfacial analysis \dotfill\ \srangelast{sec:si-interface}{end:si-interface}
  \item[\textbf{Section~S7:}] Hydraulic mobility, wall drift, stall pressure and slip \dotfill\ \srangelast{sec:si-mobility}{end:si-mobility}
  \item[\textbf{Section~S8:}] Conductance at uniform coverage \dotfill\ \srangelast{sec:si-conductance}{end:si-conductance}
  \item[\textbf{Section~S9:}] Sensitivity to the carbon--water interaction \dotfill\ \srangelast{sec:si-sigma}{end:si-sigma}
  \item[\textbf{Section~S10:}] Hydrodynamics of the patterned channel \dotfill\ \srangelast{sec:si-stokes}{sec:si-end}
\end{itemize}

Every number in the tables below is generated by script from the simulation output
and the analysis files of the campaign; none is entered by hand. Uncertainties are
95\% $t$-intervals across independent replicas unless stated otherwise.

\section{Campaign design}
\label{sec:si-matrix}

Table~\ref{tab:matrix} lists every state point. All share one cell: a
\SI{403.36}{\angstrom} $\times$ \SI{42.60}{\angstrom} graphene bilayer channel of
height $h = \SI{30}{\angstrom}$ (inner carbon plane to inner carbon plane) unless
stated, holding \num{13066} TIP4P/2005 molecules~\cite{Abascal2005}, with the wall temperature imposed
by 20 Langevin zones (four periods) or five zones per period (tier C). The hydroxyl
arrangement is fixed by a build seed that depends only on the geometry and the
replica index, so the eight offsets of a replica share an identical structure and
differ only in the placement of the thermostat pattern. Tier H, added after the
initial campaign, measures the conductance at uniform coverage
(Section~\ref{sec:si-conductance}), adds five hydroxyl arrangements at
$\xi = 0$--0.20 (Section~\ref{sec:si-replicas}) and tests the carbon--water
$\sigma_{\mathrm{CO}}$ (Section~\ref{sec:si-sigma}); the reading of each of its
outcomes, and the analysis code, were fixed before any of it ran.

\FloatBarrier
\begingroup
\footnotesize\renewcommand{\arraystretch}{0.9}
\begin{longtable}{ccccclcc}
\caption{The campaign: every state point. Tiers: A, the phase sweep; B, null controls; C, number of periods; D, drive amplitude; E, hydraulic mobility; F, robustness to cutoff, height and coverage; G, higher coverage; H, tests added after the initial campaign. $\xi$ is the hydroxyl coverage range of the cosine pattern (a single value is uniform coverage), $N$ the number of chemical periods, $\delta/\lambda$ the thermal--chemical offset and $\Delta T$ the imposed wall temperature amplitude about \SI{320}{\kelvin}. Body forces $f$ are per atom in kcal\,mol$^{-1}$\,\AA$^{-1}$. 172 runs and 3,825~ns in total, 582 node-days at the measured throughput.}\label{tab:matrix}\\
\toprule
Tier & $\xi$ & $N$ & $\delta/\lambda$ & $\Delta T$ (K) & Other & Replicas & ns per replica \\
\midrule
\endfirsthead
\multicolumn{8}{l}{\small\itshape Table~\ref{tab:matrix}, continued} \\
\toprule
Tier & $\xi$ & $N$ & $\delta/\lambda$ & $\Delta T$ (K) & Other & Replicas & ns per replica \\
\midrule
\endhead
\bottomrule
\endlastfoot
A & 0.00--0.20 & 4 & 0.000 & 20 & -- & 5 & 25 \\
A & 0.00--0.20 & 4 & 0.125 & 20 & -- & 5 & 25 \\
A & 0.00--0.20 & 4 & 0.250 & 20 & -- & 5 & 25 \\
A & 0.00--0.20 & 4 & 0.375 & 20 & -- & 5 & 25 \\
A & 0.00--0.20 & 4 & 0.500 & 20 & -- & 5 & 25 \\
A & 0.00--0.20 & 4 & 0.625 & 20 & -- & 5 & 25 \\
A & 0.00--0.20 & 4 & 0.750 & 20 & -- & 5 & 25 \\
A & 0.00--0.20 & 4 & 0.875 & 20 & -- & 5 & 25 \\
B & 0.10 & 4 & 0.250 & 20 & -- & 5 & 25 \\
B & 0.00--0.20 & 4 & 0.250 & 20 & two-leg drive & 5 & 25 \\
B & 0.00--0.20 & 4 & 0.250 & 0 & -- & 5 & 25 \\
C & 0.00--0.20 & 2 & 0.250 & 20 & 5 zones/period & 3 & 25 \\
C & 0.00--0.20 & 3 & 0.250 & 20 & 5 zones/period & 3 & 25 \\
C & 0.00--0.20 & 4 & 0.250 & 20 & 5 zones/period & 3 & 25 \\
C & 0.00--0.20 & 5 & 0.250 & 20 & 5 zones/period & 3 & 25 \\
C & 0.00--0.20 & 8 & 0.250 & 20 & 5 zones/period & 3 & 25 \\
C & 0.00--0.20 & 10 & 0.250 & 20 & 5 zones/period & 3 & 25 \\
D & 0.00--0.20 & 4 & 0.250 & 5 & -- & 3 & 20 \\
D & 0.00--0.20 & 4 & 0.250 & 10 & -- & 3 & 20 \\
D & 0.00--0.20 & 4 & 0.250 & 30 & -- & 3 & 20 \\
D & 0.00--0.20 & 4 & 0.250 & 40 & -- & 3 & 20 \\
D & 0.00--0.20 & 4 & 0.250 & 40 & -- & 5 & 50 \\
E & 0.00--0.20 & 4 & 0.250 & 0 & $f$ = $2\times10^{-5}$ & 3 & 10 \\
E & 0.00--0.20 & 4 & 0.250 & 0 & $f$ = $5\times10^{-5}$ & 3 & 10 \\
E & 0.00--0.20 & 4 & 0.250 & 0 & $f$ = $1\times10^{-4}$ & 3 & 10 \\
F & 0.00--0.20 & 4 & 0.250 & 20 & $r_c = 14$~\AA & 3 & 20 \\
F & 0.00--0.20 & 4 & 0.250 & 20 & $h = 25$~\AA & 3 & 20 \\
F & 0.00--0.10 & 4 & 0.250 & 20 & -- & 3 & 20 \\
G & 0.00--0.30 & 4 & 0.250 & 20 & -- & 5 & 25 \\
G & 0.00--0.30 & 4 & 0.750 & 20 & -- & 5 & 25 \\
H & 0.00 & 4 & 0.250 & 0 & $f$ = $5\times10^{-5}$ & 3 & 10 \\
H & 0.10 & 4 & 0.250 & 0 & $f$ = $5\times10^{-5}$ & 3 & 10 \\
H & 0.20 & 4 & 0.250 & 0 & $f$ = $5\times10^{-5}$ & 3 & 10 \\
H & 0.30 & 4 & 0.250 & 0 & $f$ = $5\times10^{-5}$ & 3 & 10 \\
H & 0.00 & 4 & 0.250 & 0 & $f$ = $1\times10^{-5}$ & 3 & 10 \\
H & 0.00--0.20 & 4 & 0.250 & 20 & structures 5--9 & 5 & 25 \\
H & 0.00--0.20 & 4 & 0.750 & 20 & structures 5--9 & 5 & 25 \\
H & 0.00--0.20 & 4 & 0.250 & 20 & $\sigma_{\mathrm{CO}} = 3.349$~\AA & 5 & 25 \\
H & 0.00--0.20 & 4 & 0.750 & 20 & $\sigma_{\mathrm{CO}} = 3.349$~\AA & 5 & 25 \\
H & 0.00--0.20 & 4 & 0.250 & 20 & $\sigma_{\mathrm{CO}} = 3.349$~\AA; structures 5--9 & 5 & 25 \\
H & 0.00--0.20 & 4 & 0.750 & 20 & $\sigma_{\mathrm{CO}} = 3.349$~\AA; structures 5--9 & 5 & 25 \\
H & 0.005 & 4 & 0.250 & 0 & $f$ = $5\times10^{-5}$ & 3 & 10 \\
H & 0.02 & 4 & 0.250 & 0 & $f$ = $5\times10^{-5}$ & 3 & 10 \\
H & 0.05 & 4 & 0.250 & 0 & $f$ = $5\times10^{-5}$ & 3 & 10 \\
\end{longtable}
\endgroup

\sectionend{end:si-matrix}
\FloatBarrier
\section{Water model, fill and cutoff}
\label{sec:si-validation}

\begin{figure}[H]
\centering
\includegraphics[width=\textwidth,alt={(a) Density of bulk TIP4P/2005 water against temperature at 1 atm, compared with experiment. (b) Normal pressure in the calibration cell against the number of water molecules, with the calibrated count marked.}]{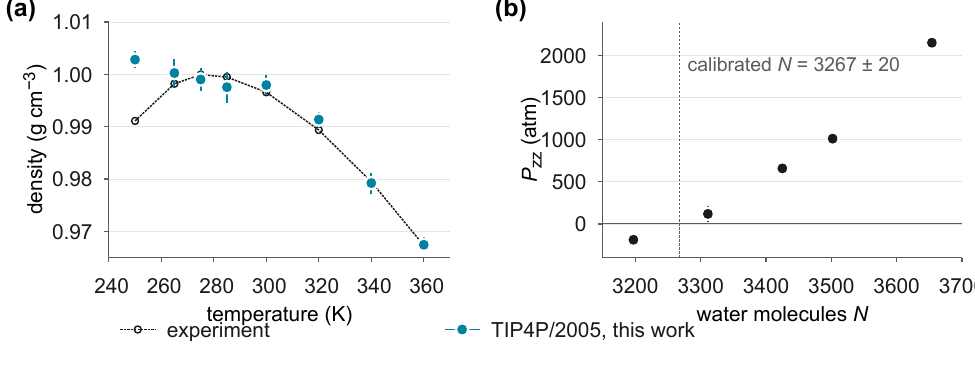}
\caption{(a) Density of bulk TIP4P/2005 against temperature at \SI{1}{\atmosphere},
with experiment. (b) Normal pressure in the calibration cell against the number of
water molecules; the dotted line marks the calibrated count, read from the
bracketing pair.}
\label{fig:validation}
\end{figure}

The bulk equation of state (Table~\ref{tab:bulk}, Figure~\ref{fig:validation}a)
reproduces the experimental density to within 0.20\% between 265 and \SI{360}{\kelvin},
with the known TIP4P/2005 overestimate in the supercooled liquid at
\SI{250}{\kelvin}~\cite{Vega2011}.
Because the loop is fully periodic and has no reservoir, the number of molecules in
a fixed box sets the pressure, so it was calibrated against the normal pressure
(Table~\ref{tab:calibration}, Figure~\ref{fig:validation}b). The Lennard-Jones cutoff
was tested by re-evaluating identical configurations, which removes run-to-run noise
from the comparison (Table~\ref{tab:cutoff}).

\begin{table}[H]
\centering
\footnotesize
\caption{Bulk TIP4P/2005 density at \SI{1}{\atmosphere}: 512 molecules, 60~ps per temperature at a 2~fs timestep, uncertainties are block errors~\cite{Flyvbjerg1989}. Experimental densities of liquid water for comparison.}
\label{tab:bulk}
\begin{tabular}{ccccc}
\toprule
$T$ target (K) & $T$ measured (K) & $\rho$ (g\,cm$^{-3}$) & $\rho_{\mathrm{exp}}$ (g\,cm$^{-3}$) & deviation (\%) \\
\midrule
250 & $247.7$ & $1.0028 \pm 0.0016$ & 0.9911 & $+1.18$ \\
265 & $264.2$ & $1.0002 \pm 0.0028$ & 0.9982 & $+0.20$ \\
275 & $274.5$ & $0.9990 \pm 0.0022$ & 0.9999 & $-0.09$ \\
285 & $285.3$ & $0.9975 \pm 0.0030$ & 0.9995 & $-0.19$ \\
300 & $299.9$ & $0.9980 \pm 0.0019$ & 0.9966 & $+0.14$ \\
320 & $319.0$ & $0.9914 \pm 0.0015$ & 0.9894 & $+0.20$ \\
340 & $341.3$ & $0.9792 \pm 0.0020$ & 0.9795 & $-0.02$ \\
360 & $361.0$ & $0.9675 \pm 0.0014$ & 0.9673 & $+0.01$ \\
\bottomrule
\end{tabular}
\end{table}

\begin{table}[H]
\centering
\footnotesize
\caption{Calibration of the water count in the calibration cell ($h$ = 30~\AA, 320~K): normal pressure $P_{zz}$ against the number of molecules. The crossing of \SI{1}{\atmosphere} is read from the bracketing pair ($N = 3197$, $-188$~atm and $N = 3311$, $119$~atm) because $P_{zz}(N)$ is convex (segment slopes 2.7, 4.74, 4.59, 7.5 atm per molecule). The resulting excluded layer, 3.500~\AA\ per wall, sets the count in every production cell; the uncalibrated estimate (1.6~\AA, 3806 molecules) would have run production near \SI{1450}{\atmosphere}.}
\label{tab:calibration}
\begin{tabular}{cc}
\toprule
$N$ & $P_{zz}$ (atm) \\
\midrule
3197 & $-188 \pm 49$ \\
3311 & $+119 \pm 96$ \\
3425 & $+660 \pm 68$ \\
3502 & $+1013 \pm 78$ \\
3654 & $+2153 \pm 70$ \\
\midrule
crossing: 3267 $\pm$ 20 & excluded layer 3.500~\AA \\
\bottomrule
\end{tabular}
\end{table}

\begin{table}[H]
\centering
\footnotesize
\caption{Lennard-Jones cutoff, by paired re-evaluation of the same 61 configurations of a functionalised, calibrated cell (3266 water molecules). The interfacial excess is in kcal\,mol$^{-1}$ per molecule; differences are against \SI{14}{\angstrom}, with the paired standard error. The criterion fixed in advance was a difference below 5\%.}
\label{tab:cutoff}
\begin{tabular}{ccccc}
\toprule
$r_c$ (\AA) & frames & excess & difference from 14~\AA & relative (\%) \\
\midrule
10 & 61 & $-0.0920$ & $+0.00362 \pm 0.00002$ & $3.8$ \\
12 & 61 & $-0.0946$ & $+0.00108 \pm 0.00001$ & $1.1$ \\
14 & 61 & $-0.0956$ & reference & -- \\
\bottomrule
\end{tabular}
\end{table}

\sectionend{end:si-validation}
\FloatBarrier
\section{Flux estimators and sampling}
\label{sec:si-flux}

The loop velocity is measured three ways (Table~\ref{tab:routes}). The
centre-of-mass and plane-crossing routes share no input: one integrates velocities,
the other counts molecules crossing planes using positions alone, and they agree to
within \SI{1}{\centi\meter\per\second} in every run of tiers A--G and within
\SI{2}{\centi\meter\per\second} on every hydroxylated wall of tier H. They part on
pristine graphene. At the larger force the water there moves \SI{198}{\angstrom}
between stored frames, \SI{500}{\pico\second} apart, against half the
\SI{403}{\angstrom} box, so the minimum-image displacement that the crossing count
relies on cannot tell forward motion from backward for a large share of the
molecules, and the count fails; the centre-of-mass and profile routes, which do not
depend on the frame interval, agree there. At the smaller force one replica differs
by \SI{0.6}{\meter\per\second} only because the equilibration detection removes the
first 27\% of its centre-of-mass series; over the same window the two routes agree to
\SI{0.004}{\meter\per\second}. The profile route averages the
axial velocity profile and is written far less often, so its intervals are wider.
The statistical inefficiency of the centre-of-mass series is close to 1.6 in every
tier but H (Table~\ref{tab:sampling}), giving several thousand effectively independent
samples per run. On pristine graphene it rises to about 100: that channel conducts
some 180 times more freely (Section~\ref{sec:si-conductance}), so the wall friction
that damps fluctuations of the centre-of-mass velocity is much weaker, the series
decorrelates far more slowly, and 40 to 60 effectively independent samples remain
per run. The
uncertainties are taken across replicas in either case. The wall drift velocity is of order \SI{e-5}{\meter\per\second}
in every thermally driven run and is discussed for the body-force runs in
Section~\ref{sec:si-mobility}.

\FloatBarrier
\begingroup
\scriptsize\setlength{\tabcolsep}{3.2pt}
\begin{longtable}{llcccccccc}
\caption{Loop velocity of every completed state point by the three estimators: the water centre-of-mass velocity (COM), the count of molecules crossing reference planes, and the mass-weighted mean of the axial velocity profile. Velocities in m\,s$^{-1}$, raw (no background subtraction), means with 95\% $t$-intervals across replicas. $\max|\Delta|$ is the largest per-run difference between the COM and crossing routes; $\bar g$ the mean statistical inefficiency of the COM series; the last column the largest wall drift velocity of any run (m\,s$^{-1}$), which is negligible except under the body forces of tiers E and H (Table~\ref{tab:mobility}). The profile route samples the flow 50 times less often and is correspondingly noisier. $^{a}$On the pristine wall at the larger force the water moves nearly half the box between stored frames, and the crossing count fails (Section~\ref{sec:si-flux}).}\label{tab:routes}\\
\toprule
Tier & State point & $n$ & ns & COM & crossing & profile & $\max|\Delta|$ & $\bar g$ & wall \\
\midrule
\endfirsthead
\multicolumn{10}{l}{\small\itshape Table~\ref{tab:routes}, continued} \\
\toprule
Tier & State point & $n$ & ns & COM & crossing & profile & $\max|\Delta|$ & $\bar g$ & wall \\
\midrule
\endhead
\bottomrule
\endlastfoot
A & $\delta/\lambda = 0.000$ & 5 & 25 & $-0.0097 \pm 0.0243$ & $-0.0087 \pm 0.0258$ & $-0.061 \pm 0.060$ & $0.0053$ & $1.69$ & $1.9\times10^{-5}$ \\
A & $\delta/\lambda = 0.125$ & 5 & 25 & $+0.0271 \pm 0.0102$ & $+0.0265 \pm 0.0109$ & $+0.034 \pm 0.084$ & $0.0025$ & $1.67$ & $1.5\times10^{-5}$ \\
A & $\delta/\lambda = 0.250$ & 5 & 25 & $+0.0203 \pm 0.0486$ & $+0.0195 \pm 0.0491$ & $+0.054 \pm 0.050$ & $0.0043$ & $1.63$ & $2.6\times10^{-5}$ \\
A & $\delta/\lambda = 0.375$ & 5 & 25 & $+0.0113 \pm 0.0395$ & $+0.0113 \pm 0.0415$ & $-0.018 \pm 0.042$ & $0.0030$ & $1.60$ & $1.4\times10^{-5}$ \\
A & $\delta/\lambda = 0.500$ & 5 & 25 & $+0.0015 \pm 0.0362$ & $+0.0017 \pm 0.0362$ & $-0.007 \pm 0.102$ & $0.0043$ & $1.58$ & $1.6\times10^{-5}$ \\
A & $\delta/\lambda = 0.625$ & 5 & 25 & $-0.0100 \pm 0.0127$ & $-0.0093 \pm 0.0125$ & $-0.005 \pm 0.072$ & $0.0018$ & $1.60$ & $1.5\times10^{-5}$ \\
A & $\delta/\lambda = 0.750$ & 5 & 25 & $-0.0389 \pm 0.0304$ & $-0.0380 \pm 0.0311$ & $-0.071 \pm 0.050$ & $0.0043$ & $1.64$ & $1.3\times10^{-5}$ \\
A & $\delta/\lambda = 0.875$ & 5 & 25 & $-0.0126 \pm 0.0526$ & $-0.0129 \pm 0.0525$ & $-0.040 \pm 0.087$ & $0.0017$ & $1.68$ & $2.1\times10^{-5}$ \\
B & uniform $\xi = 0.10$ & 5 & 25 & $+0.0115 \pm 0.0339$ & $+0.0096 \pm 0.0364$ & $-0.029 \pm 0.058$ & $0.0033$ & $1.46$ & $9.7\times10^{-6}$ \\
B & two-leg drive & 5 & 25 & $-0.0014 \pm 0.0437$ & $-0.0018 \pm 0.0445$ & $-0.001 \pm 0.087$ & $0.0016$ & $1.63$ & $1.3\times10^{-5}$ \\
B & isothermal & 5 & 25 & $+0.0116 \pm 0.0490$ & $+0.0115 \pm 0.0506$ & $+0.003 \pm 0.104$ & $0.0034$ & $1.64$ & $1.2\times10^{-5}$ \\
C & $N = 2$ & 3 & 25 & $+0.0154 \pm 0.1093$ & $+0.0141 \pm 0.1124$ & $+0.050 \pm 0.080$ & $0.0022$ & $1.77$ & $8.5\times10^{-6}$ \\
C & $N = 3$ & 3 & 25 & $+0.0089 \pm 0.0381$ & $+0.0099 \pm 0.0373$ & $+0.036 \pm 0.111$ & $0.0024$ & $1.69$ & $1.2\times10^{-5}$ \\
C & $N = 4$ & 3 & 25 & $+0.0235 \pm 0.0518$ & $+0.0207 \pm 0.0453$ & $+0.019 \pm 0.196$ & $0.0056$ & $1.63$ & $1.4\times10^{-5}$ \\
C & $N = 5$ & 3 & 25 & $+0.0301 \pm 0.0636$ & $+0.0298 \pm 0.0592$ & $+0.010 \pm 0.078$ & $0.0026$ & $1.58$ & $3.3\times10^{-6}$ \\
C & $N = 8$ & 3 & 25 & $+0.0340 \pm 0.0920$ & $+0.0328 \pm 0.0881$ & $+0.066 \pm 0.104$ & $0.0034$ & $1.55$ & $8.0\times10^{-6}$ \\
C & $N = 10$ & 3 & 25 & $+0.0639 \pm 0.0556$ & $+0.0630 \pm 0.0577$ & $-0.024 \pm 0.018$ & $0.0018$ & $1.54$ & $1.2\times10^{-5}$ \\
D & $\Delta T = 5$~K & 3 & 20 & $-0.0134 \pm 0.0220$ & $-0.0134 \pm 0.0222$ & $-0.052 \pm 0.226$ & $0.0015$ & $1.63$ & $1.7\times10^{-5}$ \\
D & $\Delta T = 10$~K & 3 & 20 & $-0.0022 \pm 0.0263$ & $-0.0028 \pm 0.0217$ & $-0.004 \pm 0.143$ & $0.0031$ & $1.64$ & $1.8\times10^{-5}$ \\
D & $\Delta T = 30$~K & 3 & 20 & $+0.0080 \pm 0.0819$ & $+0.0082 \pm 0.0791$ & $-0.005 \pm 0.227$ & $0.0019$ & $1.63$ & $1.8\times10^{-5}$ \\
D & $\Delta T = 40$~K & 3 & 20 & $+0.0339 \pm 0.0538$ & $+0.0325 \pm 0.0544$ & $-0.014 \pm 0.261$ & $0.0026$ & $1.63$ & $2.0\times10^{-5}$ \\
D & $\Delta T = 40$~K, 5 $\times$ 50~ns & 5 & 50 & $+0.0334 \pm 0.0303$ & $+0.0324 \pm 0.0303$ & $+0.066 \pm 0.074$ & $0.0013$ & $1.61$ & $3.9\times10^{-6}$ \\
E & $f = 2\times10^{-5}$ & 3 & 10 & $+0.1334 \pm 0.0350$ & $+0.1325 \pm 0.0273$ & $+0.220 \pm 0.067$ & $0.0048$ & $1.65$ & $9.7\times10^{-3}$ \\
E & $f = 5\times10^{-5}$ & 3 & 10 & $+0.2877 \pm 0.1341$ & $+0.2909 \pm 0.1442$ & $+0.261 \pm 0.123$ & $0.0099$ & $1.63$ & $2.4\times10^{-2}$ \\
E & $f = 1\times10^{-4}$ & 3 & 10 & $+0.6495 \pm 0.0428$ & $+0.6463 \pm 0.0406$ & $+0.615 \pm 0.099$ & $0.0050$ & $1.63$ & $4.9\times10^{-2}$ \\
F & $r_c = 14$~\AA & 3 & 20 & $+0.0405 \pm 0.0647$ & $+0.0384 \pm 0.0644$ & $+0.125 \pm 0.301$ & $0.0032$ & $1.63$ & $1.9\times10^{-5}$ \\
F & $h = 25$~\AA & 3 & 20 & $+0.0577 \pm 0.0113$ & $+0.0564 \pm 0.0093$ & $+0.046 \pm 0.167$ & $0.0047$ & $1.32$ & $6.1\times10^{-6}$ \\
F & $\xi = 0$--0.10 & 3 & 20 & $+0.0179 \pm 0.0837$ & $+0.0169 \pm 0.0807$ & $+0.052 \pm 0.131$ & $0.0032$ & $1.81$ & $5.0\times10^{-6}$ \\
G & $\xi = 0$--0.30, $\delta = \lambda/4$ & 5 & 25 & $+0.0368 \pm 0.0487$ & $+0.0362 \pm 0.0476$ & $+0.016 \pm 0.086$ & $0.0017$ & $1.58$ & $1.8\times10^{-5}$ \\
G & $\xi = 0$--0.30, $\delta = 3\lambda/4$ & 5 & 25 & $-0.0448 \pm 0.0150$ & $-0.0437 \pm 0.0159$ & $-0.047 \pm 0.104$ & $0.0018$ & $1.58$ & $1.4\times10^{-5}$ \\
H & uniform $\xi = 0.00$, $f = 5\times10^{-5}$ & 3 & 10 & $+39.6839 \pm 0.3292$ & $+4.2420 \pm 1.7222$$^{a}$ & $+39.815 \pm 0.461$ & $36.3315$$^{a}$ & $95.07$ & $2.6\times10^{-2}$ \\
H & uniform $\xi = 0.10$, $f = 5\times10^{-5}$ & 3 & 10 & $+0.2437 \pm 0.1308$ & $+0.2448 \pm 0.1387$ & $+0.227 \pm 0.131$ & $0.0059$ & $1.44$ & $2.4\times10^{-2}$ \\
H & uniform $\xi = 0.20$, $f = 5\times10^{-5}$ & 3 & 10 & $+0.2464 \pm 0.1160$ & $+0.2461 \pm 0.1270$ & $+0.189 \pm 0.293$ & $0.0054$ & $1.40$ & $2.3\times10^{-2}$ \\
H & uniform $\xi = 0.30$, $f = 5\times10^{-5}$ & 3 & 10 & $+0.2651 \pm 0.0672$ & $+0.2681 \pm 0.0791$ & $+0.256 \pm 0.254$ & $0.0072$ & $1.47$ & $2.1\times10^{-2}$ \\
H & uniform $\xi = 0.00$, $f = 1\times10^{-5}$ & 3 & 10 & $+7.1414 \pm 4.4111$ & $+7.3443 \pm 3.5560$ & $+7.189 \pm 4.554$ & $0.6030$ & $104.45$ & $5.2\times10^{-3}$ \\
H & $\delta = \lambda/4$, structures 5--9 & 5 & 25 & $+0.0169 \pm 0.0388$ & $+0.0153 \pm 0.0372$ & $+0.046 \pm 0.069$ & $0.0052$ & $1.63$ & $1.1\times10^{-5}$ \\
H & $\delta = 3\lambda/4$, structures 5--9 & 5 & 25 & $-0.0096 \pm 0.0370$ & $-0.0083 \pm 0.0379$ & $+0.044 \pm 0.071$ & $0.0028$ & $1.64$ & $1.6\times10^{-5}$ \\
H & $\sigma_{\mathrm{CO}} = 3.349$~\AA, $\delta = \lambda/4$ & 5 & 25 & $+0.0087 \pm 0.0364$ & $+0.0092 \pm 0.0358$ & $+0.002 \pm 0.046$ & $0.0037$ & $1.62$ & $1.2\times10^{-5}$ \\
H & $\sigma_{\mathrm{CO}} = 3.349$~\AA, $\delta = 3\lambda/4$ & 5 & 25 & $-0.0042 \pm 0.0268$ & $-0.0040 \pm 0.0252$ & $-0.030 \pm 0.061$ & $0.0035$ & $1.64$ & $1.7\times10^{-5}$ \\
H & $\sigma_{\mathrm{CO}} = 3.349$~\AA, $\delta = \lambda/4$, structures 5--9 & 5 & 25 & $+0.0399 \pm 0.0342$ & $+0.0406 \pm 0.0360$ & $+0.003 \pm 0.058$ & $0.0030$ & $1.62$ & $1.8\times10^{-5}$ \\
H & $\sigma_{\mathrm{CO}} = 3.349$~\AA, $\delta = 3\lambda/4$, structures 5--9 & 5 & 25 & $-0.0144 \pm 0.0437$ & $-0.0141 \pm 0.0441$ & $+0.021 \pm 0.061$ & $0.0016$ & $1.62$ & $1.6\times10^{-5}$ \\
H & uniform $\xi = 0.005$, $f = 5\times10^{-5}$ & 3 & 10 & $+1.8199 \pm 0.5459$ & $+1.8092 \pm 0.5492$ & $+1.641 \pm 0.585$ & $0.0163$ & $5.96$ & $2.6\times10^{-2}$ \\
H & uniform $\xi = 0.02$, $f = 5\times10^{-5}$ & 3 & 10 & $+0.6766 \pm 0.1563$ & $+0.6761 \pm 0.1575$ & $+0.741 \pm 0.093$ & $0.0011$ & $2.43$ & $2.6\times10^{-2}$ \\
H & uniform $\xi = 0.05$, $f = 5\times10^{-5}$ & 3 & 10 & $+0.3751 \pm 0.0277$ & $+0.3757 \pm 0.0167$ & $+0.283 \pm 0.080$ & $0.0057$ & $1.64$ & $2.5\times10^{-2}$ \\
\end{longtable}
\endgroup

\begin{table}[H]
\centering
\footnotesize
\caption{Sampling of the centre-of-mass velocity series. $g$ is the statistical inefficiency; the discarded fraction is the initial transient removed by automated equilibration detection~\cite{Chodera2016}; $N_{\mathrm{eff}}$ the effective number of independent samples per run. The wide range in tier H comes from its body-force runs on pristine graphene, where $g$ reaches 122 (Section~\ref{sec:si-flux}).}
\label{tab:sampling}
\begin{tabular}{cccccc}
\toprule
Tier & runs & mean $g$ & range of $g$ & discarded (\%) & mean $N_{\mathrm{eff}}$ \\
\midrule
A & 40 & $1.64$ & 1.57--1.70 & $0.0$ & 7,649 \\
B & 15 & $1.58$ & 1.44--1.65 & $0.0$ & 7,953 \\
C & 18 & $1.63$ & 1.52--1.80 & $0.0$ & 7,705 \\
D & 17 & $1.63$ & 1.59--1.66 & $0.0$ & 8,897 \\
E & 9 & $1.64$ & 1.61--1.66 & $0.0$ & 3,054 \\
F & 9 & $1.59$ & 1.31--1.82 & $0.0$ & 6,400 \\
G & 10 & $1.58$ & 1.56--1.60 & $0.0$ & 7,916 \\
H & 54 & $12.79$ & 1.37--122.23 & $0.6$ & 5,179 \\
\bottomrule
\end{tabular}
\end{table}

\sectionend{end:si-flux}
\FloatBarrier
\section{Replica- and structure-level results}
\label{sec:si-replicas}

\begin{figure}[H]
\centering
\includegraphics[width=0.88\textwidth,alt={Loop velocity against phase offset for each of the five tier A replicas, one hydroxyl arrangement per panel.}]{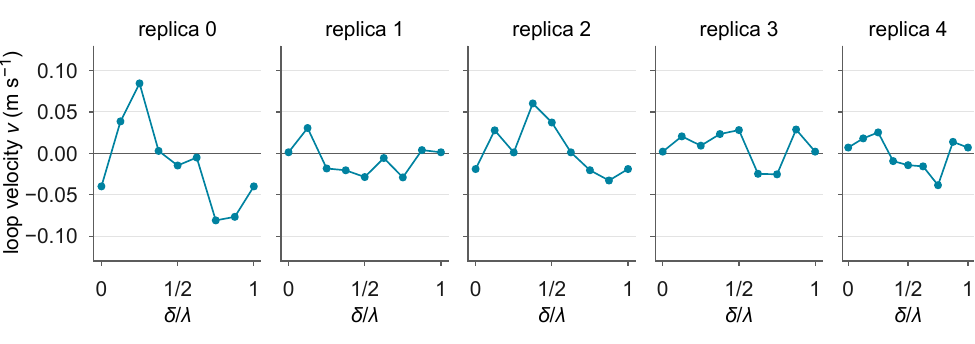}
\caption{Loop velocity against the offset for each of the five tier A replicas
(centre-of-mass route, raw), one hydroxyl arrangement per panel.}
\label{fig:replicas}
\end{figure}

Table~\ref{tab:replicas} and Figure~\ref{fig:replicas} give the tier A loop velocity
for each replica at each offset. Replica 0 carries the largest amplitude; all five
per-replica fits of $v_{\max}$ have the same sign, but their 95\% interval includes
zero, which is why the manuscript rests the rectification on the structure-level
reversal of Table~\ref{tab:structures} rather than on the tier A amplitude alone.
That table holds fifteen arrangements: the five of the phase sweep, five built
afterwards at the same coverage, and five at the higher coverage. Two of the added
arrangements reverse in the opposite sense, each by a third of the per-run standard
deviation. The pairing is by the build seed recorded in each run's manifest, never
by directory name or replica index; the added arrangements carry replica indices 5--9
so that their build seeds differ from those of the phase sweep.

\begin{table}[H]
\centering
\footnotesize
\caption{Tier A loop velocity (m\,s$^{-1}$, centre-of-mass route, raw) for each replica at each offset $\delta/\lambda$. Replica $k$ uses the same hydroxyl arrangement at every offset. The last row is eq.~(3) with a cosine term fitted to each replica's eight offsets; its 95\% $t$-interval across replicas is the uncertainty used for the stall pressure.}
\label{tab:replicas}
\begin{tabular}{cccccccc}
\toprule
$\delta/\lambda$ & $r_0$ & $r_1$ & $r_2$ & $r_3$ & $r_4$ & mean \\
\midrule
0.000 & $-0.0399$ & $+0.0013$ & $-0.0190$ & $+0.0020$ & $+0.0070$ & $-0.0097 \pm 0.0243$ \\
0.125 & $+0.0385$ & $+0.0305$ & $+0.0278$ & $+0.0205$ & $+0.0180$ & $+0.0271 \pm 0.0102$ \\
0.250 & $+0.0845$ & $-0.0183$ & $+0.0011$ & $+0.0092$ & $+0.0252$ & $+0.0203 \pm 0.0486$ \\
0.375 & $+0.0029$ & $-0.0205$ & $+0.0602$ & $+0.0233$ & $-0.0094$ & $+0.0113 \pm 0.0395$ \\
0.500 & $-0.0147$ & $-0.0286$ & $+0.0372$ & $+0.0281$ & $-0.0143$ & $+0.0015 \pm 0.0362$ \\
0.625 & $-0.0051$ & $-0.0057$ & $+0.0012$ & $-0.0247$ & $-0.0158$ & $-0.0100 \pm 0.0127$ \\
0.750 & $-0.0810$ & $-0.0291$ & $-0.0204$ & $-0.0254$ & $-0.0385$ & $-0.0389 \pm 0.0304$ \\
0.875 & $-0.0767$ & $+0.0039$ & $-0.0328$ & $+0.0287$ & $+0.0138$ & $-0.0126 \pm 0.0526$ \\
\midrule
$v_{\max}$ & $-0.0631$ & $-0.0048$ & $-0.0265$ & $-0.0157$ & $-0.0178$ & $-0.0256 \pm 0.0278$ \\
\bottomrule
\end{tabular}
\end{table}

\FloatBarrier
\begingroup
\footnotesize
\begin{longtable}{llcccc}
\caption{Structure-level reversal. Each hydroxyl arrangement, identified by the build seed recorded in its manifest, was run at $\delta = \lambda/4$ and $3\lambda/4$; the amplitude is half the difference (m\,s$^{-1}$, raw centre-of-mass route). Tier A is the phase sweep, tier H five further arrangements built at the same coverage, tier G the higher coverage. Summary rows give the mean with its 95\% $t$-interval and the two-sided one-sample $t$-test; 13 of 15 amplitudes are positive (exact sign test $p = 0.0074$).}\label{tab:structures}\\
\toprule
Tier & $\xi$ & build seed & $v(\lambda/4)$ & $v(3\lambda/4)$ & amplitude \\
\midrule
\endfirsthead
\multicolumn{6}{l}{\small\itshape Table~\ref{tab:structures}, continued} \\
\toprule
Tier & $\xi$ & build seed & $v(\lambda/4)$ & $v(3\lambda/4)$ & amplitude \\
\midrule
\endhead
\bottomrule
\endlastfoot
A & 0--0.20 & 974000 & $+0.0845$ & $-0.0810$ & $+0.0827$ \\
A & 0--0.20 & 974001 & $-0.0183$ & $-0.0291$ & $+0.0054$ \\
A & 0--0.20 & 974002 & $+0.0011$ & $-0.0204$ & $+0.0107$ \\
A & 0--0.20 & 974003 & $+0.0092$ & $-0.0254$ & $+0.0173$ \\
A & 0--0.20 & 974004 & $+0.0252$ & $-0.0385$ & $+0.0318$ \\
H & 0--0.20 & 974005 & $+0.0357$ & $-0.0037$ & $+0.0197$ \\
H & 0--0.20 & 974006 & $+0.0238$ & $-0.0160$ & $+0.0199$ \\
H & 0--0.20 & 974007 & $+0.0450$ & $-0.0473$ & $+0.0462$ \\
H & 0--0.20 & 974008 & $-0.0352$ & $-0.0162$ & $-0.0095$ \\
H & 0--0.20 & 974009 & $+0.0151$ & $+0.0352$ & $-0.0101$ \\
G & 0--0.30 & 104000 & $-0.0091$ & $-0.0608$ & $+0.0258$ \\
G & 0--0.30 & 104001 & $+0.0951$ & $-0.0420$ & $+0.0685$ \\
G & 0--0.30 & 104002 & $+0.0239$ & $-0.0362$ & $+0.0301$ \\
G & 0--0.30 & 104003 & $+0.0216$ & $-0.0317$ & $+0.0266$ \\
G & 0--0.30 & 104004 & $+0.0524$ & $-0.0536$ & $+0.0530$ \\
\midrule
A &  & 5 structures & 5 positive &  & $+0.0296 \pm 0.0389$, $p = 0.1020$ \\
H &  & 5 structures & 3 positive &  & $+0.0132 \pm 0.0293$, $p = 0.2780$ \\
A+H &  & 10 structures & 8 positive &  & $+0.0214 \pm 0.0197$, $p = 0.0361$ \\
all &  & 15 structures & 13 positive &  & $+0.0279 \pm 0.0144$, $p = 0.0010$ \\
\end{longtable}
\endgroup

\sectionend{end:si-replicas}
\FloatBarrier
\section{Subsets of the global test}
\label{sec:si-global}

The global test of the manuscript fits all 24 thermally driven state points
together, with one scale factor on $M$ and an offset consistent with zero. Table~\ref{tab:globalfits} repeats it on each sweep separately. With the
wall temperature as the driving field, every subset gives a positive $k$ consistent
with the global value; with the channel-averaged water temperature, the
period-number sweep alone gives a $k$ of the wrong sign.

\begin{table}[H]
\centering
\footnotesize
\caption{The one-parameter test $v = k\,v_{\mathrm{model}} + c$ on subsets of the thermally driven state points, with the model driven by the measured wall zone temperatures or by the channel-averaged water temperature profile, and uniform (measured) slip. Errors are from the per-run noise pooled over all runs (0.0283~m\,s$^{-1}$ at 25~ns); $p$ is the $\chi^2$ probability. The water temperature gives a negative $k$ on the period-number sweep alone and is excluded as the driving field.}
\label{tab:globalfits}
\begin{tabular}{llcccc}
\toprule
Driving field & Subset & $k$ & $c$ (m\,s$^{-1}$) & $\chi^2$/dof & $p$ \\
\midrule
wall temperature & all 24 points & $+0.0158 \pm 0.0019$ & $-0.0024 \pm 0.0032$ & 12.3/22 & 0.95 \\
 & phase offset & $+0.0156 \pm 0.0037$ & $-0.0004 \pm 0.0040$ & 4.7/8 & 0.79 \\
 & number of periods & $+0.0243 \pm 0.0100$ & $-0.0102 \pm 0.0176$ & 1.1/4 & 0.90 \\
 & drive amplitude & $+0.0185 \pm 0.0067$ & $-0.0193 \pm 0.0157$ & 0.5/3 & 0.93 \\
 & offset and coverage & $+0.0173 \pm 0.0028$ & $-0.0010 \pm 0.0037$ & 5.3/10 & 0.87 \\
\midrule
water temperature & all 24 points & $+0.0224 \pm 0.0030$ & $-0.0007 \pm 0.0031$ & 20.7/22 & 0.54 \\
 & phase offset & $+0.0229 \pm 0.0055$ & $-0.0003 \pm 0.0040$ & 4.7/8 & 0.79 \\
 & number of periods & $-0.0694 \pm 0.0646$ & $+0.0858 \pm 0.0530$ & 5.8/4 & 0.21 \\
 & drive amplitude & $+0.0274 \pm 0.0099$ & $-0.0194 \pm 0.0157$ & 0.5/3 & 0.93 \\
 & offset and coverage & $+0.0249 \pm 0.0040$ & $-0.0009 \pm 0.0037$ & 5.1/10 & 0.89 \\
\bottomrule
\end{tabular}
\end{table}

\sectionend{end:si-global}
\section{Interfacial analysis}
\label{sec:si-interface}

Table~\ref{tab:hbcriteria} shows the first-layer hydrogen-bond composition under
each of the four geometric criteria, and Table~\ref{tab:excess} the dependence of
the energy-excess coefficients on the group cutoff and on the origin of the depth
weighting.

\begin{table}[H]
\centering
\footnotesize
\caption{Hydrogen bonds per first-layer water molecule over $\lambda/4$ windows centred on the coverage maximum (rich) and minimum (poor), by partner, for the four geometric criteria (O$\cdots$O cutoff, H--O$\cdots$O angle). Differences are rich relative to poor, with 95\% $t$-intervals across five replicas. The water--water deficit and the small net gain are insensitive to the criterion.}
\label{tab:hbcriteria}
\begin{tabular}{lcccccc}
\toprule
Criterion & rich, water & rich, hydroxyl & poor, water & poor, hydroxyl & $\Delta$ total (\%) & $\Delta$ water--water (\%) \\
\midrule
3.2~\AA, 25$^{\circ}$ & $2.18$ & $0.65$ & $2.60$ & $0.07$ & $+5.8 \pm 0.8$ & $-16.1 \pm 0.9$ \\
3.4~\AA, 30$^{\circ}$ & $2.46$ & $0.82$ & $2.98$ & $0.09$ & $+6.9 \pm 0.7$ & $-17.4 \pm 0.7$ \\
3.5~\AA, 30$^{\circ}$ & $2.50$ & $0.84$ & $3.03$ & $0.09$ & $+6.9 \pm 0.7$ & $-17.6 \pm 0.7$ \\
3.5~\AA, 35$^{\circ}$ & $2.63$ & $0.96$ & $3.21$ & $0.10$ & $+8.5 \pm 0.7$ & $-18.0 \pm 0.7$ \\
\bottomrule
\end{tabular}
\end{table}

\begin{table}[H]
\centering
\footnotesize
\caption{Chemistry-locked $n = 4$ coefficient of the depth-weighted potential-energy excess of water within 1.5~nm of the wall (kcal\,mol$^{-1}$\,\AA$^{-1}$), with 95\% intervals across replicas. Upper block: the group cutoff raised from 12 to 19.5~\AA; the coefficients change by less than 0.01. Lower block, at the 12~\AA\ cutoff: $z$ measured from 2, 3 and 4~\AA\ beyond the inner carbon plane instead of from the plane; the binding part changes sign while the complete excess does not move.}
\label{tab:excess}
\begin{tabular}{lccc}
\toprule
Quantity & $r_c = 12$~\AA & 16~\AA & 19.5~\AA \\
\midrule
binding part & $-0.340 \pm 0.020$ & $-0.347 \pm 0.019$ & $-0.349 \pm 0.018$ \\
density part & $+0.279 \pm 0.048$ & $+0.280 \pm 0.049$ & $+0.280 \pm 0.049$ \\
complete excess & $-0.062 \pm 0.064$ & $-0.067 \pm 0.064$ & $-0.069 \pm 0.063$ \\
complete, uniform $T$ & $-0.115 \pm 0.115$ & $-0.122 \pm 0.118$ & $-0.125 \pm 0.119$ \\
\midrule
origin at 2~\AA & $-0.162 \pm 0.016$ (binding) & $-0.070 \pm 0.053$ (complete) &  \\
origin at 3~\AA & $-0.072 \pm 0.015$ (binding) & $-0.075 \pm 0.048$ (complete) &  \\
origin at 4~\AA & $+0.017 \pm 0.014$ (binding) & $-0.079 \pm 0.044$ (complete) &  \\
\bottomrule
\end{tabular}
\end{table}

\sectionend{end:si-interface}
\FloatBarrier
\section{Hydraulic mobility, wall drift, stall pressure and slip}
\label{sec:si-mobility}

The isothermal body-force runs of tier E push the water along the channel with
\texttt{fix addforce}, which acts on the water alone. The water drags the wall, and
the Langevin friction that holds the wall balances that drag only at a steady drift
proportional to the force. In these runs the wall moves at about 8\% of the water
velocity, whereas under a thermal drive its drift is below
\SI{e-4}{\meter\per\second}. A mobility, a slip length and a stall condition concern
the water moving relative to its wall, and Table~\ref{tab:mobility} gives them in
that frame; the laboratory-frame mobility is shown for comparison.

\begin{table}[H]
\centering
\footnotesize
\caption{Hydraulic mobility of the patterned channel ($\xi$ = 0--0.20) from isothermal body-force runs. Water velocities (m\,s$^{-1}$, mean $\pm$ standard error over three replicas) in the laboratory frame and relative to the wall, whose drift is 486~m\,s$^{-1}$ per unit per-atom force (kcal\,mol$^{-1}$\,\AA$^{-1}$). Mobilities in m\,s$^{-1}$ per unit force. The stall pressure combines the wall-relative mobility with $|v_{\max}|$ and its 95\% $t$-interval across replicas (Table~\ref{tab:replicas}). Viscosity and slip come from the force-normalised, wall-relative velocity profiles pooled over all nine runs, with jackknife standard errors and the wall at the edge of the water density; the mobility route uses that viscosity with either fluid height.}
\label{tab:mobility}
\begin{tabular}{lccc}
\toprule
Force per atom & $v$, laboratory & wall drift & $v$, relative to wall \\
\midrule
$2\times10^{-5}$ & $0.1334 \pm 0.0081$ & $0.0097$ & $0.1236 \pm 0.0081$ \\
$5\times10^{-5}$ & $0.2877 \pm 0.0312$ & $0.0243$ & $0.2634 \pm 0.0312$ \\
$1\times10^{-4}$ & $0.6495 \pm 0.0100$ & $0.0485$ & $0.6009 \pm 0.0099$ \\
\midrule
mobility, laboratory frame & $6487 \pm 85$ & $R^2 = 0.990$ & intercept $+0.0027 \pm 0.0104$ \\
mobility, relative to the wall & $6001 \pm 85$ & $R^2 = 0.989$ & intercept $+0.0027 \pm 0.0104$ \\
\midrule
stall pressure (MPa) & $-1.18 \pm 1.29$ ($h$ = 23~\AA) & $-1.11 \pm 1.21$ ($h$ = 24.5~\AA) &  \\
viscosity, pooled profile & $0.68 \pm 0.16$ mPa\,s &  &  \\
slip length (nm) & $0.19 \pm 0.11$ (profile) & 0.10--0.13 (mobility) &  \\
\bottomrule
\end{tabular}
\end{table}

The sign of the stall pressure follows the rectification direction; its magnitude is
not resolved by the five tier A replicas.

\begin{figure}[H]
\centering
\includegraphics[width=0.5\textwidth,alt={Axial velocity of water relative to the wall across the channel in the body-force runs, with standard errors and a fitted parabola; dotted lines mark the wall positions used for the slip length.}]{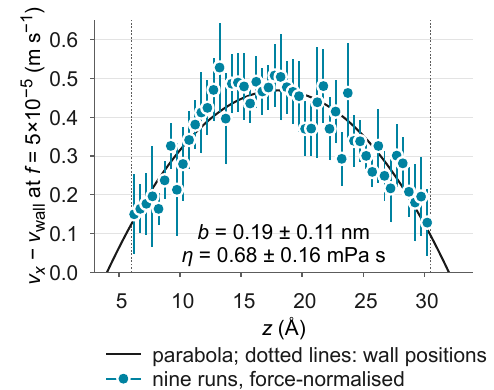}
\caption{Axial velocity of water relative to the wall across the channel in the
tier E body-force runs, each normalised to a per-atom force of
$5\times10^{-5}$~kcal\,mol$^{-1}$\,\AA$^{-1}$ and averaged over nine runs (standard
errors), with the fitted parabola. Dotted lines mark the wall positions used for the
slip length, at the edge of the water density.}
\label{fig:profile}
\end{figure}

The slip length is taken from the velocity profile itself
(Figure~\ref{fig:profile}). A single 10~ns profile does not constrain the curvature
of the parabola, so the profiles of all nine runs are divided by their force, taken
relative to the wall, averaged and fitted, and the uncertainty is a jackknife over
runs. The fitted viscosity is consistent with the value for TIP4P/2005 near
\SI{320}{\kelvin}. With the wall placed at the edge of the water density, the slip
length is a fraction of a nanometre; because moving that reference by an
{\aa}ngstr{\"o}m changes the slip length by a comparable amount, it is best read as
an upper bound of about \SI{0.2}{\nano\meter}, two orders of magnitude below the
slip lengths of the coverage mapping.

\sectionend{end:si-mobility}
\FloatBarrier
\section{Conductance at uniform coverage}
\label{sec:si-conductance}

The linear-response model maps hydroxyl coverage to slip, and hence to hydraulic
conductance, and the direction it predicts depends on how strongly the conductance
is modulated across the pattern relative to the thermo-osmotic coefficient. The
conductance was therefore measured on walls of uniform coverage with the same body
force as tier E (Table~\ref{tab:conductance}). Pristine graphene slips tens of
nanometres and conducts two orders of magnitude more than any hydroxylated wall,
and the three hydroxylated walls are indistinguishable.

\begin{table}[H]
\centering
\footnotesize
\caption{Isothermal body-force runs on walls of uniform hydroxyl coverage $\xi$ (three replicas of 10~ns, per-atom force $5\times10^{-5}$~kcal\,mol$^{-1}$\,\AA$^{-1}$). Water velocity relative to the wall with its 95\% $t$-interval; conductance relative to $\xi = 0.20$ with its standard error; Navier slip length and viscosity from the pooled, wall-relative velocity profiles (jackknife standard errors), with the wall at the edge of the water density. The pristine profile is plug-like and its curvature, hence its viscosity and slip, poorly determined; from the conductance ratio by $K \propto 1 + 3b/H$ its slip is 72~nm. On the hydroxylated walls the no-slip plane lies slightly inside the water. At $\xi = 0$ the mobility at $1\times10^{-5}$ is 1.11 $\pm$ 0.16 times that at $5\times10^{-5}$, so the friction is still linear. The patterned channel of tier E ($\xi$ = 0--0.20) conducts 1.18 (95\%: 0.56--2.57) times the $\xi = 0.20$ surface.}
\label{tab:conductance}
\begin{tabular}{ccccc}
\toprule
$\xi$ & $v - v_{\mathrm{wall}}$ (m\,s$^{-1}$) & $K/K(0.20)$ & slip (nm) & $\eta$ (mPa\,s) \\
\midrule
0.00 & $39.66 \pm 0.33$ & $177.38 \pm 21.40$ & $45 \pm 21$ & $0.45 \pm 0.17$ \\
0.10 & $0.219 \pm 0.131$ & $0.98 \pm 0.18$ & $-0.10 \pm 0.04$ & $0.60 \pm 0.04$ \\
0.20 & $0.224 \pm 0.116$ & 1 & $-0.08 \pm 0.09$ & $0.78 \pm 0.61$ \\
0.30 & $0.244 \pm 0.067$ & $1.09 \pm 0.15$ & $-0.08 \pm 0.01$ & $0.58 \pm 0.15$ \\
\bottomrule
\end{tabular}
\end{table}

With so large a pristine conductance, the way $K$ falls between $\xi = 0$ and 0.10
decides what a local model predicts, and no run lay in that gap: unless the pristine
slip is confined to coverage below about 1\%, a local conductance gives the
pre-registered direction, opposite to the one measured
(Table~\ref{tab:collapse}, computed before the gap was filled). The patterned
channel's own conductance is too uncertain to select among those interpolations. Its
small slip length, below \SI{0.2}{\nano\meter} (Table~\ref{tab:mobility}), follows
from the local conductances adding in series (Section~\ref{sec:si-stokes}) and so
does not show whether the near-pristine strips slip.

\begin{table}[H]
\centering
\footnotesize
\caption{What a local conductance implies, depending on how $K(\xi)$ falls between $\xi = 0$ and 0.10, where no run exists: the pristine excess $K(0) - K(0.10)$ is taken to vanish linearly by the coverage in the first column (0.10 is linear interpolation through the measurements). $m_R$ is the chemistry-locked first-harmonic modulation of $M/K$ over the $\xi$ = 0--0.20 pattern, with $M$ from the model's mapping, and decides the direction the local model predicts; the last two columns give the patterned channel's conductance that a local $K$ predicts (the harmonic mean over the pattern, relative to $\xi = 0.20$) and the measured value divided by it. Intervals are 95\% parametric bootstraps over the replica means. This sensitivity was added after the data and is not a pre-registered test.}
\label{tab:collapse}
\begin{tabular}{ccccc}
\toprule
$K$ excess gone by $\xi$ & $m_R$ [95\%] & direction & local series $K$ & measured/local [95\%] \\
\midrule
0.10 & $+1.21$ [$+$1.05, $+$1.37] & pre-registered & $1.95$ & $0.61$ [0.31, 1.16] \\
0.05 & $+0.75$ [$+$0.46, $+$1.03] & pre-registered & $1.47$ & $0.80$ [0.42, 1.57] \\
0.02 & $+0.39$ [$+$0.05, $+$0.74] & pre-registered & $1.24$ & $0.95$ [0.50, 1.91] \\
0.01 & $+0.23$ [$-$0.13, $+$0.60] & undecided & $1.15$ & $1.02$ [0.53, 2.04] \\
0.005 & $+0.12$ [$-$0.24, $+$0.51] & undecided & $1.10$ & $1.07$ [0.56, 2.15] \\
\bottomrule
\end{tabular}
\end{table}

Three coverages inside that gap were therefore run, chosen to be the ones the strips
actually carry: counting hydroxyls per \SI{5}{\angstrom} slice along the pattern, the
strictly hydroxyl-free region is 0.5 to \SI{1}{\nano\meter} wide, and the rest of the
least-covered strip sits at $\xi \approx 0.006$ to 0.05. The slip collapses steeply
across them but not completely (Table~\ref{tab:lowcoverage}), and carrying the
measured $b(\xi)$ into the exact Stokes flow leaves the direction reversed
(Table~\ref{tab:direction}): every interval of $V_K/V_M$ lies below $-1$. That is
the second of the three readings registered before the runs, and it excludes the
collapse of the strips' slip as the explanation of the measured direction.

\begin{table}[H]
\centering
\footnotesize
\caption{Conductance of uniform walls at the coverages the least-covered strips carry. The three realised coverages in bold were added to test the near-pristine strips; the others repeat Table~\ref{tab:conductance} on the same isothermal body-force design, three replicas of 10~ns at $5\times10^{-5}$~kcal\,mol$^{-1}$\,\AA$^{-1}$. The graft quota is rounded per 5~\AA\ slice, so the nominal 0.005, 0.02 and 0.05 realise one, three and eight hydroxyls per slice. Velocities are relative to the wall with 95\% $t$-intervals across replicas; the slip length comes from the conductance ratio by $K \propto 1 + 3b/H$ with $b(0.20) = 0$, the same relation at every coverage. One hydroxyl per slice removes 96\% of the pristine slip and eight remove 99.7\%, but at the strips' own coverages the slip is still a fifth to twice the half-height $H = 1.23$~nm. Neither pre-registered sanity condition fails: the conductance nowhere rises with coverage beyond its interval, and no low-coverage velocity exceeds the pristine one.}
\label{tab:lowcoverage}
\begin{tabular}{cccc}
\toprule
realised $\xi$ & $v - v_{\mathrm{wall}}$ (m\,s$^{-1}$) & $K/K(0.20)$ & $b$ from $K$ (nm) \\
\midrule
0.00 & $39.66 \pm 0.33$ & $177.38$ & $72.0$ \\
\textbf{0.0062} & $1.794 \pm 0.546$ & $8.02$ & $2.87$ \\
\textbf{0.0188} & $0.651 \pm 0.156$ & $2.91$ & $0.78$ \\
\textbf{0.0500} & $0.350 \pm 0.028$ & $1.57$ & $0.23$ \\
0.10 & $0.219 \pm 0.131$ & $0.98$ & $-0.01$ \\
0.20 & $0.224 \pm 0.116$ & 1 & $0.00$ \\
0.30 & $0.244 \pm 0.067$ & $1.09$ & $0.04$ \\
\bottomrule
\end{tabular}
\end{table}

\begin{table}[H]
\centering
\footnotesize
\caption{Does the measured local slip give the observed direction? $V_K/V_M$ is the ratio of the flow driven by the slip modulation to the thermo-osmotic flow, from the exact Stokes solution with the model's $M$ modulation, pooled over the ten reversal structures with 289 bootstrap draws over the conductances and the hydroxyl placements. A ratio above $-1$ would give the observed direction; every interval lies entirely below it. Converged in the slice sampling (15 and 45 placements per slice give $-7.59$ and $-7.58$). 176 of the 289 draws returned a negative slip for the hydroxylated wall, which the solver cannot take, and were clipped to zero; the tier E case, whose reference slip is positive, clips 37 and agrees.}
\label{tab:direction}
\begin{tabular}{lccc}
\toprule
local slip map & $V_K/V_M$ & 95\% interval & direction \\
\midrule
realised per-slice step (primary) & $-7.59$ & [$-9.30$, $-5.24$] & reversed \\
linear between slices & $-7.85$ & [$-9.54$, $-5.58$] & reversed \\
smooth design profile & $-7.89$ & [$-9.56$, $-5.65$] & reversed \\
\midrule
primary, $b(0.20) = 0.19$~nm & $-7.61$ & [$-9.43$, $-5.53$] & reversed \\
\bottomrule
\end{tabular}
\end{table}

\sectionend{end:si-conductance}
\section{Sensitivity to the carbon--water interaction}
\label{sec:si-sigma}

The carbon--water Lennard-Jones $\sigma_{\mathrm{CO}}$ of \SI{3.28}{\angstrom} used
throughout was fitted rather than mixed; the geometric mean of the carbon and water
values is \SI{3.349}{\angstrom}. The two extrema were repeated at the geometric mean on
ten arrangements of the phase sweep -- the five of the first test and five added to
resolve it -- keeping every other parameter, including the number of water molecules,
unchanged, and paired arrangement by arrangement against the production value
(Table~\ref{tab:sigma}). Five readings were fixed before the second set of runs
(robust in direction and magnitude; direction robust; sign change; suppressed;
otherwise unresolved), and the one that applies is the last of them: unresolved at
ten arrangements, with the limitation kept as written. Six of ten keep their sign,
against the eight, with the ratio of mean amplitudes inside [1/2,~2], that the
robustness criterion requires; the reversal at the geometric mean,
\SI{+0.017+-0.017}{\meter\per\second}, does not clear zero ($p = 0.053$); and the
paired difference, \SI{-0.005+-0.031}{\meter\per\second} ($p = 0.74$), resolves
neither a suppression nor a sign change. Doubling the arrangements tightened that
difference's interval by a factor of 1.9 without moving any threshold, which is why
the manuscript points to a free-energy calibration of $\sigma_{\mathrm{CO}}$
rather than to more arrangements. The water count, not recalibrated, shifts the
pressure by a few tens of bar at most.

\begin{table}[H]
\centering
\footnotesize
\caption{Reversal amplitude (m\,s$^{-1}$) on ten phase-sweep arrangements with the production carbon--water $\sigma_{\mathrm{CO}} = 3.28$~\AA\ and with the geometric-mean value 3.349~\AA, paired by build seed: the five arrangements of the first test and the five added to resolve them. The ratio of mean amplitudes is 0.78 with a bootstrap 95\% interval of [$0.07$, $3.38$]; 7 of ten paired differences and 8 of ten amplitudes at the geometric mean are positive. The larger $\sigma_{\mathrm{CO}}$ changes the core water density from 0.9160 to 0.9155~g\,cm$^{-3}$, a pressure change of $-13$~bar at the TIP4P/2005 compressibility.}
\label{tab:sigma}
\begin{tabular}{cccc}
\toprule
build seed & 3.28~\AA & 3.349~\AA & difference \\
\midrule
974000 & $+0.0827$ & $-0.0052$ & $-0.0879$ \\
974001 & $+0.0054$ & $+0.0094$ & $+0.0040$ \\
974002 & $+0.0107$ & $+0.0332$ & $+0.0224$ \\
974003 & $+0.0173$ & $+0.0242$ & $+0.0069$ \\
974004 & $+0.0318$ & $-0.0295$ & $-0.0613$ \\
974005 & $+0.0197$ & $+0.0494$ & $+0.0297$ \\
974006 & $+0.0199$ & $+0.0354$ & $+0.0155$ \\
974007 & $+0.0462$ & $+0.0048$ & $-0.0413$ \\
974008 & $-0.0095$ & $+0.0387$ & $+0.0482$ \\
974009 & $-0.0101$ & $+0.0073$ & $+0.0173$ \\
\midrule
mean, 95\% & $+0.0214 \pm 0.0197$ & $+0.0168 \pm 0.0171$ & $-0.0046 \pm 0.0314$ \\
\bottomrule
\end{tabular}
\end{table}

\sectionend{end:si-sigma}
\FloatBarrier
\section{Hydrodynamics of the patterned channel}
\label{sec:si-stokes}

The least-covered strips are two to three nanometres wide, and their
hydroxyl-free cores narrower still, against a pristine slip length of tens of
nanometres, so the local (lubrication) description of the loop flow was
checked against the exact Stokes flow of a periodic slit with slip varying along
the wall. With the stream function odd about the mid-plane and no net pressure drop
around the loop, a Fourier mode of wavenumber $k$ in the wall shear rate $C_k$
produces a wall velocity $\gamma_k C_k$ with
\begin{equation}
\gamma_k = \frac{\sinh(2kH)/2 - kH}{2k\sinh^2(kH)},
\end{equation}
which tends to $H/3$ for $kH \to 0$ (lubrication) and to $1/2k$ for $kH \to \infty$
(the unconfined stripe limit), $H$ being the half-height. The Navier condition with
local slip $b(x)$ and thermo-osmotic slip velocity $u_{\mathrm{s}}(x)$ then reads
\begin{equation}
V + \sum_{k \neq 0} C_k\,\bigl[\gamma_k + b(x_j)\bigr]\,e^{ikx_j} = u_{\mathrm{s}}(x_j),
\end{equation}
imposed at 1601 collocation points per period for the mean loop velocity $V$ and the
$C_k$. The solver reproduces uniform slip exactly, slip-corrected Poiseuille flow,
the lubrication model at long period, and the shear-free stripe results of
Philip~\cite{Philip1972} and of Lauga and Stone~\cite{LaugaStone2003} in the
unconfined and thin-channel limits, and converges in
the number of modes. The hydrodynamic walls are placed at the edges of the water
density in the tier E profile.

Table~\ref{tab:stokes} gives the direction and the patterned conductance for each
way the slip can fall between $\xi = 0$ and 0.10; Table~\ref{tab:direction} repeats
the test with that fall measured rather than assumed. The channel is close to the
lubrication limit, the slip contrast reverses the direction in every case, and the
predicted patterned conductance is consistent with the measured one throughout.
Amplification of the drive by $(1 + b/\ell)$ restores the observed direction for
$\ell$ below $\ell^*$. Table~\ref{tab:periodamp} shows that it also makes the flow per
period fall with $N$, which the tier C sweep does not show.

\begin{table}[H]
\centering
\scriptsize\setlength{\tabcolsep}{3.2pt}
\caption{Exact two-dimensional Stokes flow in the patterned channel ($\xi$ = 0--0.20, $\delta = \lambda/4$, $h = 2.45$~nm between hydrodynamic walls) with local Navier slip $b(\xi)$, every value taken from the measured conductance by $K \propto 1 + 3b/H$ (72~nm on pristine graphene for $b_{\mathrm{cov}} = 0$), falling linearly to $b_{\mathrm{cov}}$ by the coverage in the second column. $V_K/V_M$ is the loop velocity driven by the slip contrast acting on the mean thermo-osmotic drive, relative to that driven by the modulation of $M$ in the model's mapping; below $-1$ the pre-registered direction results. The lubrication column is the local model. $K_{\mathrm{pattern}}$ is the predicted pressure-driven conductance relative to uniform $\xi = 0.20$ (measured: 1.18, 95\%: 0.56--2.57) and $b_{\mathrm{eff}}$ the uniform slip that gives the same conductance. $\ell^*$ is the depth of the enthalpy excess below which slip amplification $(1 + b/\ell)$ of the drive gives the observed direction, with the model's contrast in $M$ and with the measured intrinsic contrast ($+2.4\%$, range over its 95\% interval). With the hydrodynamic walls 2.30~nm apart $V_K/V_M$ lies between $-12.8$ and $-1.1$; over all cases $K_{\mathrm{pattern}}$ lies between 1.05 and 1.85. This calculation was made after the data.}
\label{tab:stokes}
\begin{tabular}{cccccccc}
\toprule
$b_{\mathrm{cov}}$ (nm) & $b$ gone by $\xi$ & $V_K/V_M$ & lubrication & $K_{\mathrm{pattern}}$ & $b_{\mathrm{eff}}$ (nm) & $\ell^*$, model (nm) & $\ell^*$, measured (nm) \\
\midrule
0.00 & 0.10 & $-12.8$ & $-12.7$ & $1.74$ & $0.30$ & $0.44$ & $0.36$ [0.35, 0.38] \\
0.00 & 0.05 & $-8.8$ & $-10.3$ & $1.36$ & $0.15$ & $0.46$ & $0.36$ [0.34, 0.38] \\
0.00 & 0.02 & $-3.9$ & $-5.9$ & $1.16$ & $0.06$ & $0.56$ & $0.35$ [0.32, 0.38] \\
0.00 & 0.01 & $-2.0$ & $-3.7$ & $1.09$ & $0.03$ & $0.82$ & $0.33$ [0.29, 0.38] \\
0.00 & 0.005 & $-1.0$ & $-2.4$ & $1.05$ & $0.02$ & $>1$ & $0.30$ [0.24, 0.38] \\
\bottomrule
\end{tabular}
\end{table}

\FloatBarrier
\begingroup
\scriptsize\renewcommand{\arraystretch}{0.9}\setlength{\tabcolsep}{3.2pt}
\begin{longtable}{ccccccccc}
\caption{Period-number dependence of the amplified flow: loop velocity per period at $N$ periods relative to $N = 4$, from the exact Stokes flow with local slip ($b_{\mathrm{cov}} = 0$) and a drive amplified by $(1 + b/\ell)$, for the model's contrast in $M$ and for measured contrasts over $+2.4 \pm 2.5\%$. Uniform slip gives 1 at every $N$. The last column is the rise in $\chi^2$ of the global test when the tier C predictions are multiplied by these factors, against 12.3 for 22 degrees of freedom; both fits have two parameters, and the comparison was made after the data.}\label{tab:periodamp}\\
\toprule
contrast & $\ell$ (nm) & $b$ gone by $\xi$ & $N = 2$ & 3 & 5 & 8 & 10 & $\Delta\chi^2$ \\
\midrule
\endfirsthead
\multicolumn{9}{l}{\small\itshape Table~\ref{tab:periodamp}, continued} \\
\toprule
contrast & $\ell$ (nm) & $b$ gone by $\xi$ & $N = 2$ & 3 & 5 & 8 & 10 & $\Delta\chi^2$ \\
\midrule
\endhead
\bottomrule
\endlastfoot
mapping & $0.10$ & 0.10 & $1.13$ & $1.07$ & $0.93$ & $0.73$ & $0.60$ & $4.5$ \\
mapping & $0.10$ & 0.02 & $1.29$ & $1.14$ & $0.88$ & $0.61$ & $0.49$ & $6.5$ \\
mapping & $0.26$ & 0.10 & $1.19$ & $1.10$ & $0.89$ & $0.54$ & $0.33$ & $9.1$ \\
mapping & $0.26$ & 0.02 & $1.28$ & $1.14$ & $0.88$ & $0.58$ & $0.44$ & $7.3$ \\
mapping & $0.36$ & 0.10 & $1.36$ & $1.20$ & $0.78$ & $0.08$ & $-0.37$ & $24.8$ \\
mapping & $0.36$ & 0.02 & $1.27$ & $1.14$ & $0.87$ & $0.54$ & $0.37$ & $8.5$ \\
$+0.0\%$ & $0.10$ & 0.10 & $1.13$ & $1.07$ & $0.93$ & $0.71$ & $0.59$ & $4.7$ \\
$+0.0\%$ & $0.10$ & 0.02 & $1.31$ & $1.15$ & $0.87$ & $0.57$ & $0.45$ & $7.3$ \\
$+0.0\%$ & $0.26$ & 0.10 & $1.24$ & $1.13$ & $0.86$ & $0.42$ & $0.15$ & $12.8$ \\
$+0.0\%$ & $0.26$ & 0.02 & $1.44$ & $1.22$ & $0.80$ & $0.34$ & $0.11$ & $14.1$ \\
$+2.4\%$ & $0.10$ & 0.10 & $1.13$ & $1.07$ & $0.93$ & $0.71$ & $0.58$ & $4.8$ \\
$+2.4\%$ & $0.10$ & 0.02 & $1.32$ & $1.16$ & $0.86$ & $0.56$ & $0.43$ & $7.6$ \\
$+2.4\%$ & $0.26$ & 0.10 & $1.26$ & $1.14$ & $0.85$ & $0.37$ & $0.07$ & $14.4$ \\
$+2.4\%$ & $0.26$ & 0.02 & $1.53$ & $1.26$ & $0.76$ & $0.20$ & $-0.08$ & $18.6$ \\
$+4.9\%$ & $0.10$ & 0.10 & $1.13$ & $1.07$ & $0.93$ & $0.71$ & $0.58$ & $4.9$ \\
$+4.9\%$ & $0.10$ & 0.02 & $1.33$ & $1.16$ & $0.86$ & $0.55$ & $0.41$ & $7.9$ \\
$+4.9\%$ & $0.26$ & 0.10 & $1.28$ & $1.15$ & $0.83$ & $0.31$ & $-0.02$ & $16.5$ \\
$+4.9\%$ & $0.26$ & 0.02 & $1.70$ & $1.34$ & $0.69$ & $-0.06$ & $-0.42$ & $27.0$ \\
\end{longtable}
\endgroup

\label{sec:si-end}

\renewcommand{\bibfont}{\small}
\setlength{\bibsep}{2pt}
\bibliographystyle{achemso}
\bibliography{manuscript-references}